\documentclass[12pt,authoryear,review]{elsarticle}

\usepackage{hyperref}
\usepackage{textcomp}
\usepackage{mathtools}
\usepackage{lineno}
\usepackage{booktabs}
\usepackage{epstopdf}
\usepackage{subcaption}
\usepackage{amsfonts}
\usepackage{bm}
\usepackage{tikz}
\usetikzlibrary{shapes,arrows}
\usepackage[space]{grffile}
\definecolor{darkred}{rgb}{0.55, 0.0, 0.0}
\definecolor{darkblue}{rgb}{0.0, 0.0, 0.55}

\usepackage[margin=2.0cm]{geometry}

\newcommand{\figref}[1]{Fig.~\ref{#1}}
\newcommand{\figsref}[1]{Figs.~\ref{#1}}
\newcommand{\tabref}[1]{Table~\ref{#1}}
\renewcommand{\eqref}[1]{Eq.~\ref{#1}}
\newcommand{\eqsref}[1]{Eqs.~\ref{#1}}
\newcommand{\secref}[1]{Sec.~\ref{#1}}
\newcommand{\secsref}[1]{Secs.~\ref{#1}}
\renewcommand{\d}{\,\Delta}					
\renewcommand{\vec}[1]{\bm{\mathrm{#1}}}	
\newcommand{\mat}[1]{\bm{\mathrm{#1}}}		

\journal{International Journal of Solids and Structures}

\begin{document}

\graphicspath{{./figs/}}
\epstopdfsetup{outdir=./figs/}

\begin{frontmatter}

\title{Elastic wave propagation in dry granular media: effects of probing characteristics and stress history}

\author[addr1]{Hongyang Cheng\corref{cor}}
\cortext[cor]{Corresponding author}
\ead{h.cheng@utwente.nl}

\author[addr1]{Stefan Luding}
\ead{s.luding@utwente.nl}

\author[addr2]{Kuniyasu Saitoh}
\ead{kuniyasu.saitoh.c6@tohoku.ac.jp}

\author[addr1]{Vanessa Magnanimo}
\ead{v.magnanimo@utwente.nl}

\address[addr1]{Multiscale Mechanics (MSM), Faculty of Engineering Technology, MESA+, University of Twente, P.O. Box 217, 7500 AE Enschede, The Netherlands}
\address[addr2]{Research Alliance Center for Mathematical Sciences (RACMaS), Tohoku University, Japan}

\begin{abstract}

Elastic wave propagation provides a noninvasive way to probe granular materials.
The discrete element method using particle configuration as input, allows a micromechanical interpretation on the acoustic response of a given granular system.
This paper compares static and dynamic numerical probing methods, from which wave velocities are either deduced from elastic moduli or extracted from the time/frequency-domain signals.
The dependence of wave velocities on key characteristics, i.e., perturbation magnitude and direction for static probing, and maximum travel distance and inserted signals for dynamic probing, is investigated.
It is found that processing the frequency-domain signals obtained from dynamic probing leads to reproducible wave velocities at all wavenumbers, irrespective of the perturbation characteristics, whereas the maximum travel distance and input signals for the time domain analysis have to be carefully chosen, so as to obtain the same long-wavelength limits as from the frequency domain.
Static and dynamic probes are applied to calibrated representative volumes of glass beads, subjected to cyclic oedometric compression.
Although the perturbation magnitudes are selected to reveal only the elastic moduli, the deduced wave velocities are consistently lower than the long-wavelength limits at various stress states, and thus sensitive to sample size.
While the static probes investigate the influence of stress history on modulus degradation, dynamic probing offers insights about how dispersion relations evolve during cyclic compression.
Interestingly, immediately after each load reversal the incremental behavior is reversibly elastoplastic, until it becomes truly elastic with further unload/reload.
With repeating unload/reload, the P- or S-wave dispersion relations become increasingly scalable with respect to their long-wavelength limits.

\end{abstract}


\end{frontmatter}

\section{Introduction}

Understanding elastic wave propagation in granular media is essential for many industrial, geotechnical and geophysical applications, such as oil exploration \citep{Batzle1992,Kelder1997}, geophysical tomography \citep{Daily1983,Justice1988} and non-destructive inspection of composite materials \citep{Diamanti2005,Kim2015}.
Key mechanisms involved are the nonlinear, stress- and fabric-dependent elasticity and dispersion of granular materials, fabric anisotropy and strain softening/hardening \citep{Santos2011,Mouraille2006,Pal2013,Waymel2017}.
In the last decades, the characterization of elastic properties of granular materials mostly relied on laboratory experiments, such as resonant column and triaxial testing (e.g., \cite{LoPresti1997,Santamarina1996}), in which the samples are probed with very small strains ranging from $ 10^{-4}\% $ to $ 10^{-1}\% $.
However, determining the appropriate ranges of strain probes for the rock/sand samples at various states of stress and void ratio is time-consuming and tedious \citep{CLAYTON2011}.
Bender elements made of piezoelectric materials \citep{Shirley1978} has recently gained popularity in non-destructive testing of granular materials like weak rocks and soils.
The use of bender elements is now a well-established technique both for academic research and engineering applications.
Nevertheless, the interpretation of received signals from bender elements still remains elusive, and many questions regarding the nature of the wave propagation are still unanswered \citep{Arroyo2003,Lee2005,Arroyo2006,Camacho-Tauta2015,Alvarado2011}.

The increasing computational power allows individual solid grains to be modeled, as well as their motions and the interactions among neighbors.
The nonlinear, history-dependent, elastoplastic behavior of granular materials, which is rooted in the micromechanics at contacts and irreversible rearrangements of the microstructure, can be reproduced by the discrete element method (DEM) \citep{Cundall1979}.
Virtual static/dynamic probing with the help of DEM is convenient, because i) static probes with strain levels unreachable in resonant column/triaxial tests are possible; ii) the complete space-time evolution of particle motion is available for dynamic analysis, in contrast to bender element experiments in which, often, the signal is received by a single transducer near the surface.
Moreover, in DEM simulations, it is possible to investigate the propagation of elastic waves at wavelengths close to the particle sizes, so that the dispersion and attenuation properties, etc. can be better understood.
In previous works, DEM simulations of static and dynamic probing were conducted for studying the relationship between small-strain moduli and strain magnitudes \citep{Gu2017,Kumar2014} and the frequency-dependent wave propagation/attenuation \citep{Mouraille2006,Mouraille2006a,ODonovan2016}, respectively.
The former is relevant to more explore the elastoplastic constitutive behavior of granular materials \citep{Sibille2007a, Magnanimo2008b,Kumar2014, Nicot2014}, whereas the latter helps to investigate the dispersive properties \citep{Merkel2017}.

In this work, the DEM simulations of both static and dynamic probing are performed to measure the wave velocities of a DEM representative volume at stress states over a cyclic oedometric stress path.
The packing configuration is reconstructed from 3D X-ray tomography images of a glass bead sample \citep{Cheng2017};
the micromechanical parameters are inferred from the stress--strain behavior of the sample using a sequential Bayesian inference scheme \citep{Cheng2018a,Cheng2017a}.
From the methodological point of view, the goal is to understand the suitability of static and dynamic probing for estimating  elastic wave velocities, including the sensitivity to key perturbation characteristics, such as strain magnitude and inserted wave frequency.
Both the time and frequency domain techniques are employed to investigate the dependence of wave velocities on the maximum wavelength allowed for a virtual sample, and the input waveform and frequency of wave signals.
The wave velocities resulting from static probing and dynamic probing are compared at various stress states along the cyclic oedometric path.
Moreover, the effect of stress history on the small-strain moduli and dispersive relations of the  modeled granular material is discussed.

The remainder of this paper is organized as follows.
\secref{sec:DEM} introduces the basics of DEM, including the contact laws in \secref{sec:contactLaws} and the macroscopic quantities in \secref{sec:macro}.
\secref{sec:calib} briefly explains the calibration approach and the stress path.
The static and dynamic probing methods and the key perturbation characteristics are detailed in \secref{sec:probe}, and the effects of the respective probing methods are investigated in \secsref{sec:staModuli} and \ref{sec:dynaModuli}.
\secref{sec:compareProbs} show the comparison of the evolutions of wave velocities estimated in three different ways during the cyclic oedometric loading.
Conclusions are drawn in \secref{sec:conclude}.

\section{DEM modeling}
\label{sec:DEM}

The open-source DEM package YADE \citep{Smilauer2015} is applied to perform 3D DEM simulations of dense granular materials.
DEM represents granular materials as packings of solid particles with simplified geometries (e.g., spheres) and vanishingly small interparticle overlaps.
Governed by springs, dashpots and sliders upon collision, the kinematics of the particles are updated within the explicit time integration scheme, based upon the net forces and moments resulting from interparticle forces.
The time step in DEM simulations is kept sufficiently small in order to resolve collisions between contacting particles and ensure that the interparticle forces on each particle is contributed only from its neighbors.
For DEM modeling of quasistatic shear, a background damping is generally adopted to stabilize the numerical simulations.

\subsection{Contact laws}
\label{sec:contactLaws}

The interparticle forces between two contacting solid particles, i.e., $ \vec{F}_c = \vec{F}_n + \vec{F}_s$, can be described by contact level force--displacement laws in normal and tangential directions, as defined in \eqsref{eq:fdLawn}--\ref{eq:fdLaws} \citep{Cheng2016d,Cheng2017a}.
To mimic the role of surface roughness without complicating the particle geometry, rolling/twisting resistance is typically needed in addition to the normal and tangential stiffnesses.
Both interparticle tangential forces and contact moments are bounded by Coulomb type yield criteria.
For two contacting spheres with a normal overlap $ \vec{u}_n $, a relative tangential displacement $ \d\vec{u}_s $ and a relative rotational angle $ \vec{\theta}_c $ at the contact point, the interparticle normal force $ \vec{F}_n $, tangential force $ \d\vec{F}_s $ and contact moment $ \vec{M}_c $ are calculated as
\begin{eqnarray}
	\vec{F}_n = \frac{2E_c\vec{u}_n}{3(1-\nu_c^2)}\sqrt{R^*|\vec{u}_n| }
	\label{eq:fdLawn}\\
	\d\vec{F}_s = \frac{2E_c\d\vec{u}_s}{(1+\nu_c)(2-\nu_c)}\sqrt{R^*|\vec{u}_n| }
	\label{eq:fdLaws}\\
	|\vec{F}_s| \leq \tan{\mu} |\vec{F}_n|\\
	\label{eq:coulomb}
	\vec{M}_c = k_m \vec{\theta}_c\\
	\label{eq:k_m}
	|\vec{M}_c| \leq \eta_m |\vec{F}_n|(R_1+R_2)/2
	\label{eq:eta_m}
\end{eqnarray}
where $ E_c $ and $ \nu_c $ are the Young's modulus and Poisson's ratio of the solid particles, $ \mu $ is the interparticle friction angle, $R^*$ is the equivalent radius defined as $1/(1/R_1+1/R_2) $, $ R_1 $ and $ R_2 $ are the radii of the two spherical particles in contact, $ k_m $ is the rolling stiffness, and $ \eta_m $ is the rolling friction coefficient, which controls the plastic limit of $\vec{M}_c $.

\subsection{Macroscopic quantities}
\label{sec:macro}

Starting from the contact network defined by the microstructural configuration, the effective stress tensor $ \mat{\sigma}' $ can be computed as
\begin{equation}
\mat{\sigma}' = \frac{1}{V} \sum_{c \in N_c} \vec{F}_c \otimes \vec{d}_c
\label{eq:love}
\end{equation}
where $ V $ is the total volume of the granular packing including the pore space,
$ N_c $ is the total number of contacts;
$ \vec{d}_c $ is the branch vector connecting the centers of the particles.
From the effective stress tensor $ \mat{\sigma}' $, the mean effective stress $ p' $ and the deviatoric stress $ q $ are obtained as $\frac{1}{3}\text{tr}(\mat{\sigma}')$ and $\sqrt{\frac{3}{2}\mat{\sigma}'_{dev} : \mat{\sigma}'_{dev}}$ respectively, where $\mat{\sigma}'_{dev}$ is the traceless part of $\mat{\sigma}'$.


\section{Bayesian calibration against stress--strain response}
\label{sec:calib}

\begin{figure} [htp!]
	\centering
	\includegraphics[height=0.4\textwidth]{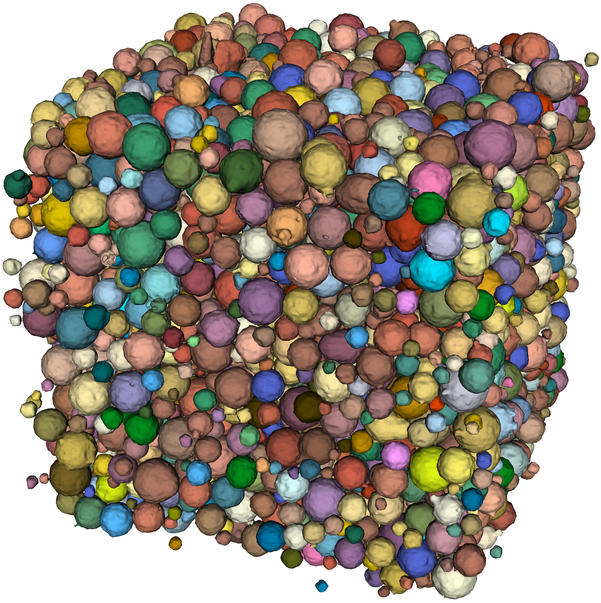}
	\caption{Reconstruction of individual glass beads in the representative volume of the glass bead specimen.}
	\label{fig:isosurf}
\end{figure}

The granular packing modeled here is made of spherical glass beads with diameters ranging from 40 to 80 $ \mathrm{\mu m} $.
A glass bead specimen is initially confined in a triaxial cell under a hydrostatic pressure of 5 MPa, at which 3D X-ray computed tomography images (3DXRCT) are captured.
A cyclic oedometric stress path at a strain rate of $ 2.0 \times 10^{-4}\%\cdot\text{s}^{-1}$ follows, with axial strain $ \varepsilon_a $ increasing from 0\% to 1.75\% and then varying between 1.75\% and approximately 1.0\% for two cycles.
To facilitate the packing generation, a core of the glass bead specimen is extracted from the 3DXRCT images (see \figref{fig:isosurf}), and the estimated particle positions and radii are used as the initial guess for the packing configuration.

After the packing has been generated, the relevant contact parameters are retrieved using a newly developed iterative Bayesian filter \citep{Cheng2018a,Cheng2017a}.
The calibration is conducted against the stress--strain response obtained from the oedometric experiment on the glass bead specimen.
The micromechanical parameters relevant to the oedometric stress path are contact level Young's modulus $ E_c $, interparticle friction $ \mu $, rolling stiffness $ k_m $ and rolling friction $ \eta_m $, as in \eqsref{eq:fdLawn}--\ref{eq:eta_m}.
The DEM simulations are strain-controlled in a quasistatic manner, following the same axial strain increments as in the experiment.
During each incremental loading, a global damping ratio of 0.2 is adopted.
The ratio is raised to 0.9 during the subsequent relaxation stage, in order to dissipate kinetic energy and extract the quasistatic macroscopic quantities.
The time step $ \d{t} $ is fixed to 1/10 of the critical time step \citep{Smilauer2015}.

\begin{figure} [b!]
	\centering
	\includegraphics[width=\textwidth]{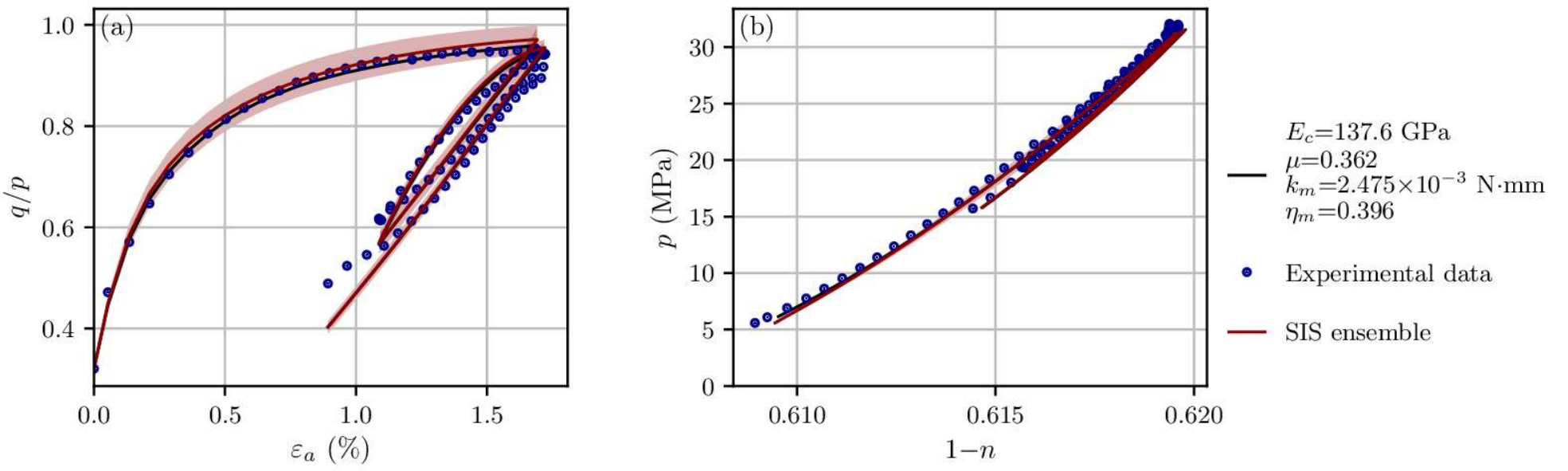}
	\caption{Comparison of experimental data and numerical predictions given by the sequential importance sampling (SIS) ensemble and the parameter sets associated to the top three highest posterior probabilities. The red lines and the red shaded areas indicate respectively the weight-averaged ensemble predictions and an uncertainty range of $ \pm 2 \sigma$.}
	\label{fig:IPs}
\end{figure}

The iterative Bayesian filter allows to search for the optimal sets of the parameters from low to high resolution in the parameter space.
After a sufficient number of iterations, an excellent agreement between the experimental results and the numerical predictions is obtained, as shown in \figref{fig:IPs}.
Interested readers are referred to \citep{Cheng2018a,Cheng2017a} for details of the iterative Bayesian filter.

\section{Static and dynamic probing}
\label{sec:probe}

Two probing methods are employed here for estimating  the elastic responses of granular materials using DEM.
Static probing consists in perturbing a DEM packing with an uniform strain field, and computing the resultant elastic modulus after stress relaxation \citep{Magnanimo2008b,Makse2004}.
Dynamic probing, on the other hand, is performed by agitating one side of the DEM packing, and then recording the motion of individual particle, while the wave travels across the packing \citep{Lemrich2017a,Otsubo2017,Gu2018}.
In this work, the effect of the probing methods on the prediction of the elastic response is investigated by varying the key perturbation characteristics.
The goal is to assess the predictive capability of the probing methods as well as the computational efficiency.

\subsection{Static probing: calculating small-strain moduli}
\label{sec:staProb}

In order to calculate the moduli, small-strain perturbations are applied to the equilibrated DEM packing at the stress states belonging to the cyclic oedometric stress path.
Specifically, shear strain increments $ \d \varepsilon_{xz} $ and $ \d \varepsilon_{yz} $ are applied for measuring the shear moduli and $ \d \varepsilon_{zz} $ for the oedometric modulus, where $ z $ is the axial direction and $ x $ and $ y $ are the radial directions.
After perturbation, a long relaxation follows until the granular packing recovers equilibrium \cite{Magnanimo2008b,Kumar2014}.
The resulting stress components before and after the probing can be obtained from \eqref{eq:love}, and subsequently the stress increments $ \d \tau_{xz} $, $ \d \tau_{yz} $ and $ \d \sigma_{zz} $ calculated.
The small-strain shear moduli $ G_{xz} $ and $ G_{yz} $ and the compressional modulus $ M_z $ are given by
\begin{equation}
G_{xz} = \frac{\d \tau_{xz}}{2\d \varepsilon_{xz}}, \quad G_{yz} = \frac{\d \tau_{yz}}{2\d \varepsilon_{yz}}, \quad \text{and} \quad M_z = \frac{\d \sigma_{zz}}{\d \varepsilon_{zz}},
\label{eq:staModuli}
\end{equation}
where the magnitude of strain increments ($\xi = |\pm\d \varepsilon_{xz}|, |\pm\d \varepsilon_{yz}|, |\pm\d \varepsilon_{zz}|$) range from $10^{-8}$ to $10^{-4}$.

One of the well-known questions pertaining to static probing is how small $ \xi $ should be for a DEM packing at a given stress state.
The appropriate strain increment that keeps the DEM packing with the true elastic regime is generally unknown beforehand, and depends not only on the void ratio and fabric anisotropy of the packing, but the stress history as well.
To identify the limits of the strain increments, one typically scans the small-strain moduli resulting from a wide range of $ \xi $.
The first plateau during which the moduli remain constant, defines the elastic regime for the DEM packing.
Within those \emph{small} $ \xi $, $ G_{xz} $, $ G_{yz} $ and $ M_z $ can be readily deduced to the shear (S)- and compressional (P)-wave velocities $ v_s $ and $ v_p $ in the long wavelength limit.
Note that strain increments are applied, with both positive and negative signs, i.e., in two opposite directions at each stress state, with the goal to understand the dependence of the elastic regime/moduli on the perturbation direction.

\subsection{Dynamic probing: propagating elastic waves in long granular columns}
\label{sec:dynaProbe}

Instead of converting wave velocities from elastic moduli, the physical process of mechanical waves propagating in a granular medium can be directly simulated using DEM.
In this type of simulations, after one end of the granular packing is agitated, the space-time evolution of the states of the constituent particles or force chains is analyzed in the time- or frequency-domain to infer the propagation velocity.
While less computational time is needed in the dynamic approach, in comparison with the static one, the packing length has to be sufficiently long for these simulations, so that the long-wavelength behavior can be reproduced.
The major difference between dynamic and static probing is that the transient response is used in the former, whereas only the steady states, before and after relaxation, are considered in the latter.
For dynamic probing, the key perturbation characteristics difficult to determine are the input waveform and frequency of the inserted wave, rather than perturbation magnitudes and directions as for static probing.

\subsubsection{Preparation of long granular columns}
\label{sec:columnLens}

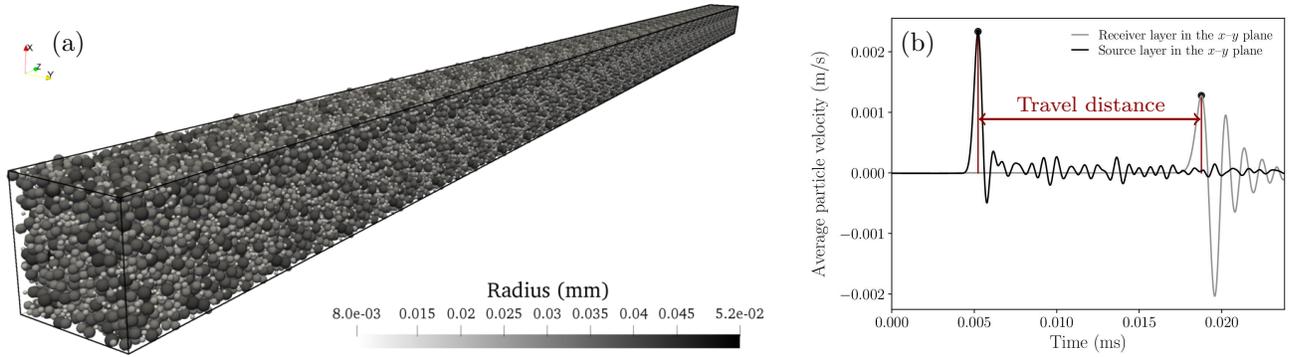
\begin{figure} [htp!]
	\centering
	\begin{subfigure}{0.6\textwidth}
		\begin{tikzpicture}
			\node (image) {\includegraphics[width=\textwidth]{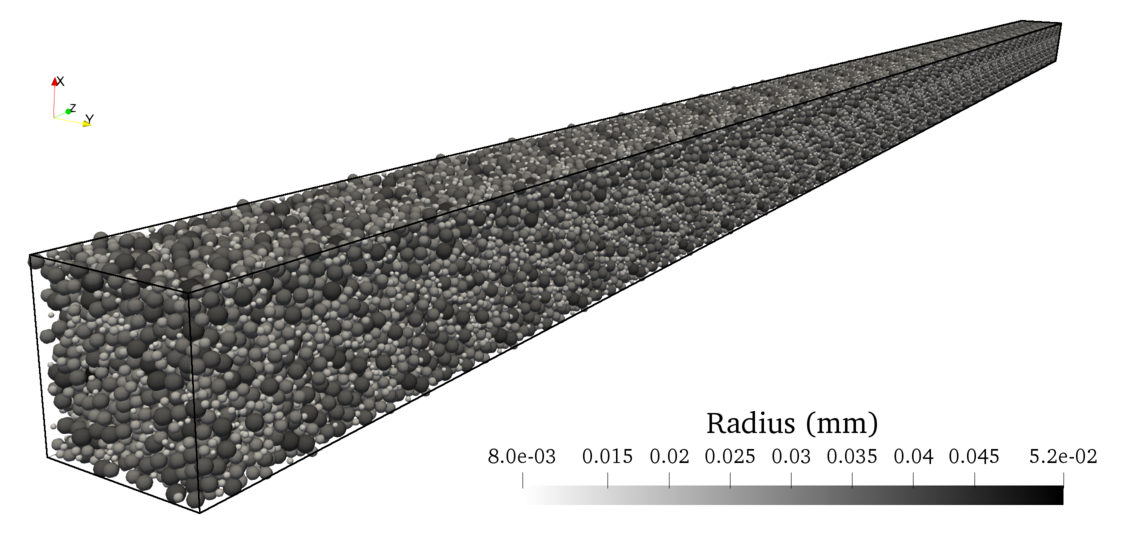}};
			\begin{scope}
			\node at (-4.2,1.85) {\footnotesize{(a)}};
			\end{scope}
		\end{tikzpicture}		
	\end{subfigure}
	\begin{subfigure}{0.39\textwidth}
		\begin{tikzpicture}
			\node (image) {\includegraphics[width=\textwidth]{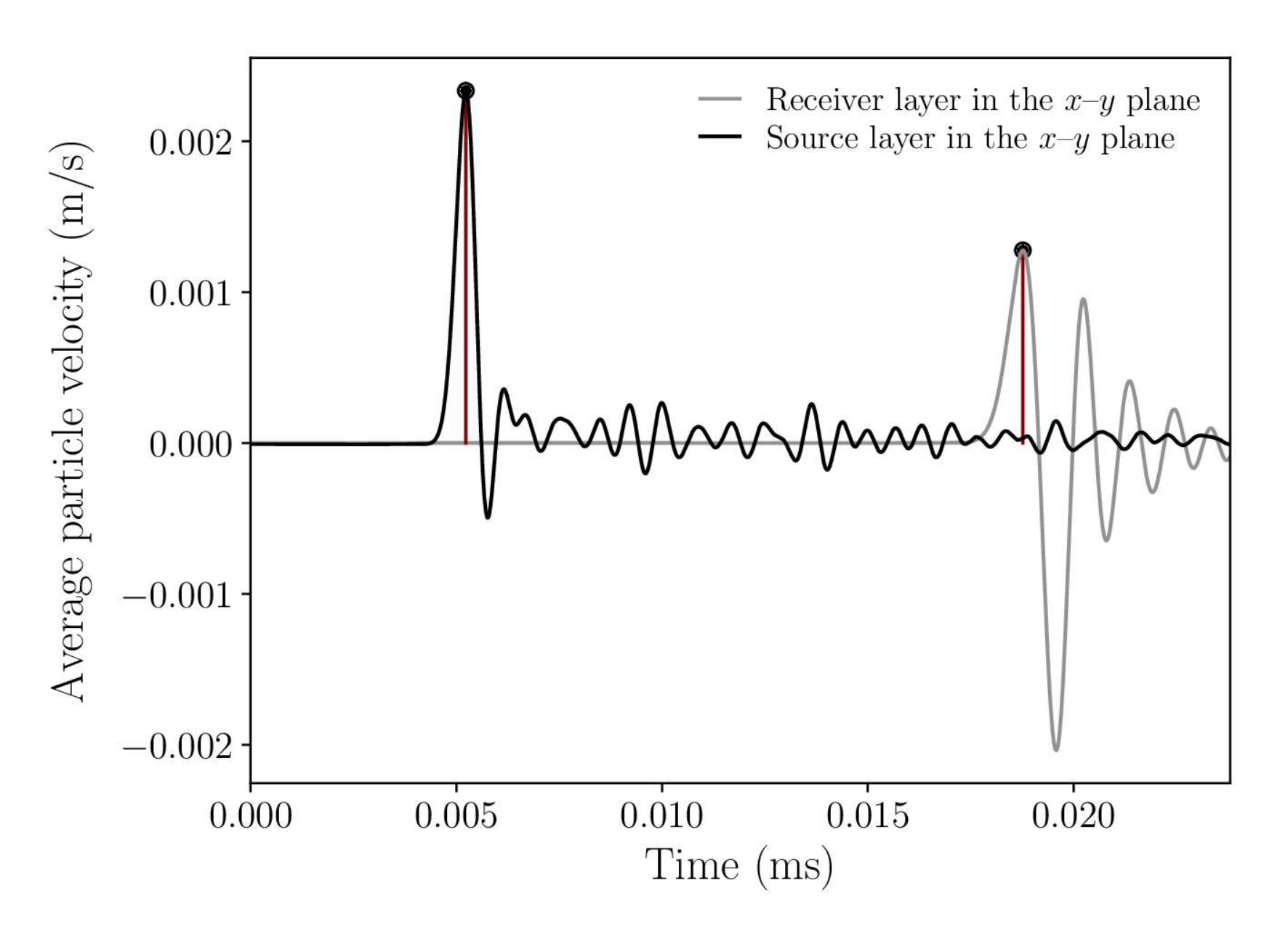}};
			\begin{scope}
			\node at (-1.75,1.85) {\footnotesize{(b)}};
			\draw[darkred,thick,<->] (-0.94,0.85) -- (2.01,0.85);
			\node[darkred] at (0.55,1.05) {\scriptsize{Travel distance}};
			\end{scope}
		\end{tikzpicture}
	\end{subfigure}
	\caption{DEM simulation of dynamic probing: (a) a long granular column generated by stacking 33 copies of the calibrated representative volume, along the propagation direction ($z$ axis), and (b) determination of travel time by the peak-to-peak method.}
	\label{fig:long3}
\end{figure}

In order to allow for long-wavelength propagation, the DEM packing calibrated in \secref{sec:calib}, referred to as the representative volume (RV), is copied and stacked one after another to create a long granular column as shown in \figref{fig:long3}.
To assure consistency, the force networks at the interfaces between neighboring RVs are restored, based upon the periodicity of the RV boundaries.
This technique allows to create large packings from RVs, with the consistent constitutive behavior and no discontinuities at the interfaces.
For the current analyses, the long granular columns are constructed with copies of the calibrated RV stacked along the $ z $ axis, as shown in \figref{fig:long3}.
Although not shown here, a preliminary study shows that, for the propagation of plane elastic waves, copying the RV along the other two axes $ x $ and $ y $ lead to negligible difference in the space-time evolution of particle velocities.

\subsubsection{Wave agitation at the source}
\label{sec:inputSigs}

The plane waves are agitated by perturbing the first $x$--$y$ layer of particles at the left end of the granular column, as shown schematically in \figref{fig:long3}.
For each $x$--$y$ layer of the column, a width equal to the mean particle diameter, $\bar{d}=45.916 \mathrm{\mu m}$, is considered.
P- and S-wave pulses are applied by moving the particles within the first $x$--$y$ layer from their original positions $ \vec{r}_i^0 $ to new positions $ \vec{r}_i $, along the longitudinal and transverse directions, respectively.
Depending on the input waveform and frequency, the perturbation persists for various time steps $t_n$, emitting a variety of single-period pulses.

\subsubsection{Evaluation of wave velocity in time and frequency domains}
\label{sec:evalWaves}

In order to calculate the wave velocity via the dynamic approach, the transient motion of individual particles in space and time are recorded.
The elastic wave velocities are then inferred from the signals, e.g., particle velocities, in either the time domain or the frequency domain.
The time domain analysis involves the determination of the time needed for a wave to travel from a \emph{source} to \emph{receivers} at various locations.
The wave velocities can be readily estimated as the ratios of the source-receiver distances to the travel times.
Time domain techniques are widely applied for both bender element experiments \citep{Gu2015,Camacho-Tauta2015,DaFonseca2009} and numerical simulations \citep{Mouraille2006,Zamani2011,ODonovan2016,Otsubo2017}.
In experiments, only a few number of transducers or bender/extender elements are attached to the surface of the specimen, which makes the interpretation of wave signals a difficult task.
However, in DEM simulations, particle motions at the degrees of freedom are available at every particle position and time step, which means that each particle can be regarded as a receiver.
As mentioned in \secref{sec:inputSigs}, in this work, the \emph{source} is chosen to be the first $x$--$y$ layer of particles.
For the time domain analyses, the \emph{receivers} are the remaining $x$--$y$ layers, with the widths equal to the mean particle diameter.
Given a receiver layer, at each time step the linear velocities of the particles within the layer are averaged according to the momentum conservation law.
The variation of averaged particle velocity in the granular column is monitored over time, as shown schematically in \figref{fig:long3}b.
The change in the average velocity of the receiver layer defines the arrival and further propagation of the wave.
The peak-to-peak method \citep{ODonovan2015} is employed to determine the travel time in the time domain.
In turn, the wave velocity is given by the travel time $ t_a $ and the source-receiver distance $ L $, as $ v_p $ or $ v_s = L/t_a $, depending on if a P- or S-wave is monitored.

Alternatively, the P- and S-wave velocities can be extracted from the respective dispersion relations, i.e., the angular frequency $\omega$ or the wave velocity (phase velocity) $ \omega/k $ as a function of wavenumber $k$ \citep{Mouraille2006,Saitoh2018}.
From the time evolution of particle velocity at each position $\{\vec{v}_i(t)\}$ $(i=1,..., N)$ with $ N $ the number of particles, the Fourier transforms are calculated as
\begin{equation}
\vec{v_k}(t) = \sum_{i=1}^{N}\vec{v}_i(t)e^{-I\vec{k}\vec{r}_i(t)},
\label{eq:dft}
\end{equation}
where $ \vec{r}_i(t) $ is the position of each particle and $ I $ the imaginary unit.
By introducing a unit wave vector $ \hat{\vec{k}}\equiv\vec{k}/|\vec{k}| $ parallel to the propagation direction, i.e., $\hat{\vec{k}}=\{0,0,1\}^{\textnormal{T}}$, the longitudinal and transverse particle velocities in the wavenumber-time domain are defined as
\begin{equation}
\vec{v_k}^{\parallel}(t) = (\vec{v_k}(t)\cdot\hat{\vec{k}})\hat{\vec{k}} \quad \textnormal{and} \quad
\vec{v_k}^{\perp}(t) = \vec{v_k}(t)-\vec{v_k}^{\parallel}(t).
\label{eq:LTdecompos}
\end{equation}
The Fourier transforms of $ \vec{v_k}^{\parallel}(t) $ and $ \vec{v_k}^{\perp}(t) $, namely,
\begin{equation}
\vec{\tilde{v}_k^{\parallel}}(\omega) = \int_{0}^{\infty}\vec{v_k}^{\parallel}(t)e^{I\omega t}\textnormal{d}t \quad \textnormal{and} \quad \vec{\tilde{v}_k^{\perp}}(\omega) = \int_{0}^{\infty}\vec{v_k}^{\perp}(t)e^{I\omega t}\textnormal{d}t
\label{eq:2dft}
\end{equation}
give the power spectra of longitudinal and transverse particle velocities, $ S_l(k,\omega) = |\vec{\tilde{v}_k^{\parallel}}(\omega)|^2 $ and $ S_t(k,\omega) = |\vec{\tilde{v}_k^{\perp}}(\omega)|^2 $, with $ k\equiv|k| $.
The P-wave and S-wave dispersion branches can be identified from the power spectra $ S_l(k,\omega) $ and $ S_l(k,\omega) $, respectively.
The long-wavelength wave velocity $ v_p^0 $ or $ v_s^0 $ is defined as the slope $\lim_{k\to0}\omega/k $ and the group velocity $ \d \omega / \d k $.
In contrast to the time domain analysis, neither local averaging of particle motions within $x$--$y$ receiver layers nor the determination of signal peaks is needed in the frequency domain analysis.
In the following sections, we will describe and compare the results from both static and dynamic probing, and both time and frequency domain techniques will be employed to analyze the space-time data from dynamic probing.

\section{Small-strain moduli along the cyclic oedometric path}
\label{sec:staModuli}

\begin{figure} [t!]
	\begin{subfigure}{0.5\textwidth}
		\centering
		\includegraphics[width=5.2cm]{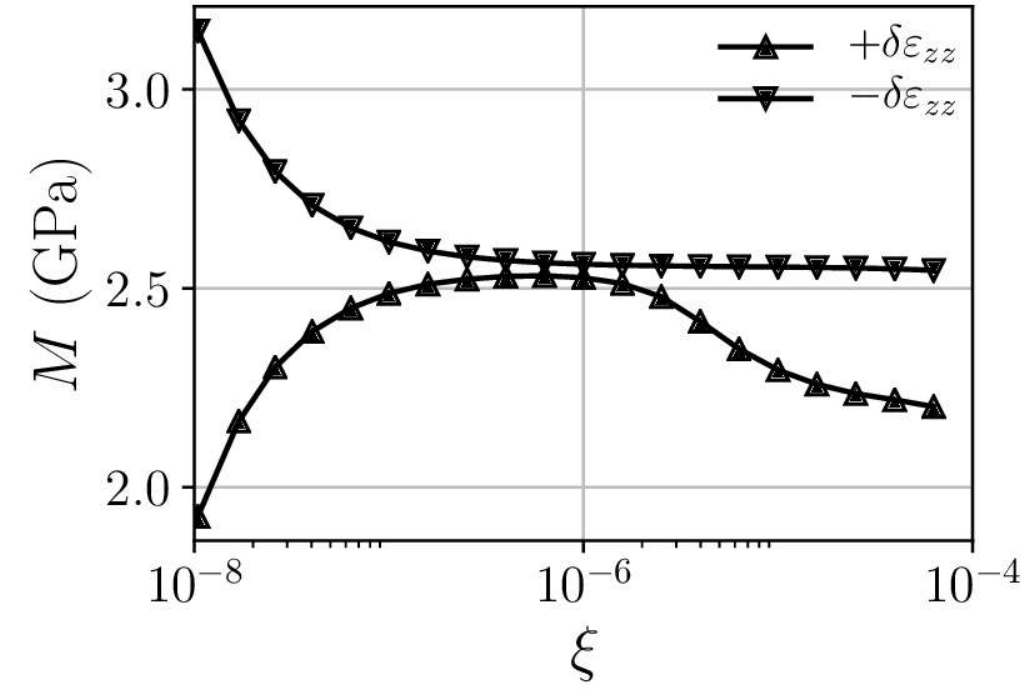}
		\caption{Oedometric modulus}
		\label{fig:M_5MPa}
	\end{subfigure}
	\begin{subfigure}{0.5\textwidth}
		\centering
		\includegraphics[width=5.2cm]{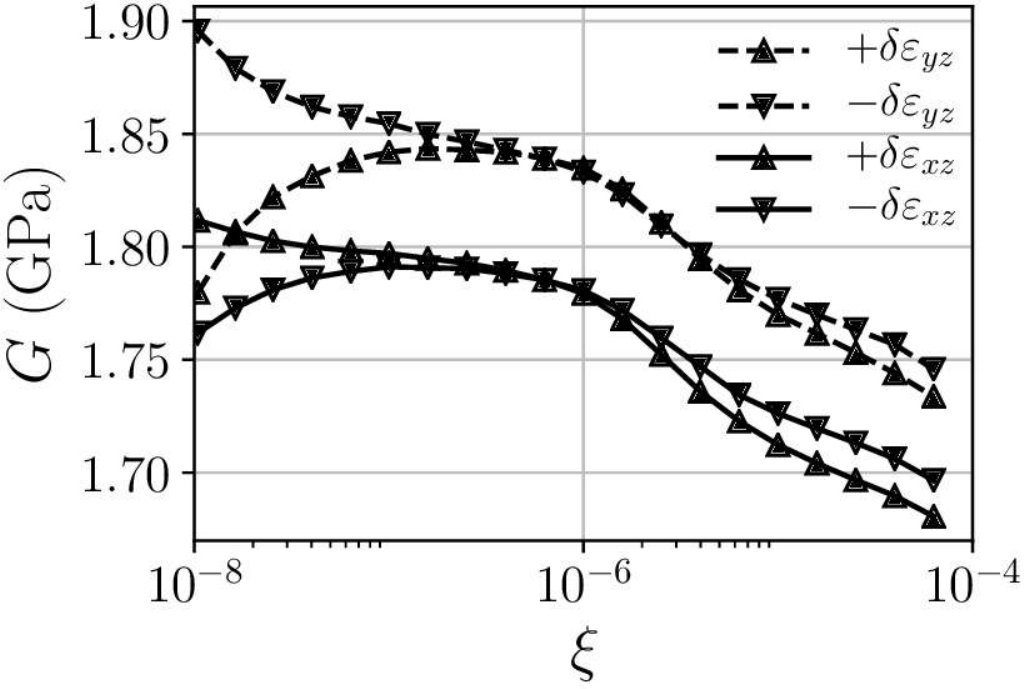}
		\caption{Shear moduli}
		\label{fig:G_5MPa}
	\end{subfigure}
	\caption{Small-strain moduli of the DEM packing at $ \varepsilon_a=0.11\%$, probed along opposite directions at various perturbation magnitudes.}
	\label{fig:perbMag}
\end{figure}

Using the procedure described in \secref{sec:staProb}, the elastic moduli $ M_z $ and $ G_{xz} $, $ G_{yz} $ are extracted for several states along the oedometric path in \figref{fig:IPs}.
For each state, the moduli are evaluated at a wide range of perturbation magnitudes $ \xi $, which results in the so-called degradation curves.
\figref{fig:perbMag} shows the degradation curve for the DEM representative packing at $ \varepsilon_a=0.11\%$.
Moreover, strain increments are applied in both　``positive'' and ``negative'' loading directions, which means, specifically, for the oedometric modulus $ M_z $, the packing is loaded ``forward'' and ``backward'' along the oedometric path.
The results from the positive (forward) and negative (backward) probing are plotted in \figref{fig:perbMag}.

It can be observed that the elastic regimes, suggested by the first plateau, lies in between $ 10^{-7} $ and $ 10^{-6} $ for both the oedometric modulus and the shear moduli .
The oedometric modulus $ M_z $ obtained from backward probing remains almost constant for $\xi> 10^{-6.8}$, while the backward modulus decreases after $\xi$ succeeds $10^{-6}$ (see \figref{fig:M_5MPa}).
This is associated to the previous oedometric compression experienced by the packing in the same direction as the forward probing $ \d\varepsilon_{zz} $.
The difference between the shear moduli obtained with the positive and negative shear strain increments is less significant, as shown in \figref{fig:G_5MPa}.
The values of the two shear moduli $ G_{xz} $ and $ G_{yz} $ are very close within the current range of $\xi$, which confirms that the granular packing under oedometric compression is transverse isotropic.
Therefore, only the shear modulus $G_{xz} $ is presented in the following, along with $ M_zz $, in order to show the variation of the small-strain moduli with respect to stress history.

\begin{figure} [t!]
	\begin{subfigure}{0.33\textwidth}
		\centering
		\includegraphics[width=6cm]{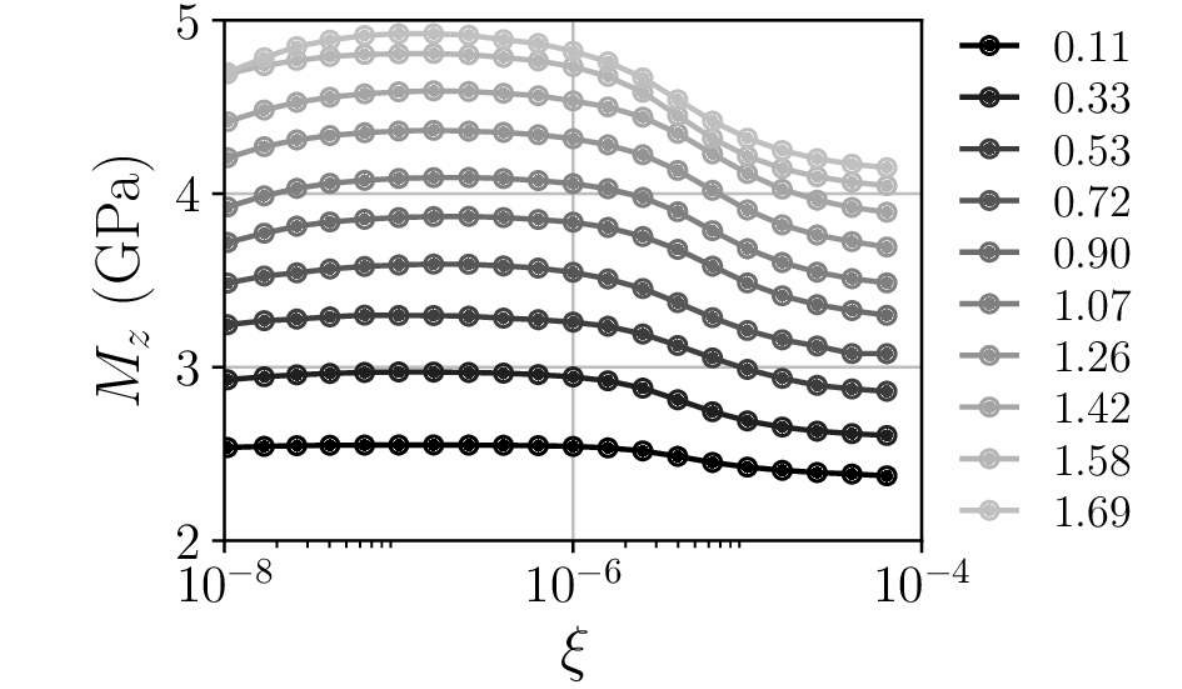}
		\caption{$ M_z $: first load}
		\label{fig:M1stLoad}
	\end{subfigure}
	\begin{subfigure}{0.33\textwidth}
		\centering
		\includegraphics[width=6cm]{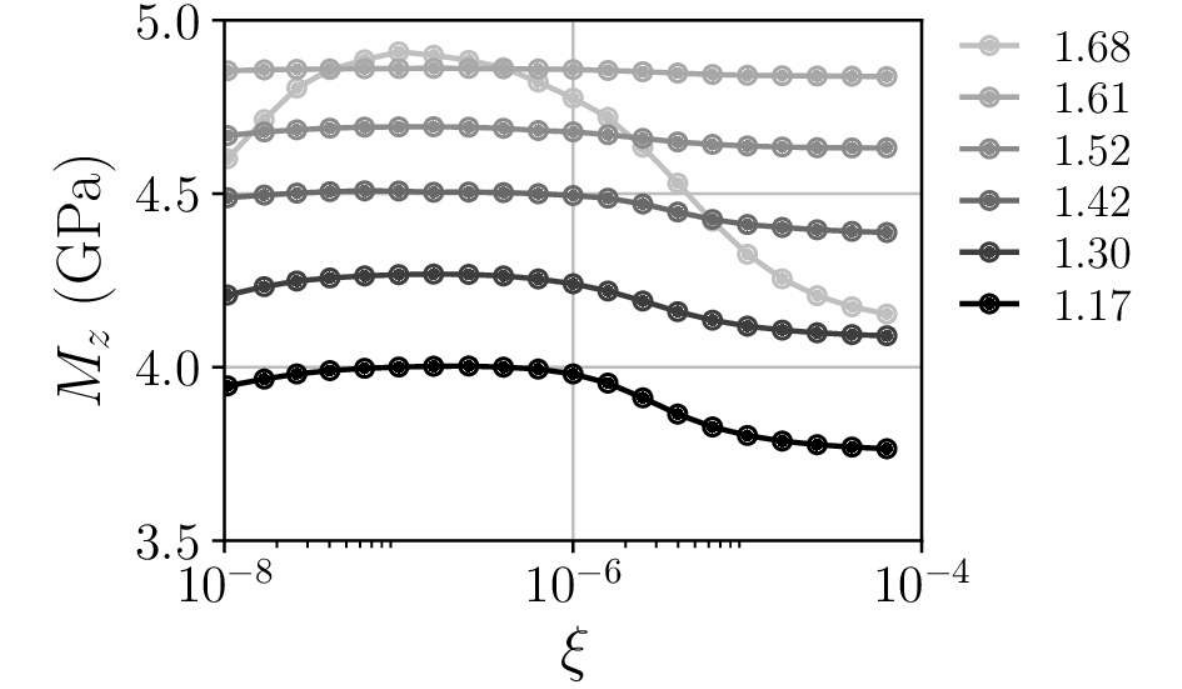}
		\caption{$ M_z $: first unload}
		\label{fig:M1stUnload}
	\end{subfigure}
	\begin{subfigure}{0.33\textwidth}
		\centering
		\includegraphics[width=6cm]{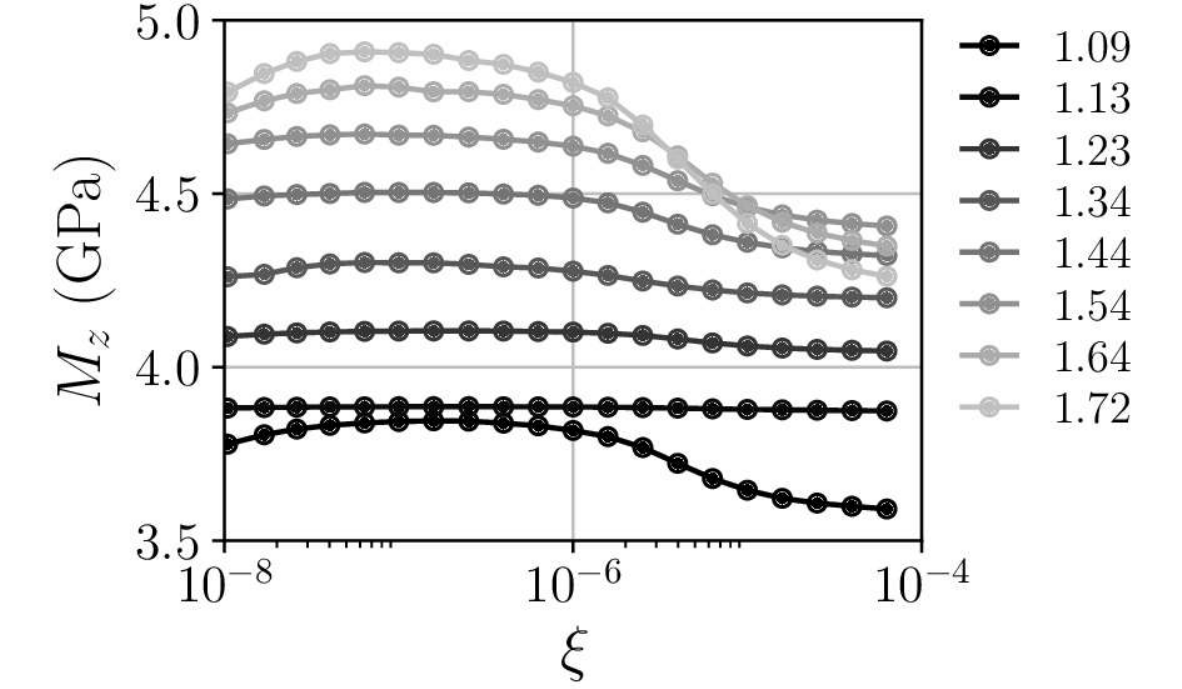}
		\caption{$ M_z $: first reload}
		\label{fig:MsecondLoad}
	\end{subfigure} \\
	\begin{subfigure}{0.33\textwidth}
		\centering
		\includegraphics[width=6cm]{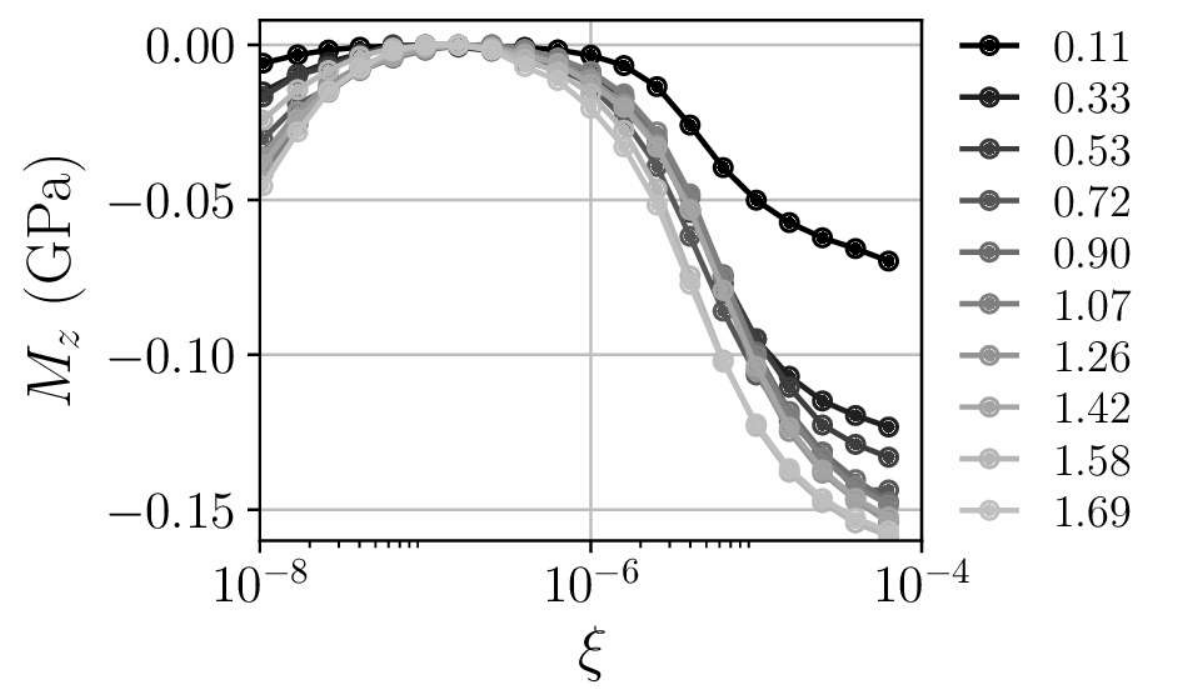}
		\caption{$\frac{\Delta M_z}{M_z^e} $: first load}
		\label{fig:M1stLoadScaled}
	\end{subfigure}
	\begin{subfigure}{0.33\textwidth}
		\centering
		\includegraphics[width=6cm]{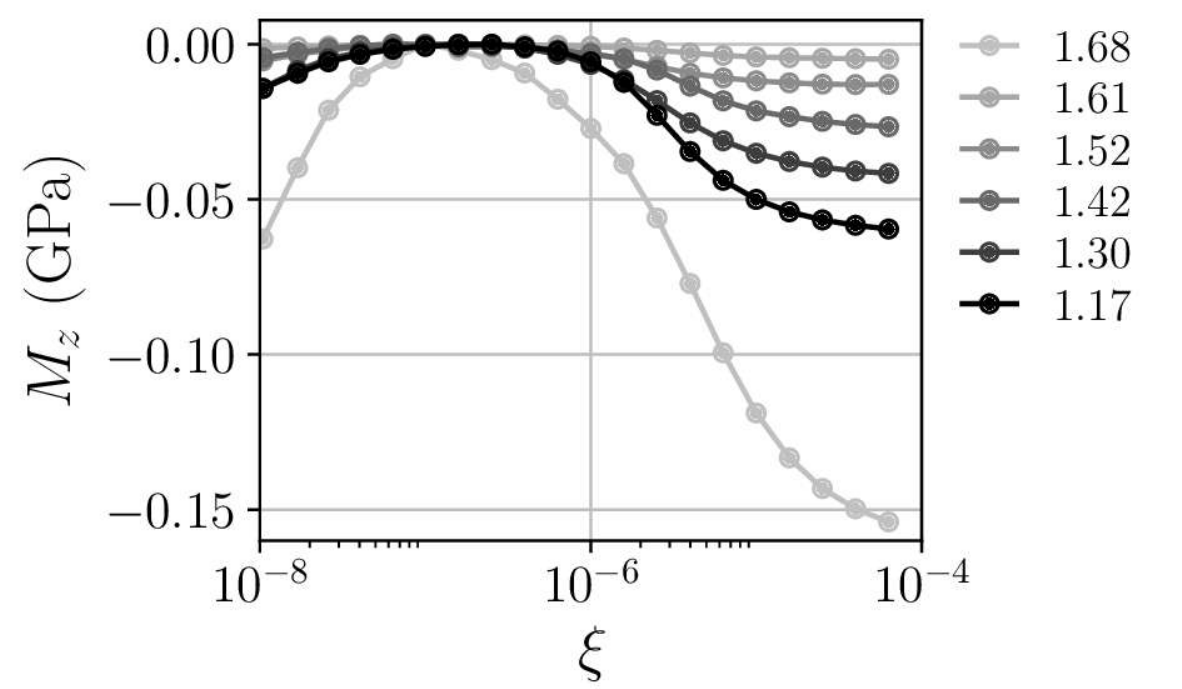}
		\caption{$ \frac{\Delta M_z}{M_z^e} $: first unload}
		\label{fig:M1stUnloadScaled}
	\end{subfigure}
	\begin{subfigure}{0.33\textwidth}
		\centering
		\includegraphics[width=6cm]{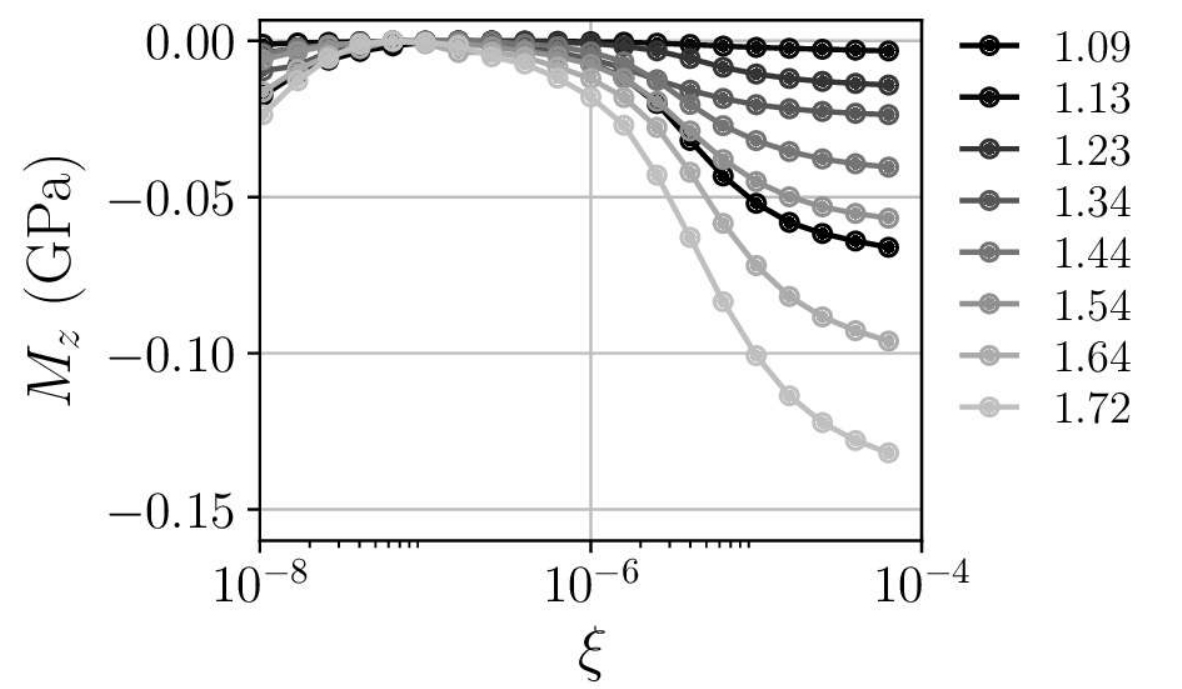}
		\caption{$ \frac{\Delta M_z}{M_z^e} $: first reload}
		\label{fig:MsecondLoadScaled}
	\end{subfigure}
	\caption{Small-strain oedometric moduli $ M_z $ (a--c) and the normalized degradation $\frac{\Delta M_z}{M_z^e}$ (d--f) as functions of the perturbation magnitude $\xi$ during the oedometric load-unload cycles. Different levels of axial strain $\varepsilon_a$ in percentage are indicated by the different gray levels.}
	\label{fig:M22CycOedo}
\end{figure}

\begin{figure} [htp!]
	\begin{subfigure}{0.33\textwidth}
		\centering
		\includegraphics[width=6cm]{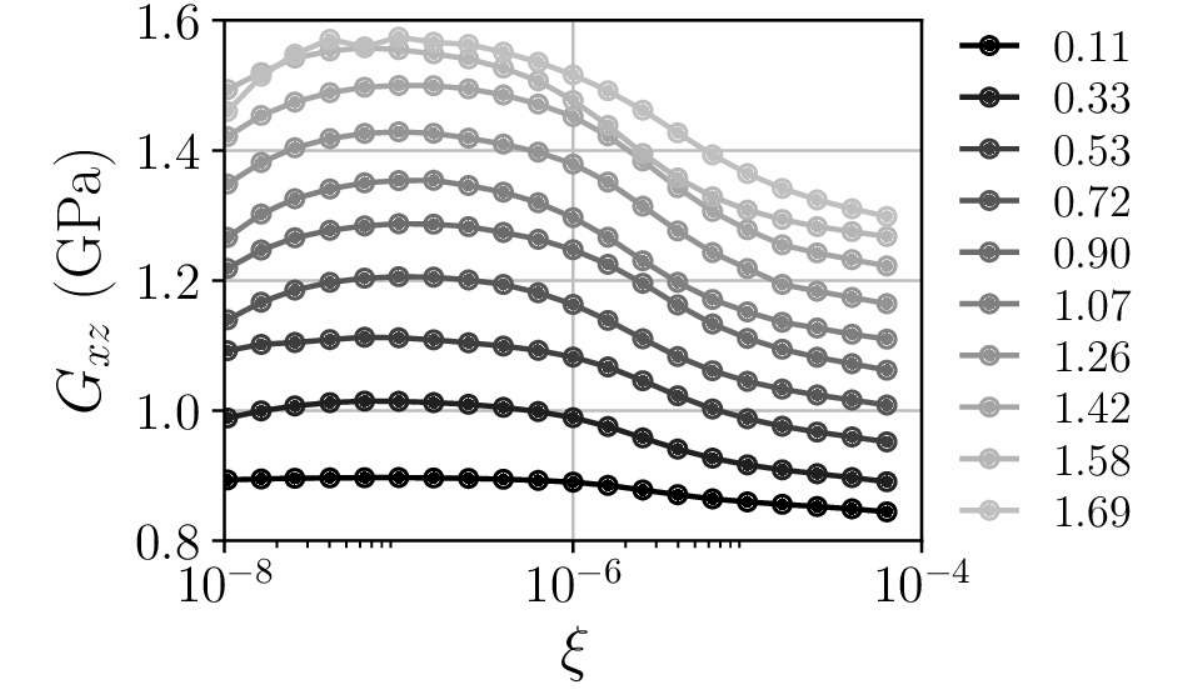}
		\caption{$ G_{xz} $: first load}
		\label{fig:G1stLoad02}
	\end{subfigure}
	\begin{subfigure}{0.33\textwidth}
		\centering
		\includegraphics[width=6cm]{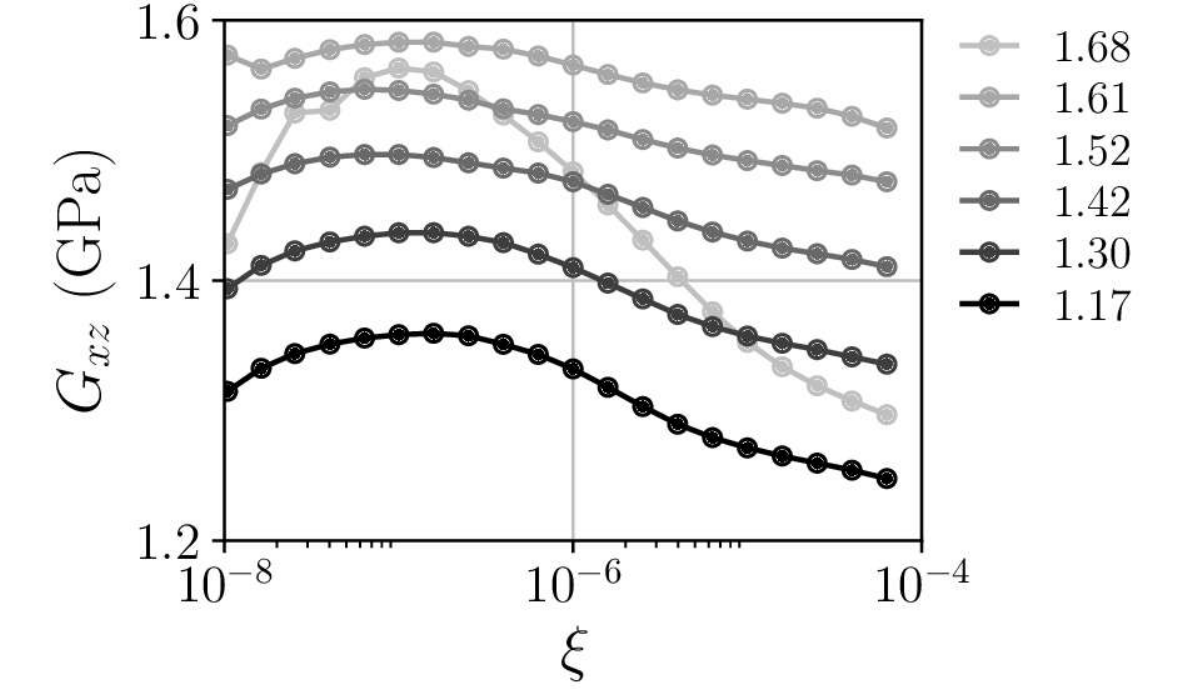}
		\caption{$ G_{xz} $: first unload}
		\label{fig:G1stUnload02}
	\end{subfigure}
	\begin{subfigure}{0.33\textwidth}
		\centering
		\includegraphics[width=6cm]{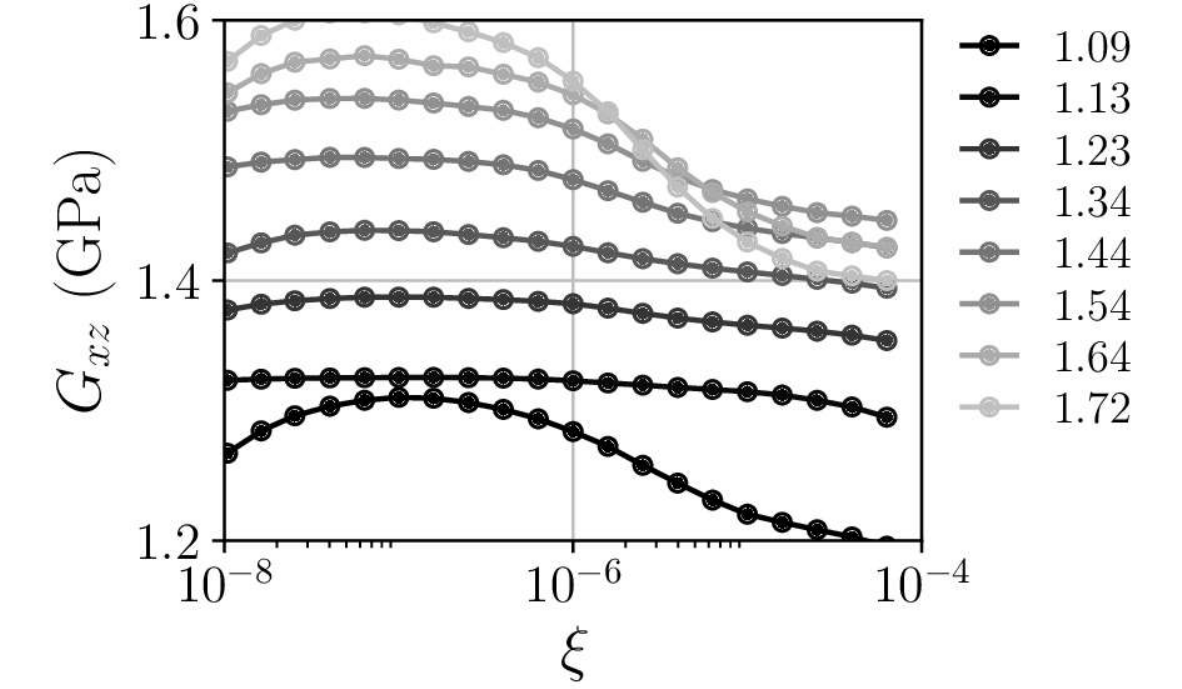}
		\caption{$ G_{xz} $: first reload}
		\label{fig:GsecondLoad02}
		\end{subfigure} \\
	\begin{subfigure}{0.33\textwidth}
		\centering
		\includegraphics[width=6cm]{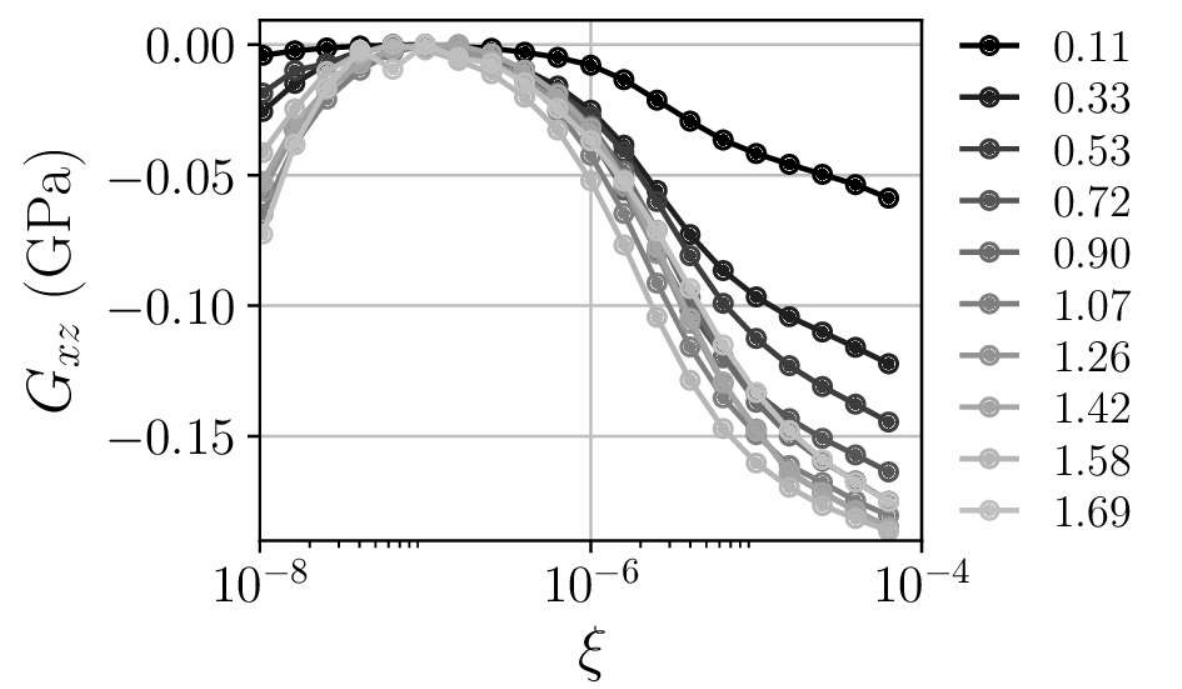}
		\caption{$ \frac{\Delta G_{xz}}{G_{xz}^e} $: first load}
		\label{fig:G1stLoad02Scaled}
	\end{subfigure}
	\begin{subfigure}{0.33\textwidth}
		\centering
		\includegraphics[width=6cm]{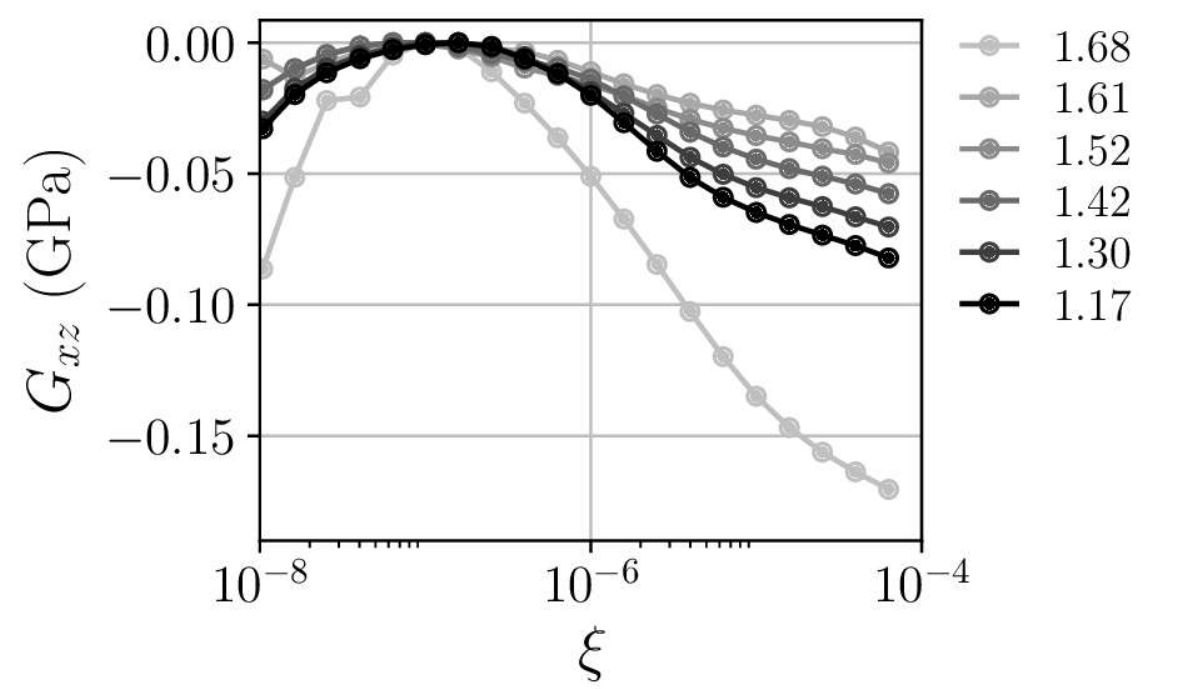}
		\caption{$ \frac{\Delta G_{xz}}{G_{xz}^e} $: first unload}
		\label{fig:G1stUnload02Scaled}
	\end{subfigure}
	\begin{subfigure}{0.33\textwidth}
		\centering
		\includegraphics[width=6cm]{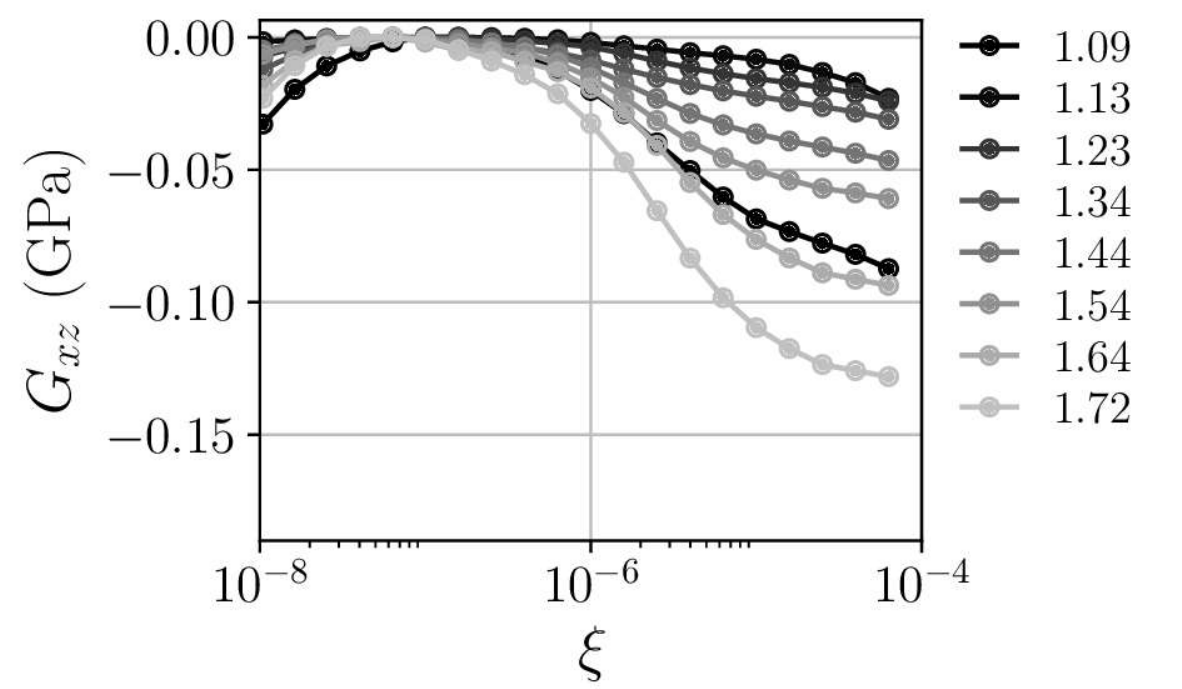}
		\caption{$ \frac{\Delta G_{xz}}{G_{xz}^e} $: first reload}
		\label{fig:GsecondLoad02Scaled}
	\end{subfigure}
	\caption{Small-strain shear moduli $ G_{xz} $ (a--c) and the normalized degradation $\frac{\Delta G_{xz}}{G_{xz}^e}$ (d--f) as functions of the perturbation magnitude $\xi$ during the oedometric load-unload cycles. Different levels of axial strain $\varepsilon_a$ in percentage are indicated by the different gray levels.}
	\label{fig:G02CycOedo}
\end{figure}

The variation of the small-strain oedometric modulus $M_z$ as a function of the perturbation magnitude $\xi$ during cyclic oedometric compression is plotted in \figsref{fig:M1stLoad}--\ref{fig:MsecondLoad}.
To clearly show the degradation of $M_z$ with increasing perturbation magnitude, the reduction of $M_z(\xi)$ with respect to the oedometric modulus in the elastic regime $M_z^e$ is normalized by $M_z^e$, as $\frac{\Delta M_z}{M_z^e}$.
The normalized degradation curves of $M_z$ are plotted in \figref{fig:M1stLoadScaled}--\ref{fig:MsecondLoadScaled}.
Similarly, the variation of the small-strain modulus $G_{xz}$ with increasing $ \xi $ is plotted in \figsref{fig:G1stLoad02}--\ref{fig:GsecondLoad02} and the normalized degradation curves of $G_{xz}$ in \figsref{fig:G1stLoad02Scaled}--\ref{fig:GsecondLoad02Scaled}.
Note that with increasing axial strain $\varepsilon_a$, the mean effective stress $ p' $, the deviatoric stress $ q $, the stress ratio $ q/p' $ and the fabric anisotropy of the granular packing increase accordingly.

As shown in \figsref{fig:M1stLoad}, \ref{fig:G1stLoad02}, \ref{fig:M1stLoadScaled} and \ref{fig:G1stLoad02Scaled}, during the first load, the degradation of both oedometric and shear moduli becomes larger, as $ \varepsilon_a $ increases from 0.11\% to 1.69\%.
The shrinkage of the elastic regime is more pronounced for the shear modulus than the oedometric modulus.
During the subsequent unloading-reloading cycles, the perturbation magnitudes associated to the maximum small-strain moduli $ M_z^e $ and $ G_{xz}^e $ hardly change and the degradation of both $ M_z $ and $ G_{xz} $ is smaller than during the first load, as shown in \figsref{fig:M1stUnload}, \ref{fig:G1stUnload02}, \ref{fig:MsecondLoad} and \ref{fig:GsecondLoad02}.
The suppressed degradation during the unloading-reloading cycles is expected, because the accumulation of plastic deformation after a load reversal should be less than during the initial load.
However, immediately after each reversal, at $\varepsilon_a=1.68\% $ of the first unload (see \figsref{fig:M1stUnload}, \ref{fig:G1stUnload02}, \ref{fig:M1stUnloadScaled}, \ref{fig:G1stUnload02Scaled}) and at $\varepsilon_a=1.09\% $ of the first reload (see \figsref{fig:MsecondLoad}, \ref{fig:GsecondLoad02}, \ref{fig:MsecondLoadScaled}, \ref{fig:GsecondLoad02Scaled}), the decrease of both $ M_z $ and $ G_{xz} $ with increasing $\xi$ is dramatic, compared with the degradation curves at the succeeding strain levels, namely, $\varepsilon_a=1.61\% $ of the first unload and $\varepsilon_a=1.13\% $ of the first reload.
After $\varepsilon_a$ increases after the load reversal, the degradation of the small-strain moduli beyond the elastic regimes is negligible at first, and then becomes increasingly pronounced, as the oedometric loading/unloading continues monotonically.
It can be argued that although the load is reversed, the incremental behavior of the granular packing immediately after the load reversal is still elastoplastic and resembles the degradation curve at the state preceding the reversal.
This means that the elastoplastic behavior of the granular packing at the states in the vicinity of the load reversal is ``reversible''.
As indicated in \figsref{fig:M22CycOedo} and \ref{fig:G02CycOedo}, the incremental response becomes \emph{truly} ``elastic'' after the packing is strained sufficiently away from the load reversal.
The degradation of the small-strain moduli further develops thereafter, until the load is reversed again.
This interesting phenomenon could be macroscopically related to cyclic mobility and instability at yield surface, which is outside the scope of this work.

\section{Evaluation of elastic wave velocities via dynamic probing}
\label{sec:dynaModuli} 

\begin{figure} [t!]
	\centering
	\begin{subfigure}{0.9\textwidth}
		\begin{tikzpicture}
			\node (image) {\includegraphics[width=\textwidth]{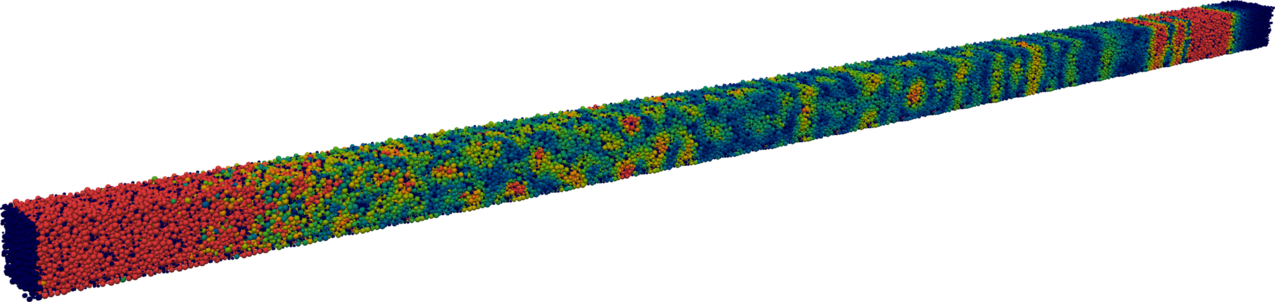}};
			\node at (-7.5,1.3) {\footnotesize{(a)}};
			\node at (5,-0.8) {\includegraphics[width=6cm]{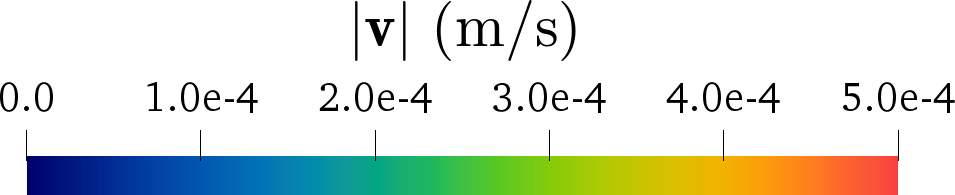}};
		\end{tikzpicture}
	\end{subfigure}
	\begin{subfigure}{0.9\textwidth}
		\begin{tikzpicture}
			\node (image) {\includegraphics[width=\textwidth]{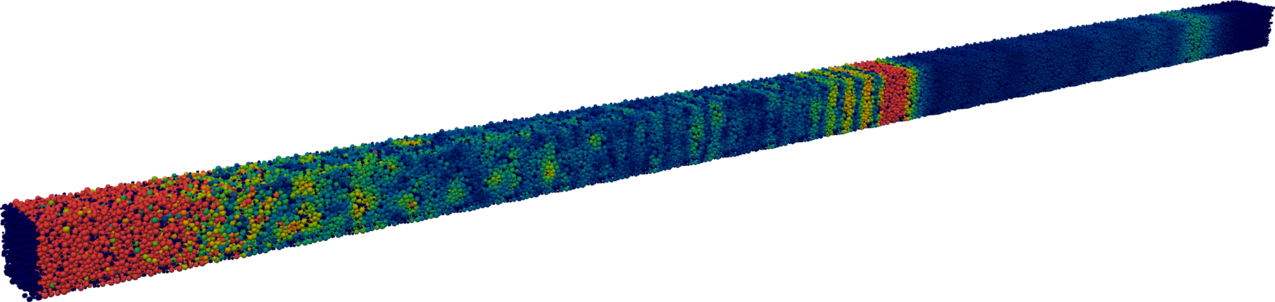}};
			\node at (-7.5,1.3) {\footnotesize{(b)}};
			\node at (5,-0.8) {\includegraphics[width=6cm]{longLegend.png}};
		\end{tikzpicture}
	\end{subfigure}
	\caption{Snapshot of the distribution of particle velocity in the 33$\times$RV granular column, agitated by longitudinal and transverse impulses respectively.}
	\label{fig:long3Wave}
\end{figure}

In the current work, four granular columns with 11, 22, 33 and 44 RVs stacked along the $ z $ direction are considered to investigate the effect of packing length on dynamic probing.
\figref{fig:long3Wave} shows a snapshot of the distribution of particle velocity in the 11$\times$RV granular column, from which both P- and S-wave fronts, agitated respectively by longitudinal and transverse impulses, can be clearly observed.
A wide range of amplitudes $A$ are tested to determine the maximum level of $A$ below which no opening and closing of contacts occur (data not shown here).
Based on preliminary investigations, $10^{-2}$ times the mean normal overlaps at equilibrium $\overline{|\vec{u}_n|}$, before the probing starts, are chosen as the amplitudes of all input signals listed in \tabref{tab:source}.
In the following subsections, the granular columns at axial strain $ \varepsilon_a=0.11\%$ are taken as the reference cases and the wave velocities are determined from both time and frequency domains.
The effects of packing length, input waveform and frequency on wave velocities are investigated and compared, using both time- and frequency-domain techniques.
After the appropriate packing length, input waveform and frequency are selected, the DEM simulations of dynamic probing are performed to obtain the P- and S-wave velocities of the granular material at various stress states along the cyclic oedometric path.

\begin{table} [t!]
	\caption{Input waveforms and frequencies of enforced displacement signals.}
	\label{tab:source}
	\centering
	\begin{tabular}{ll}
		\toprule		
		\cmidrule{1-2}
		Frequency $\omega/2\pi$ (MHz) & Waveform ($t \leq t_n \d t$, $A=10^{-2}\overline{|\vec{u}_n|}$)\\ 
		\midrule
		$1/\d t$ & Impulse: $ \vec{r}_i = \vec{r}_i^0 + A $\\ 
		\midrule
		0.75 & Cosine: $ \vec{r}_i = \vec{r}_i^0 + A (1-\cos(\omega t)) $\\
		0.38 & Cosine: $ \vec{r}_i = \vec{r}_i^0 + A (1-\cos(\omega t)) $\\
		0.25 & Cosine: $ \vec{r}_i = \vec{r}_i^0 + A (1-\cos(\omega t)) $\\
		0.19 & Cosine: $ \vec{r}_i = \vec{r}_i^0 + A (1-\cos(\omega t)) $\\
		\midrule
		0.75 & Sine: $ \vec{r}_i = \vec{r}_i^0 + A \sin(\omega t) $\\
		0.38 & Sine: $ \vec{r}_i = \vec{r}_i^0 + A \sin(\omega t) $\\
		0.25 & Sine: $ \vec{r}_i = \vec{r}_i^0 + A \sin(\omega t) $\\
		0.19 & Sine: $ \vec{r}_i = \vec{r}_i^0 + A \sin(\omega t) $\\
		\bottomrule
	\end{tabular}\\[6pt]
	\small
	The subscript $i$ takes $x$ or $y$ for S-waves and $z$ for P-waves. 
\end{table}

\subsection{Effect of packing length in the propagation direction}
\label{sec:lengthEffect}

As mentioned in \secref{sec:dynaProbe}, the packing length in the propagation direction may affect wave velocities estimated by dynamic probing.
To study such influence on both time- and frequency-domain analyses, granular columns that consist of 11, 22, 33 and 44 copies of the calibrated RV are considered.
Each DEM simulation continues for a sufficient number of time steps, until the P- or S-wave reaches the last $x$--$y$ layer of particles, opposite to the source layer.

\subsubsection{Time domain analysis}
\label{sec:lenEffectTime}

\begin{figure} [t!]
	\begin{subfigure}{0.5\textwidth}
		\centering
		\includegraphics[width=8cm]{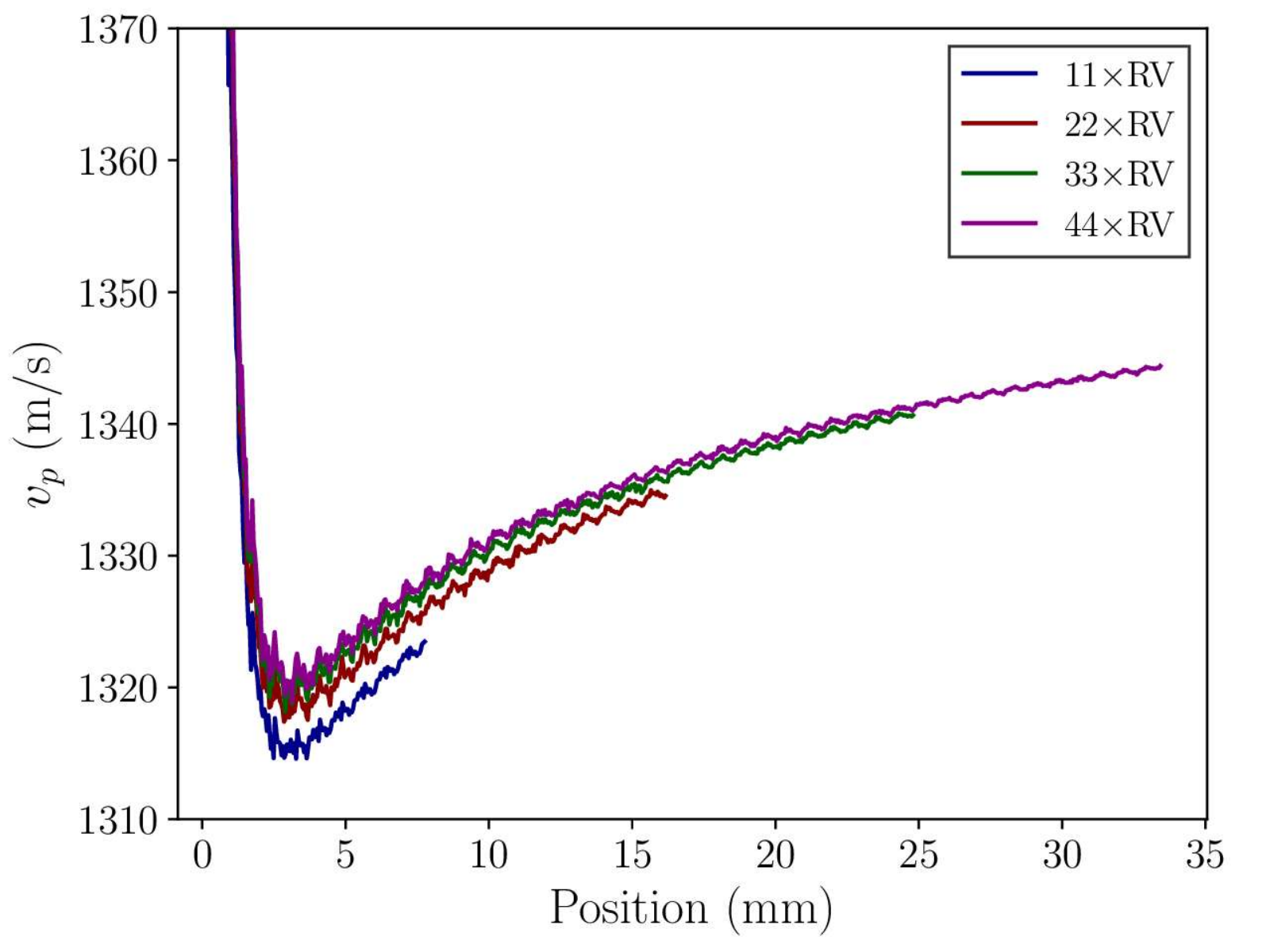}
		\caption{P-wave velocity}
		\label{fig:diffLenVpTime}
	\end{subfigure}
	\begin{subfigure}{0.5\textwidth}
		\centering
		\includegraphics[width=8cm]{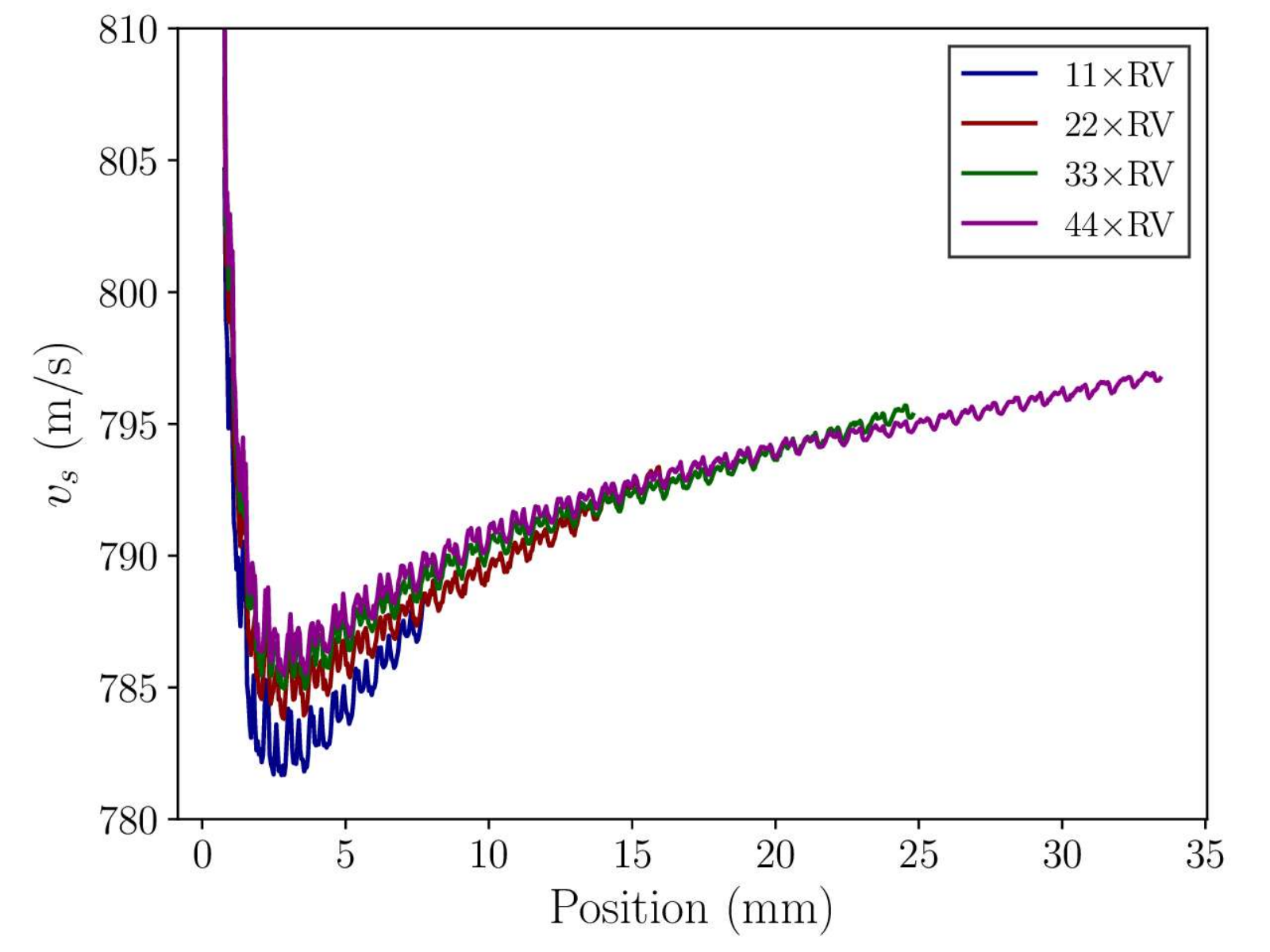}
		\caption{S-wave velocity}
		\label{fig:diffLenVsTime}
	\end{subfigure}
	\caption{Effect of the packing length in the propagation direction on wave velocities calculated from deduced travel time and travel distance. P-and S-waves are agitated by longitudinal and transverse impulse prescribed to the source layer. The legends indicate the numbers of the representative volumes that the granular column contains.}
	\label{fig:diffLenTime}
\end{figure}

With the travel time determined by the peak-to-peak method, the variation of the wave velocities with respect to the source-receiver distance are quantified for the four packing lengths.
The evolutions of the P- and S-wave velocities as functions of travel distance are plotted in \figsref{fig:diffLenVpTime} and \ref{fig:diffLenVsTime}, respectively.
As the P- and S-waves propagate away from the source, the wave velocities obtained from the time domain analysis first decrease dramatically until the distance $ \vec{r}_z\approx2.5 $ mm, independent of the packing lengths, and then increase seemingly towards the constant ``long-wavelength limits''.
Therefore, to obtain the P- and S-wave velocities that approximate the long-wavelength limits, the maximum travel distance allowed in the DEM packing has to be sufficiently long.
How close the wave velocities approach the long-wavelength limits, also depends on the inserted wavelengths and particle sizes (see \secref{sec:sigEffect}).
The dependence of the wave velocities on the travel distance can be attributed to i) the nonlinear Hertz-Mindlin contact law (see \eqsref{eq:fdLawn} and \ref{eq:fdLaws}) which causes the overestimation close to the source, and ii) the dispersion nature of the dry granular material, i.e, wave velocity decreases with the decrease of wavelength.
Note that to avoid the effect of the high-frequency noise generated close to the source (see \figref{fig:long3Wave} and supplementary video A), the first RV that contains roughly 17 layers, including the source layer, are not included for the time domain analysis.

\subsubsection{Frequency-domain analysis}
\label{sec:lenEffectFreq}

\begin{figure} [t!]
	\begin{subfigure}{0.246\textwidth}
		\centering
		\includegraphics[width=\textwidth]{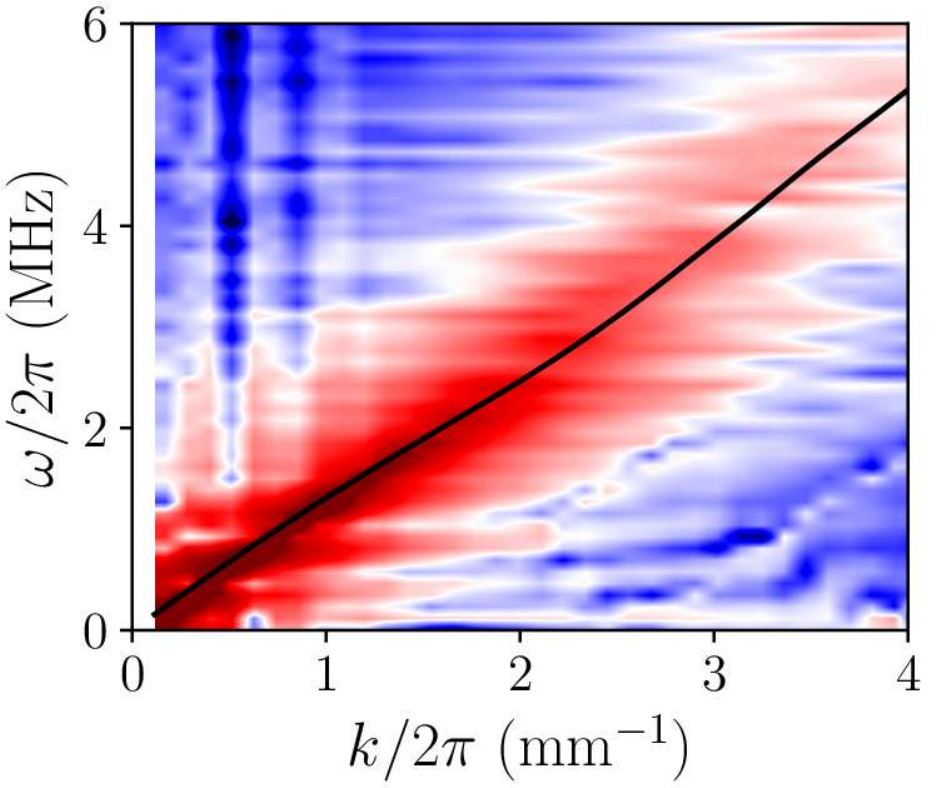}
		\caption{P-wave: 11$\times$RV}
		\label{fig:M11RV}
	\end{subfigure}
	\begin{subfigure}{0.246\textwidth}
		\centering
		\includegraphics[width=\textwidth]{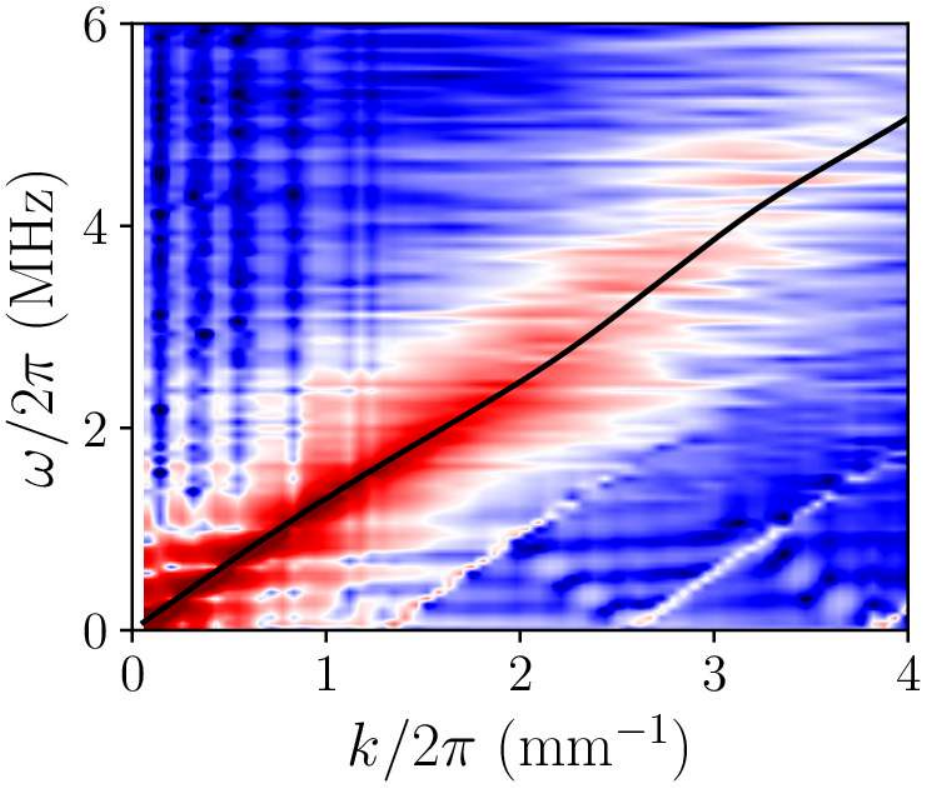}
		\caption{P-wave: 22$\times$RV}
		\label{fig:M22RV}
	\end{subfigure}
	\begin{subfigure}{0.246\textwidth}
		\centering
		\includegraphics[width=\textwidth]{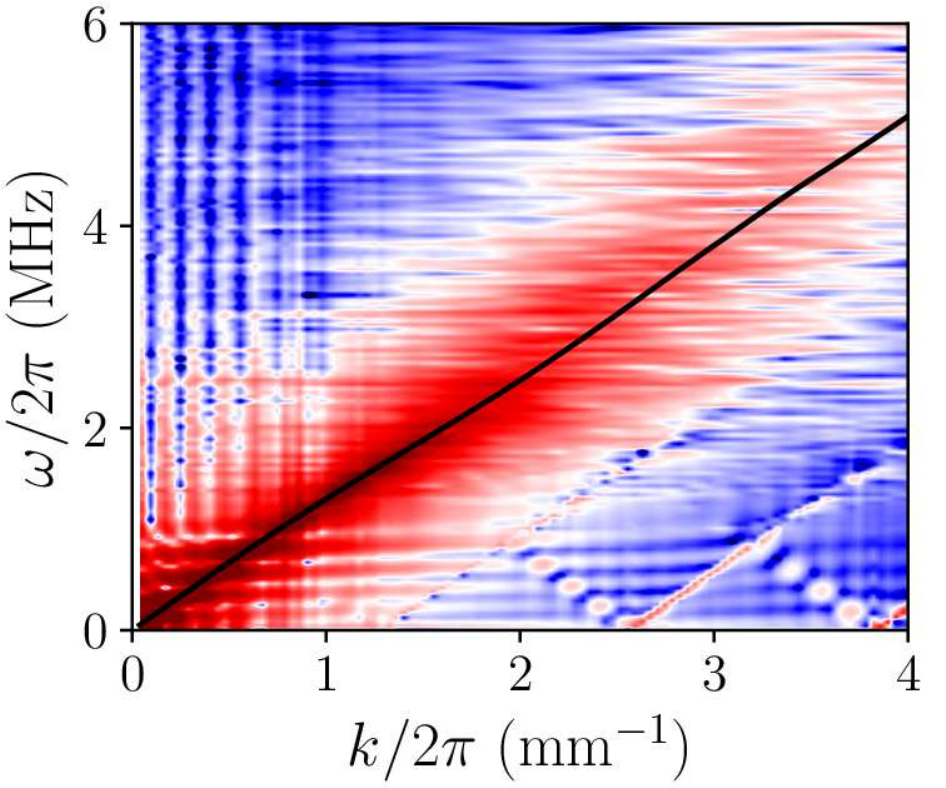}
		\caption{P-wave: 33$\times$RV}
		\label{fig:M33RV}
	\end{subfigure}
	\begin{subfigure}{0.246\textwidth}
		\centering
		\includegraphics[width=\textwidth]{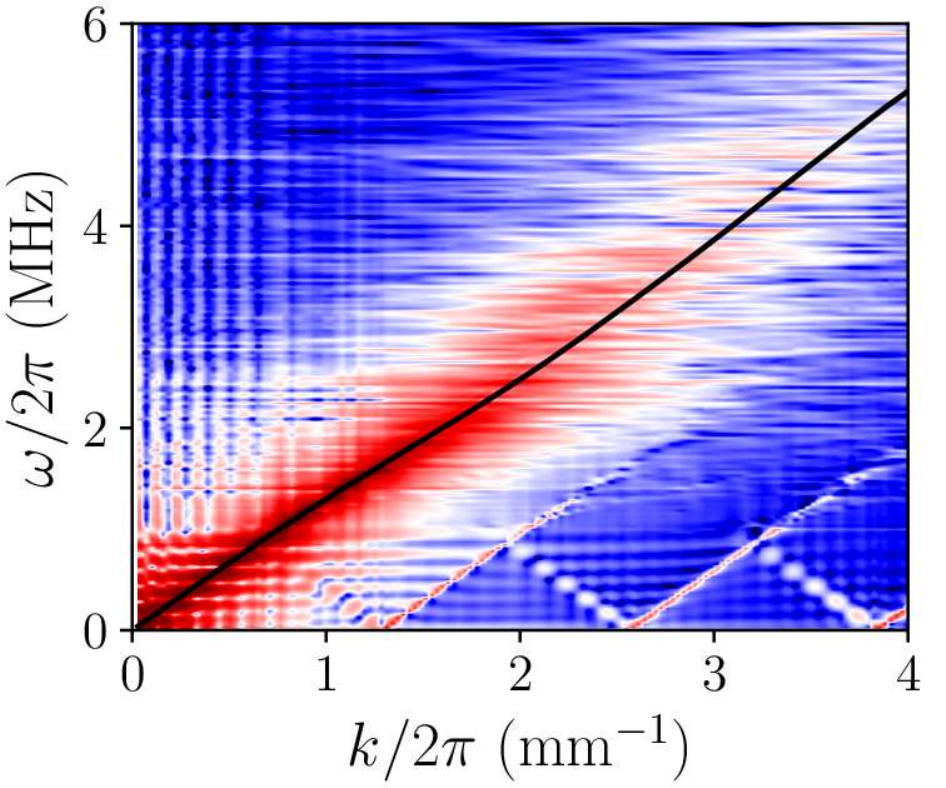}
		\caption{P-wave: 44$\times$RV}
		\label{fig:M44RV}
	\end{subfigure} \\
	\begin{subfigure}{0.246\textwidth}
		\centering
		\includegraphics[width=\textwidth]{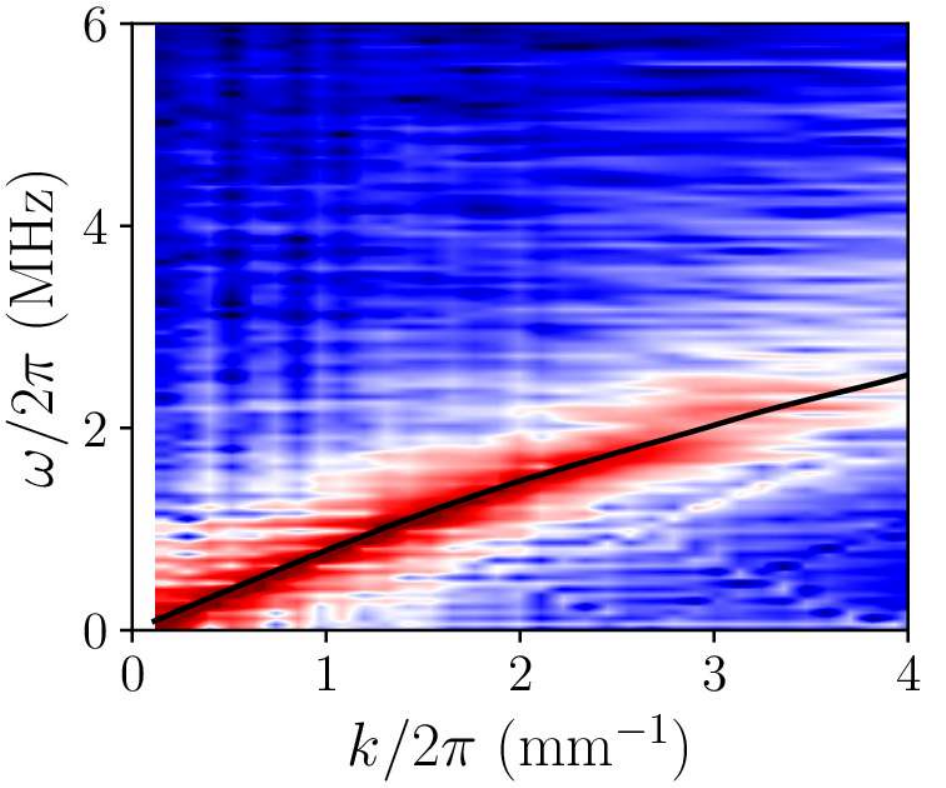}
		\caption{S-wave: 11$\times$RV}
		\label{fig:G11RV}
	\end{subfigure}
	\begin{subfigure}{0.246\textwidth}
		\centering
		\includegraphics[width=\textwidth]{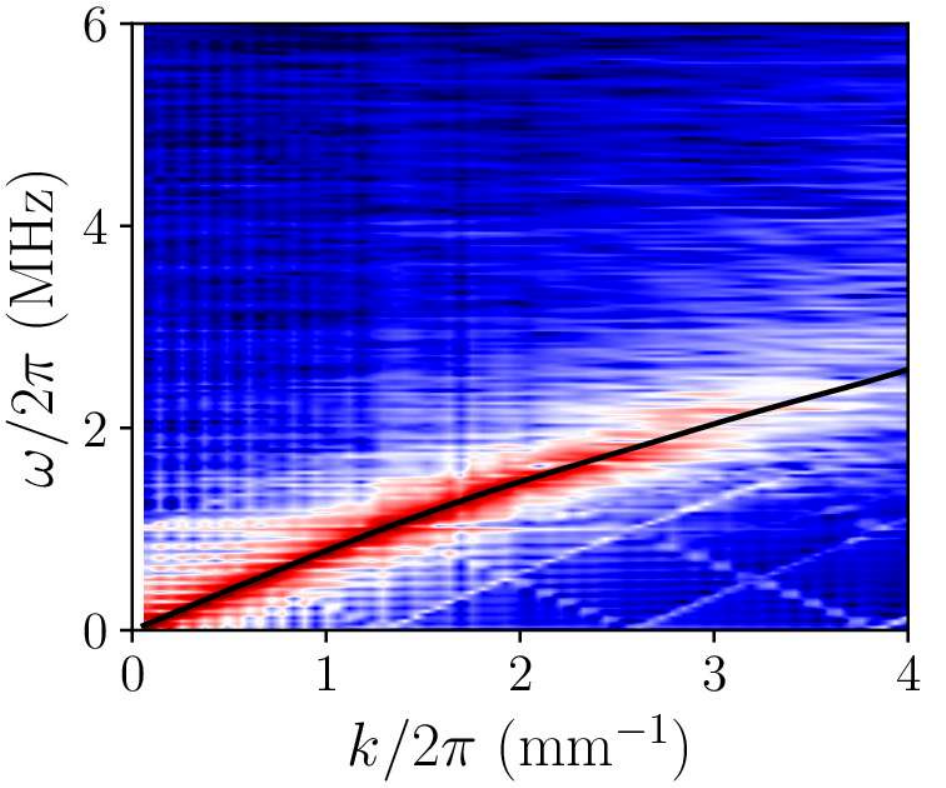}
		\caption{S-wave: 22$\times$RV}
		\label{fig:G22RV}
	\end{subfigure}
	\begin{subfigure}{0.246\textwidth}
		\centering
		\includegraphics[width=\textwidth]{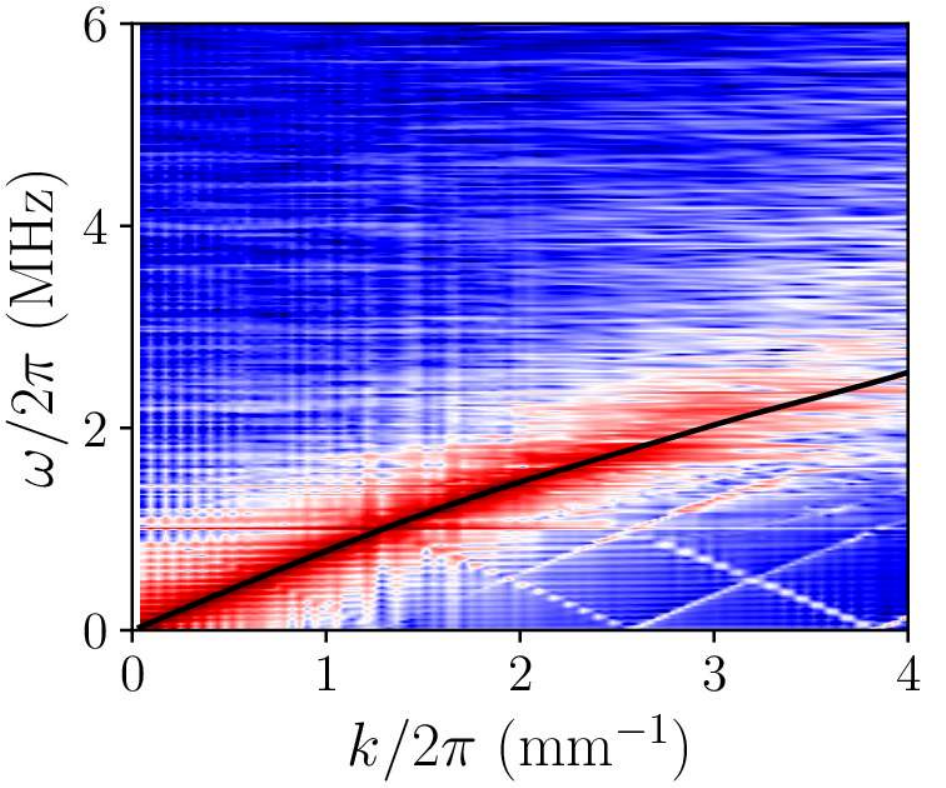}
		\caption{S-wave: 33$\times$RV}
		\label{fig:G33RV}
	\end{subfigure}
	\begin{subfigure}{0.246\textwidth}
		\centering
		\includegraphics[width=\textwidth]{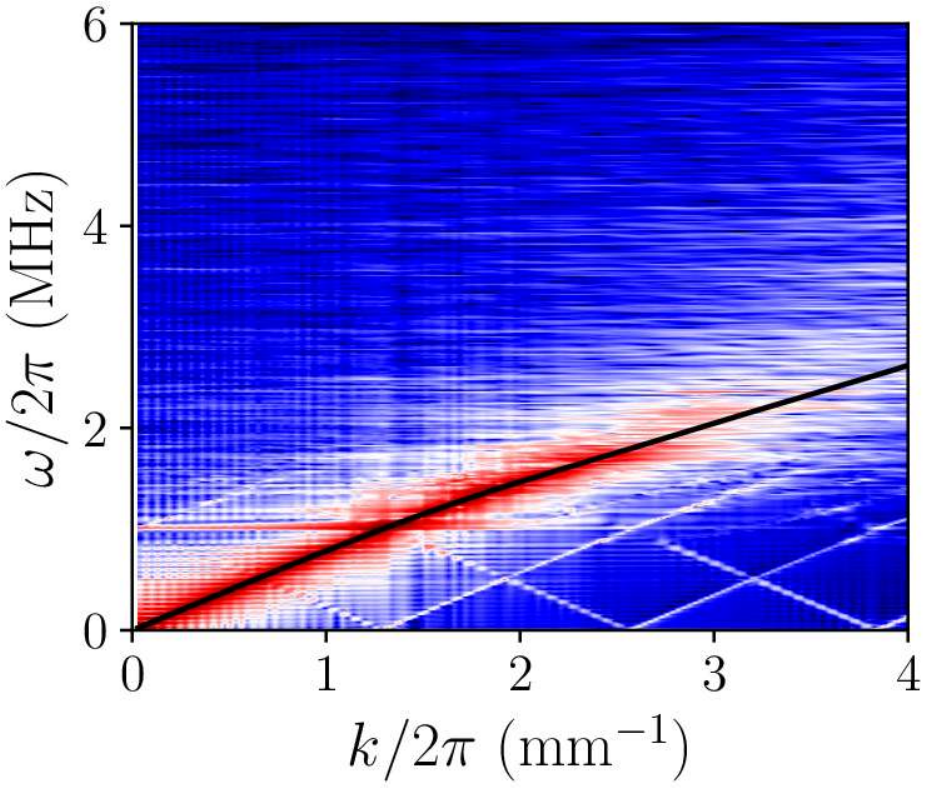}
		\caption{S-wave: 44$\times$RV}
		\label{fig:G44RV}
	\end{subfigure}
	\caption{Dependence of the P-wave dispersion relation (a--d) and the S-wave dispersion relation (e--h) on the packing length in the propagation directions. Colors in each subplot show the power spectrum given by the discrete Fourier transformation.}
	\label{fig:diffLenFreq}
\end{figure}

To reveal the dispersive properties of granular materials, the time domain responses are transformed to the frequency domain, using \eqsref{eq:dft}--\ref{eq:2dft}.
\figsref{fig:M11RV}--\ref{fig:M44RV} and \ref{fig:G11RV}--\ref{fig:G44RV} respectively show the logarithms of the power spectra $ S_t(k,\omega) $ and $ S_t(k,\omega) $ for the four granular columns with different packing lengths.
The power spectra at a given wavenumber $k$ are fitted with the Lorentzian $S(k,\omega) = \frac{S_0\gamma(k)}{(\omega-\omega(k))^2+\gamma(k)^2}$, with $S_0$ the peak at the dominant frequency $ \omega(k) $ and $ \gamma(k) $ the attenuation coefficient due to scattering.
The solid lines in \figref{fig:diffLenFreq} represent the dispersion relations, quantified by the fitted $ \omega(k) $ at each wavenumber.

\begin{figure} [t!]
	\begin{subfigure}{0.5\textwidth}
		\centering
		\includegraphics[width=7.5cm]{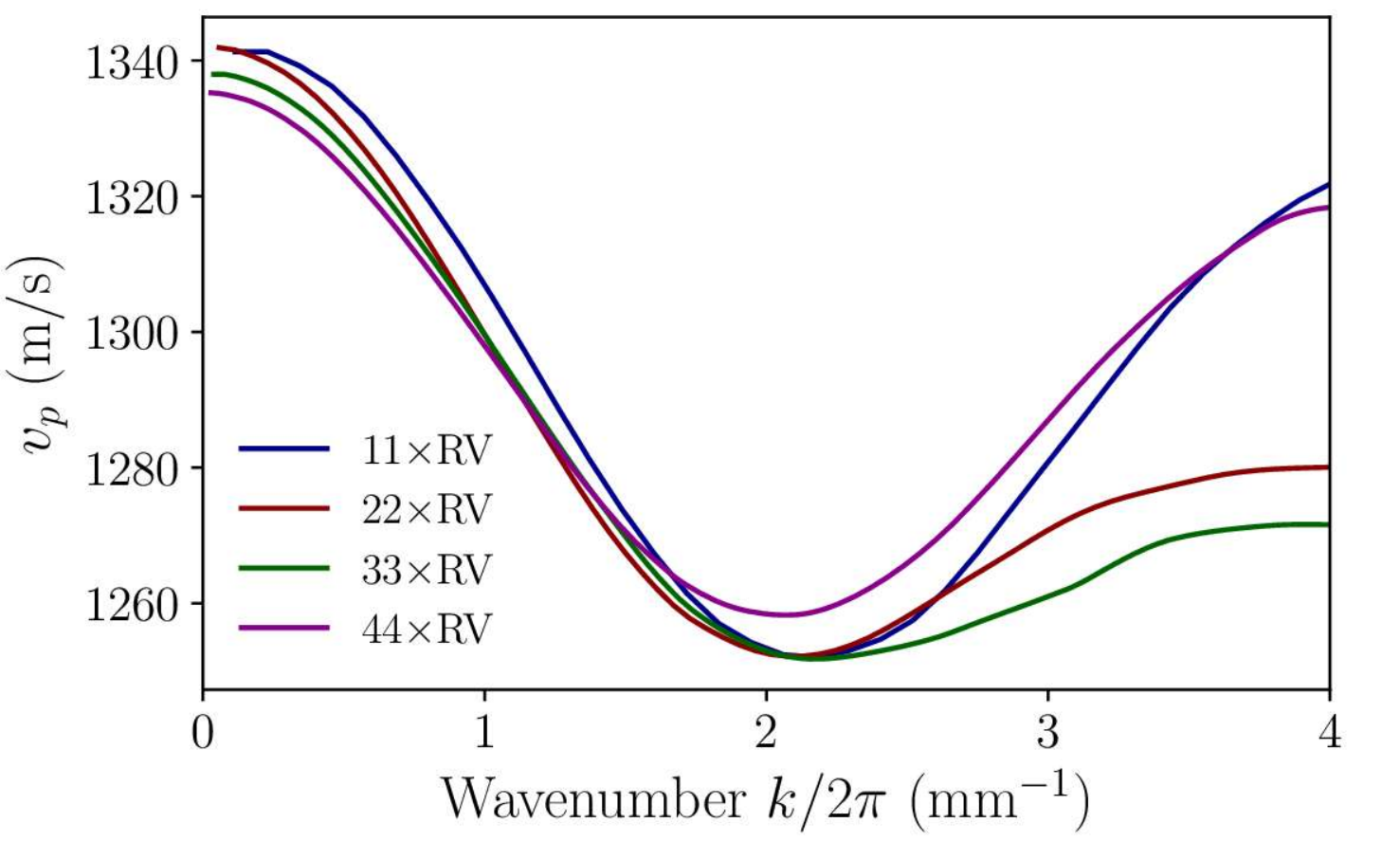}
		\caption{P-wave velocity versus wavenumber}
		\label{fig:diffLenVpFreq}
	\end{subfigure}
	\begin{subfigure}{0.5\textwidth}
		\centering
		\includegraphics[width=7.5cm]{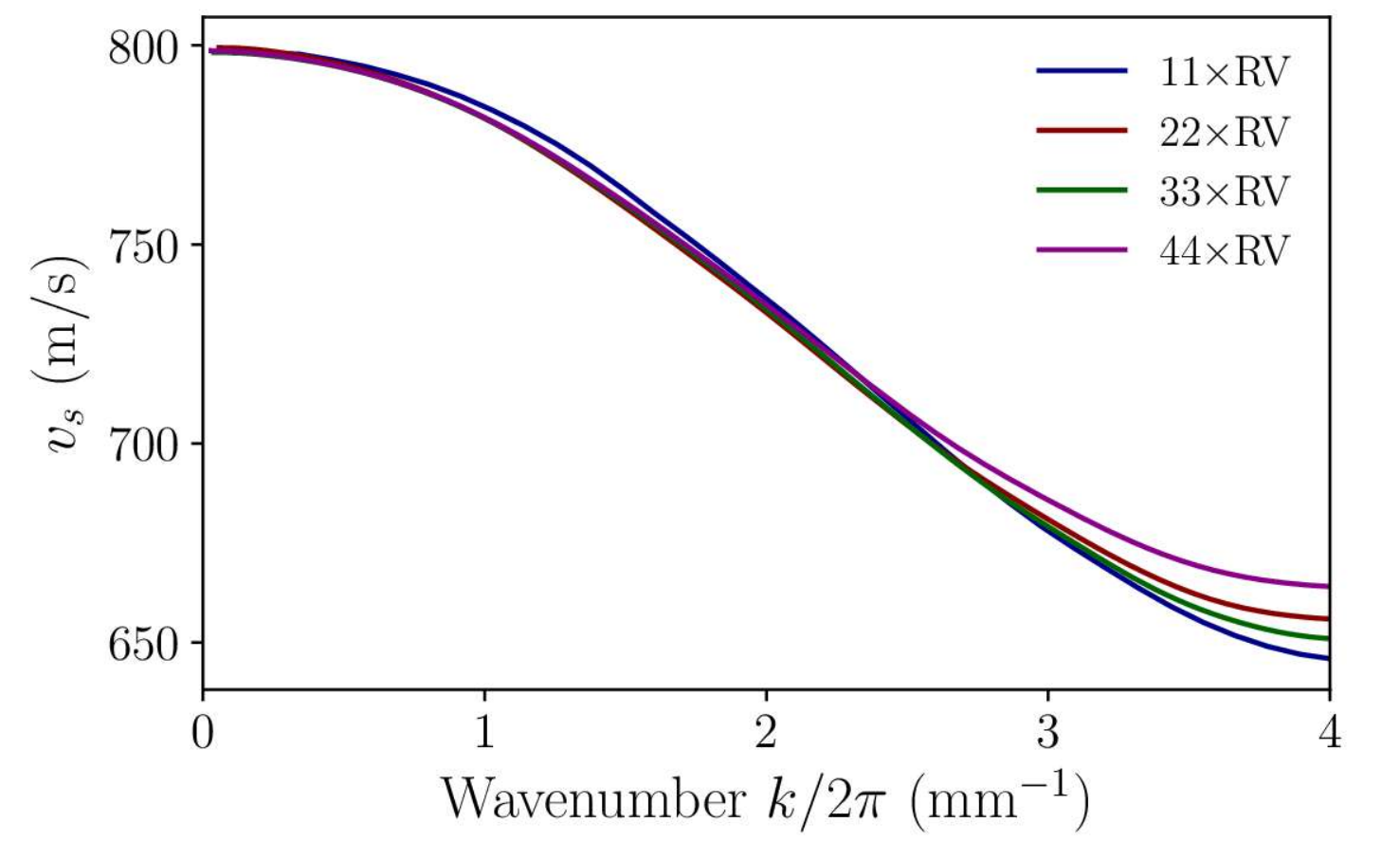}
		\caption{S-wave velocity versus wavenumber}
		\label{fig:diffLenVsFreq}
	\end{subfigure}
	\caption{P-wave velocities (a) and S-wave velocities (b) as functions of wavenumber. The legends indicate the numbers of the representative volumes that the granular column contains.}
	\label{fig:diffLenWaveFreq}
\end{figure}

As expected, with the increase of the packing length, the dispersion relations are fitted with higher resolution in the frequency domain and less noise.
The wave velocities extracted from the fitted dispersion relations at a given wavenumber are plotted in \figref{fig:diffLenWaveFreq}.
It appears that the wave velocities are only slightly affected by the packing length, when $ k/2\pi>2.0$ mm$^{-1}$ for the P-wave and $ k/2\pi>4.0$ mm$^{-1}$ for the P-wave.
In particular, the good agreement between the wave velocities for $ k/2\pi<4.0$ mm$^{-1}$ suggests that the 11$\times$RV granular column is sufficient for estimating the long-wavelength wave velocities in the frequency domain.
However, to ensure that sufficient statistics are available for the fitting and to accurately characterize the dispersion relations, including details like the secondary increase of the tangent to the P-wave dispersion relation at $ k/2\pi\approx2.0$ mm$^{-1}$, the 33$\times$RV granular column is selected for the following analyses.
Compared with the time domain analysis which shows a strong dependence of the wave velocities on packing length, the frequency domain analysis provides a more accurate and robust approach for the determination of long-wavelength wave velocities.

\subsection{Effect of input waveform and frequency}
\label{sec:sigEffect}

Following \secref{sec:inputSigs}, a variety of input waveforms and frequencies (see \tabref{tab:source}) are adopted to perform dynamic probes, in order to investigate their effects on the numerical prediction of elastic wave velocities.
The granular column consisting 33 calibrated RVs aligned in the propagation direction is selected for all the dynamic probes here.
Similar to \secref{sec:lengthEffect}, the constituent RV is at the first stress state ($ \varepsilon_a=0.11\%$) on the oedometric stress path.

\subsubsection{Time domain analysis}
\label{sec:sigEffectTime}

Irrespective of the input signals, the estimated P- and S-wave velocities tend to saturate towards the respective long-wavelength limits of the P- and S-waves.
Interestingly, the evolutions of the wave velocities versus the receiver position obtained with the impulse signals lie below those with the cosine signals.
As the input frequency of cosine signals decreases, the minimum wave velocities at approximately $ \vec{r}_z=2.5 $ mm are elevated, whereas the long-wavelength limits seem to be unaffected.
From the P- and S-wave velocities given by the cosine signals at the end ($ \vec{r}_z=25 $ mm), one finds that the wave velocities decrease with increasing input frequency, as would be normally expected for dry granular materials.
The P- and S-wave velocities obtained with the sine signals, however, show the opposite trend, that is the wave velocities increase as the input frequency increases.
The reason for the opposite trends may be that the sine signals activate more high frequency contents, as suggested by the horizontal bands in \figsref{fig:Msin40d}--\ref{fig:Msin160d} and \figsref{fig:Gsin40d}--\ref{fig:Gsin160d}.

Near the source, the wave velocities resulting from the sine input signals are smaller than the long-wavelength limits, whereas those from the cosine input signals are larger.
As mentioned in \secref{sec:lenEffectTime}, the overestimation of wave velocities caused by the cosine input signals can be attributed to the nonlinear elasticity at grain contacts.
The underestimation of wave velocities under sinusoidal wave agitations, on the other hand, are seemingly dominated by the activated high-frequency contents, as can be observed in \figsref{fig:Msin40d}--\ref{fig:Msin160d} and \figsref{fig:Gsin40d}--\ref{fig:Gsin160d}.
Therefore, it can be deduced that the wave propagation velocities close to the source are affected by the interplay between two mechanisms, namely nonlinear elasticity at contacts and high-frequency dispersion activated by the source.

Although the P- or S-wave velocities received at the end of the granular column, agitated by different signals, seemingly converge to the same long-wavelength limit, the packing length is not sufficient long to confirm the convergence.
Therefore, the input signal which shows a stagnant profile of wave velocity along the granular column, i.e., the cosine signal with $\omega/2\pi=0.38$, is selected for estimating the evolution of the \emph{long-wavelength} wave velocities during cyclic oedometric compression.

\begin{figure} [t!]
	\begin{subfigure}{0.5\textwidth}
		\centering
		\includegraphics[width=8cm]{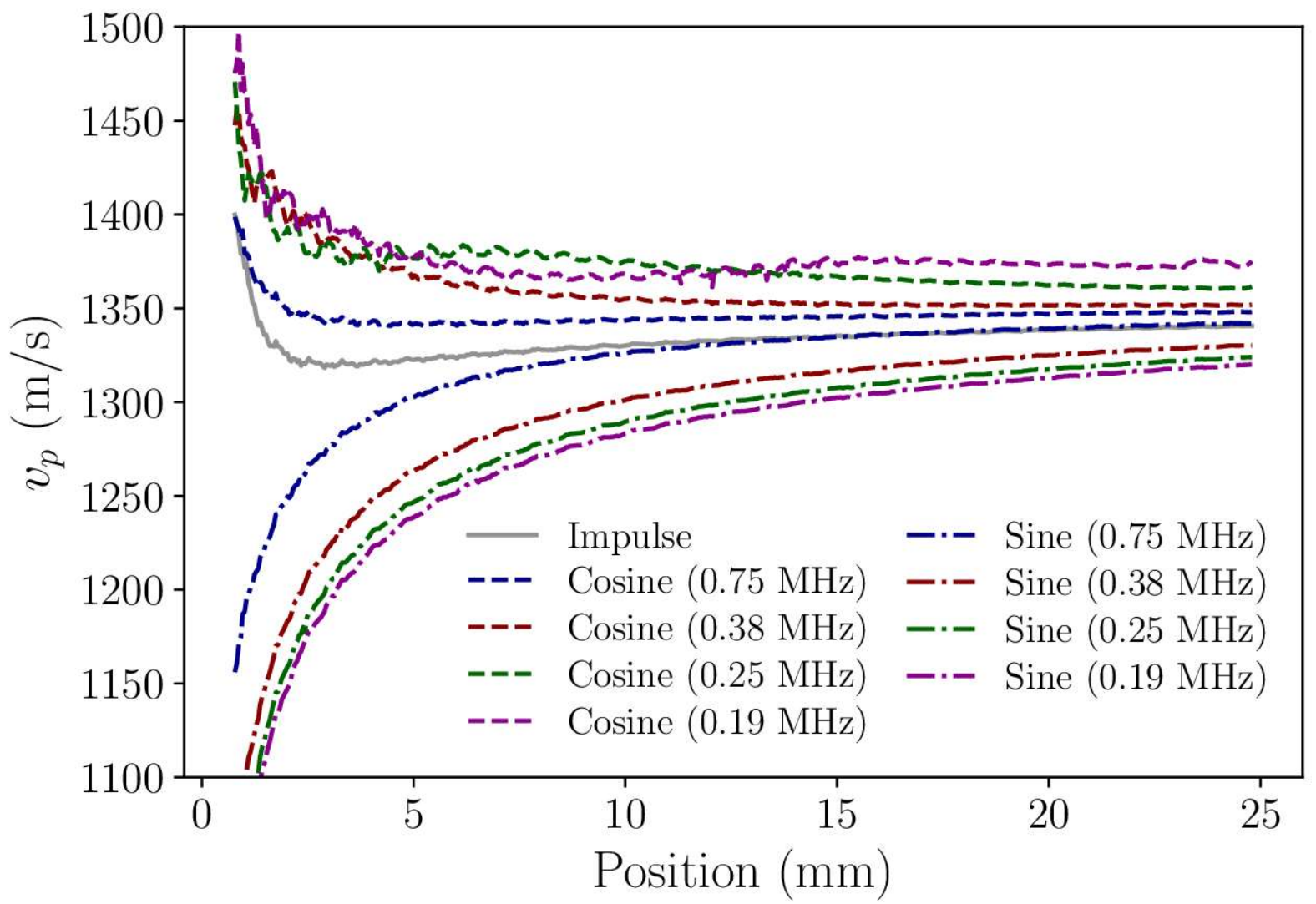}
		\caption{P-wave velocity}
		\label{fig:diffSigVpTime}
	\end{subfigure}
	\begin{subfigure}{0.5\textwidth}
		\centering
		\includegraphics[width=8cm]{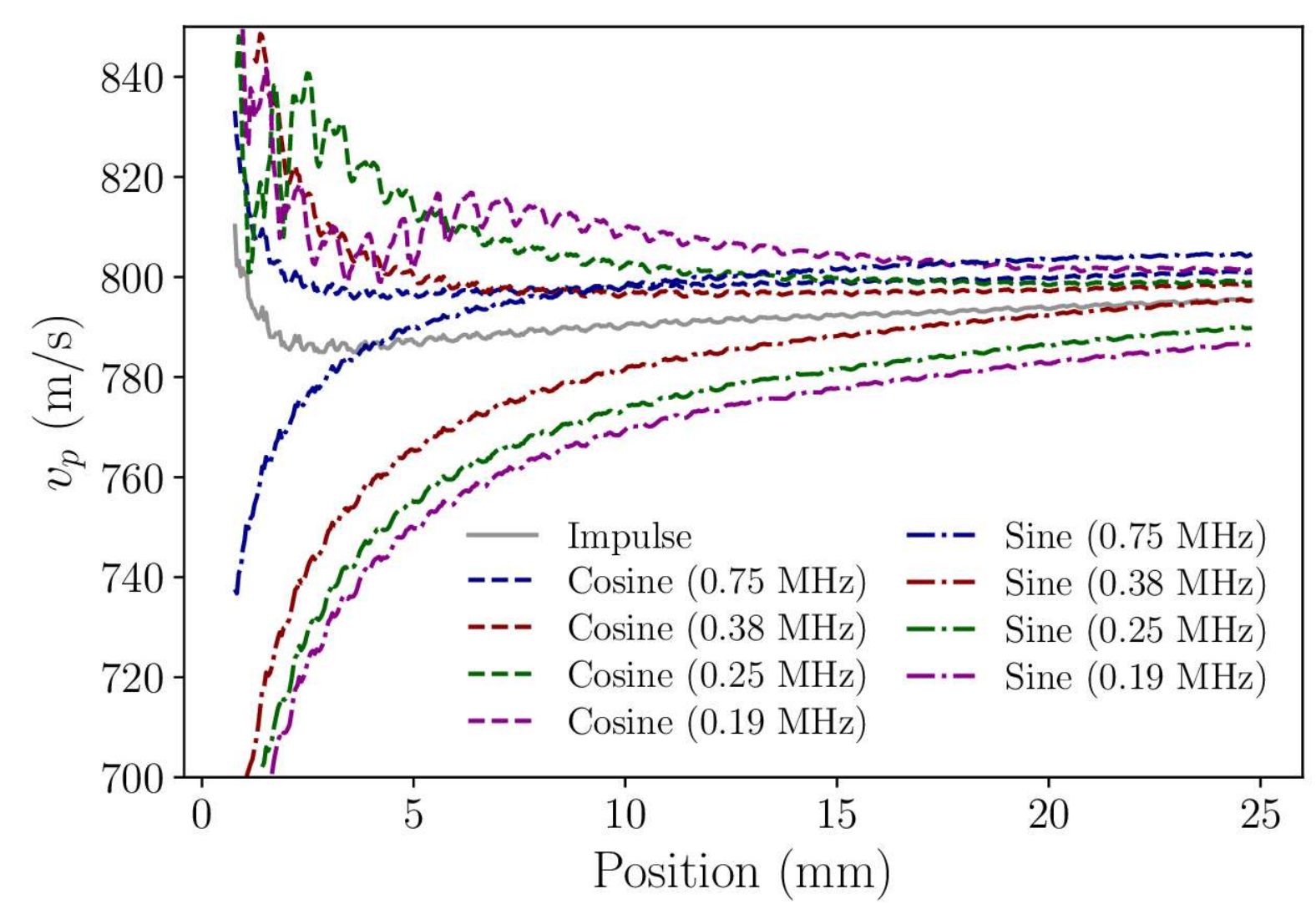}
		\caption{S-wave velocity}
		\label{fig:diffSigVsTime}
	\end{subfigure}
	\caption{Effect of input waveform and frequency on wave velocities calculated from deduced travel time and travel distance. The input waveforms and frequencies of single-period signals are indicated by the legends.}
	\label{fig:diffSigTime}
\end{figure}

\subsubsection{Frequency domain analysis}
\label{sec:sigEffectFreq}

\begin{figure} [t!]
	\begin{subfigure}{0.246\textwidth}
		\centering
		\includegraphics[width=\textwidth]{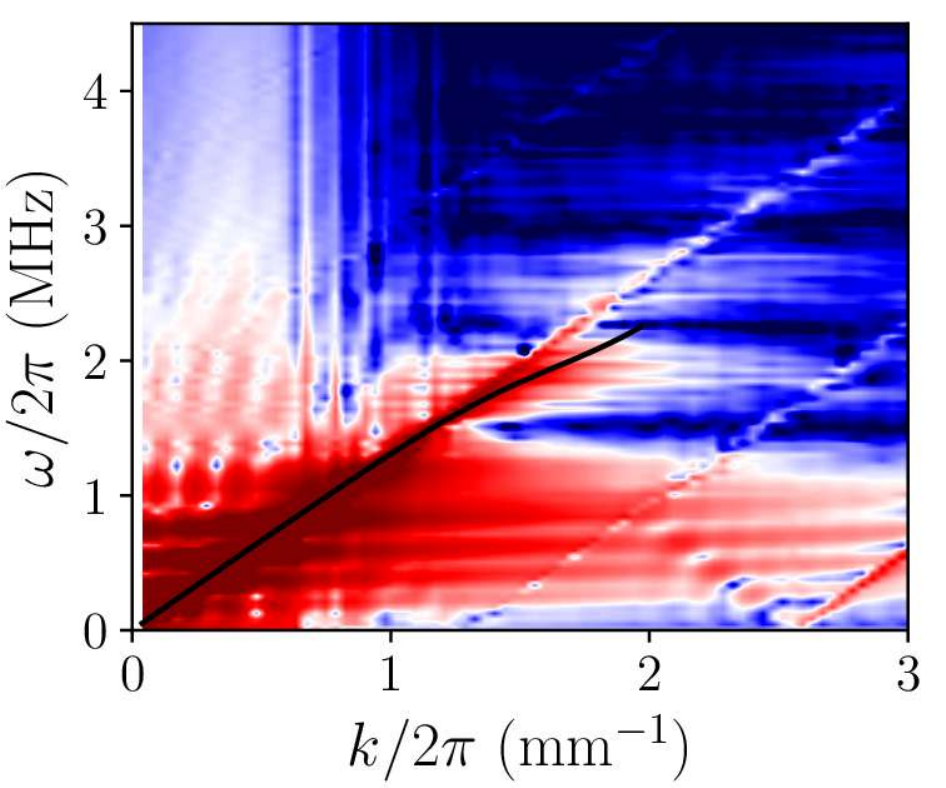}
		\caption{Cosine (0.75 MHz)}
		\label{fig:Mcos40d}
	\end{subfigure}
	\begin{subfigure}{0.246\textwidth}
		\centering
		\includegraphics[width=\textwidth]{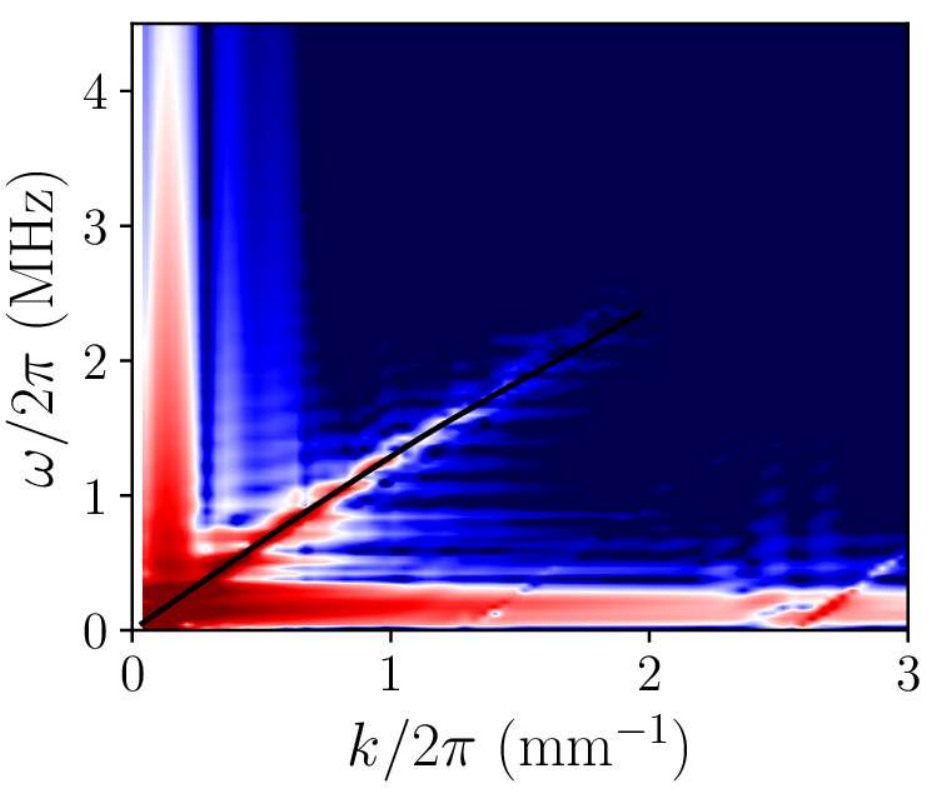}
		\caption{Cosine (0.19 MHz)}
		\label{fig:Mcos160d}
	\end{subfigure}
	\begin{subfigure}{0.246\textwidth}
		\centering
		\includegraphics[width=\textwidth]{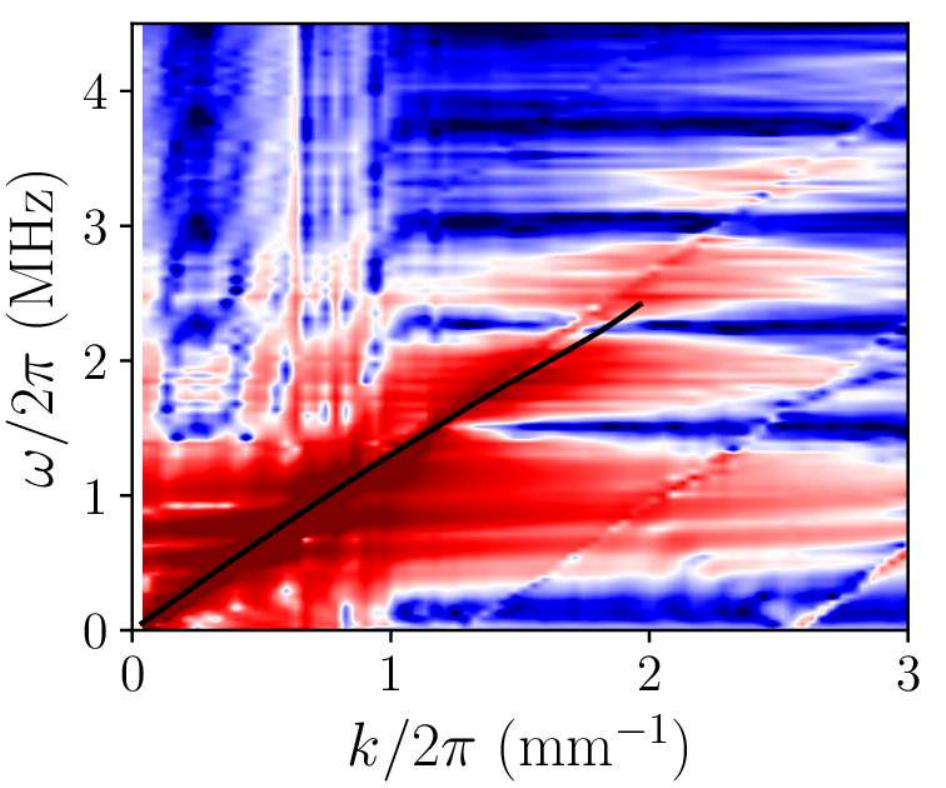}
		\caption{Sine (0.75 MHz)}
		\label{fig:Msin40d}
	\end{subfigure}
	\begin{subfigure}{0.246\textwidth}
		\centering
		\includegraphics[width=\textwidth]{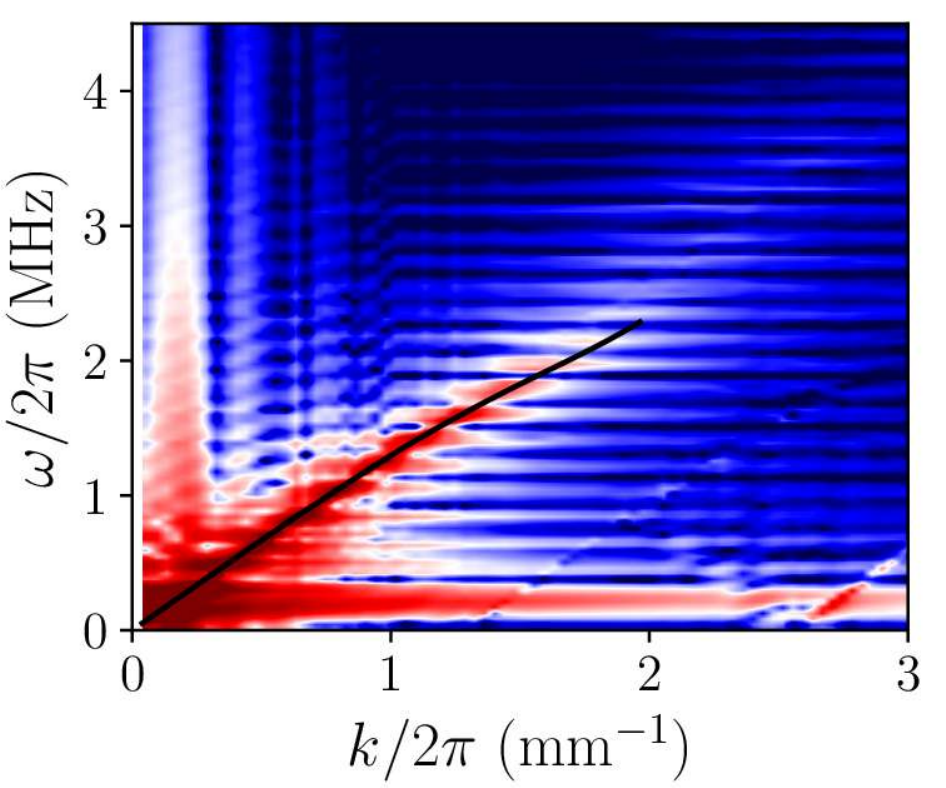}
		\caption{Sine (0.19 MHz)}
		\label{fig:Msin160d}
	\end{subfigure}
	\caption{Dependence of the P-wave dispersion relation on input waveforms and frequencies. Colors in each subplot show the power spectrum given by the discrete Fourier transformation.}
	\label{fig:diffSigVpFreq}
\end{figure}

\begin{figure} [t!]
	\begin{subfigure}{0.246\textwidth}
		\centering
		\includegraphics[width=\textwidth]{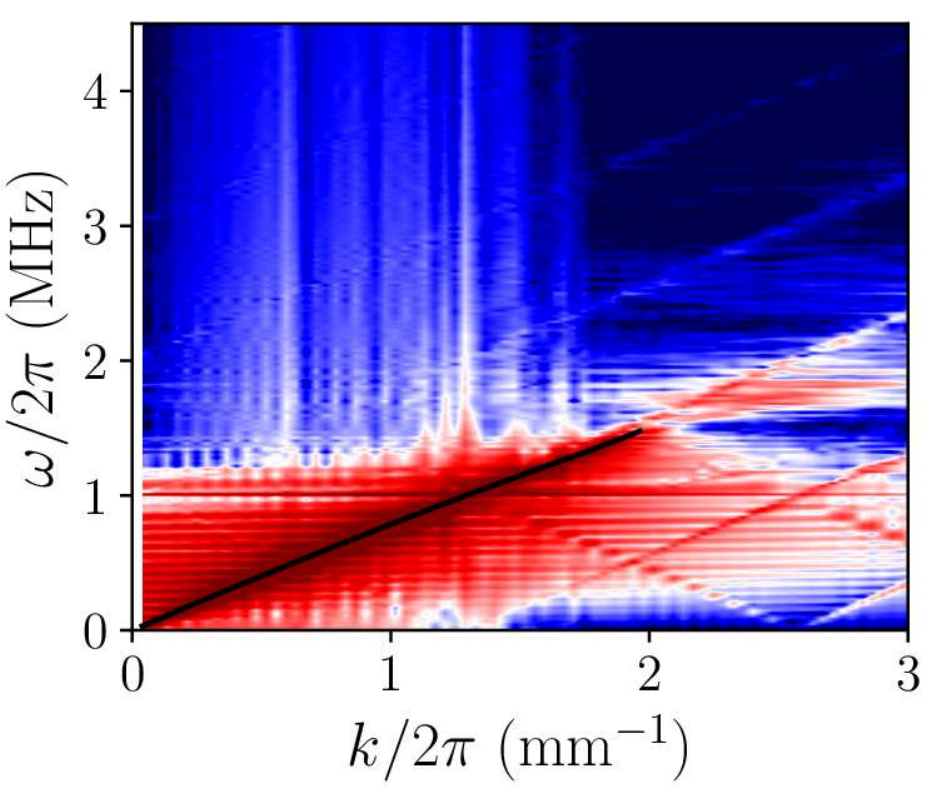}
		\caption{Cosine (0.75 MHz)}
		\label{fig:Gcos40d}
	\end{subfigure}
	\begin{subfigure}{0.246\textwidth}
		\centering
		\includegraphics[width=\textwidth]{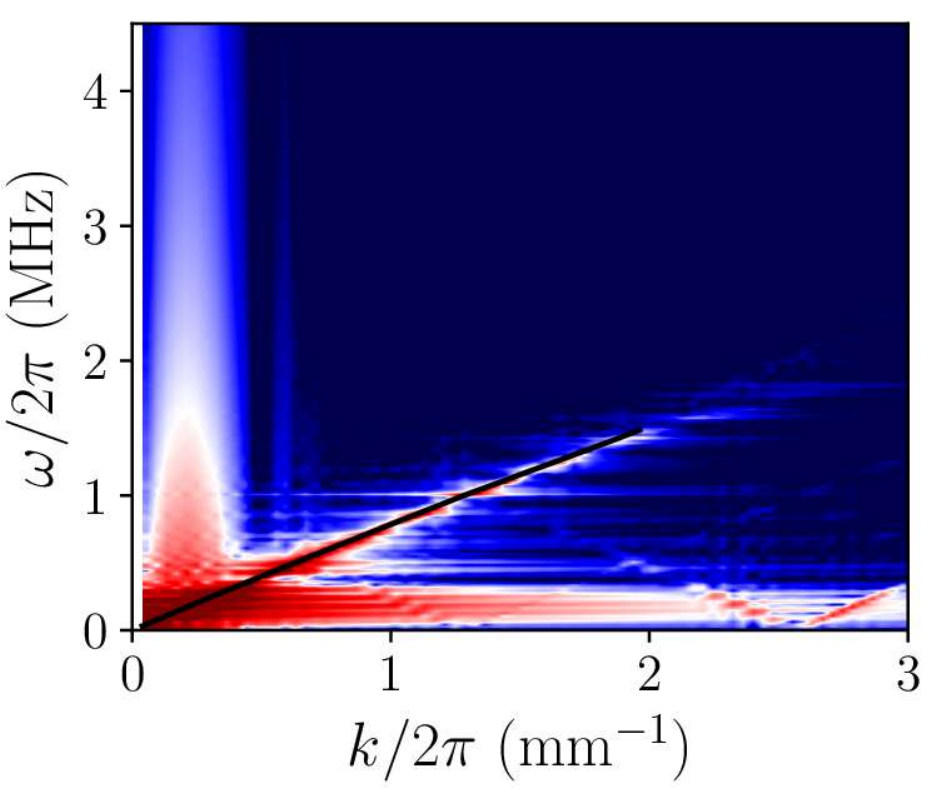}
		\caption{Cosine (0.19 MHz)}
		\label{fig:Gcos160d}
	\end{subfigure}
	\begin{subfigure}{0.246\textwidth}
		\centering
		\includegraphics[width=\textwidth]{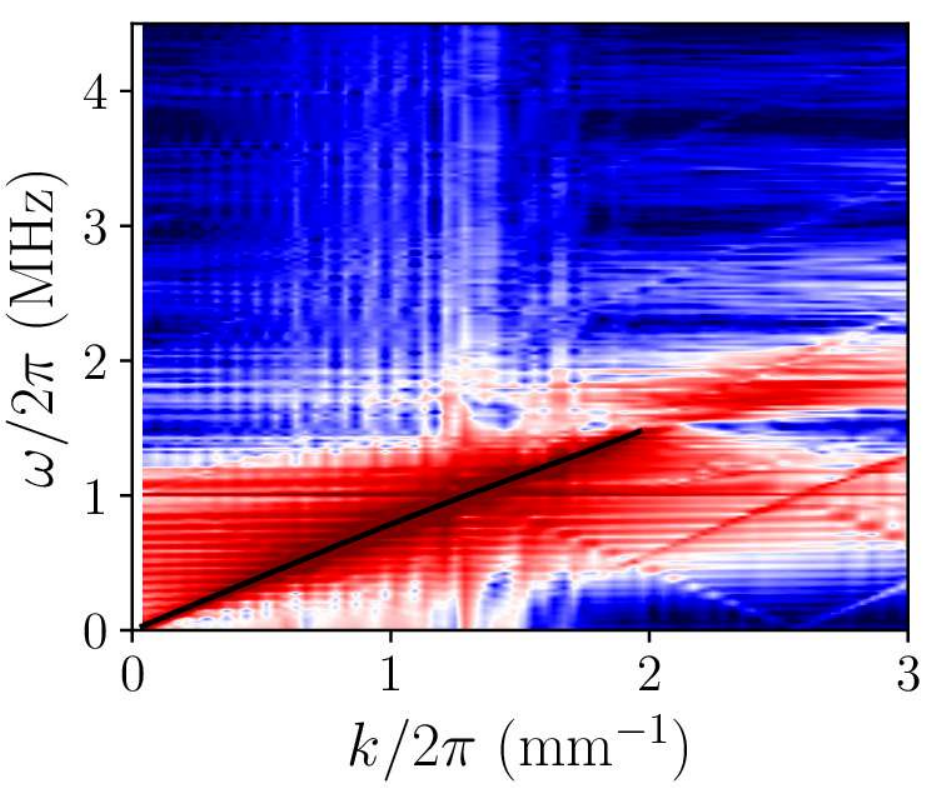}
		\caption{Sine (0.75 MHz)}
		\label{fig:Gsin40d}
	\end{subfigure}
	\begin{subfigure}{0.246\textwidth}
		\centering
		\includegraphics[width=\textwidth]{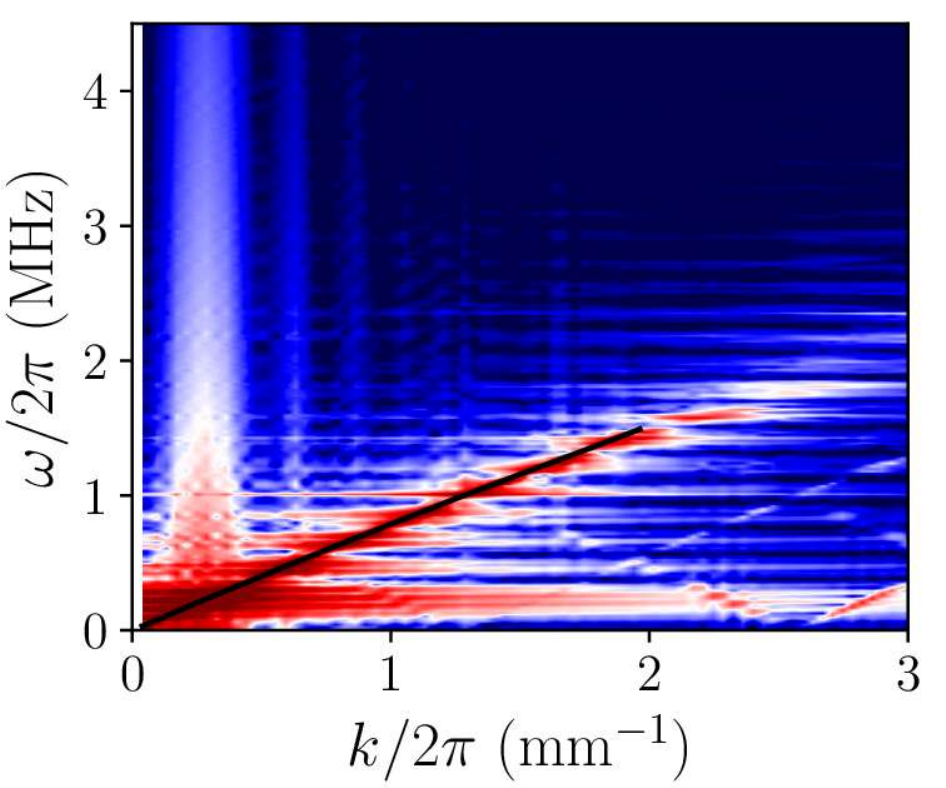}
		\caption{Sine (0.19 MHz)}
		\label{fig:Gsin160d}
	\end{subfigure}
	\caption{Dependence of the S-wave dispersion relation on input waveforms and frequencies. Colors in each subplot show the power spectrum given by the discrete Fourier transformation.}
	\label{fig:diffSigVsFreq}
\end{figure}

The dependence of the frequency domain responses on the input waveforms and frequencies for the P-waves and the S-waves is investigated using the same space-time data in the previous section.
The power spectra and the P- and S-wave dispersion relations are plotted in \figref{fig:diffSigVpFreq} and \ref{fig:diffSigVsFreq}, respectively.
For the sake of compactness, only the results obtained with the highest (0.75 MHz) and lowest input (0.19MHz) frequencies of the cosine and sine signals are shown in the paper
(the remaining plots in supplementary material B).
As the input frequency of the cosine signal decreases, the P- and S-wave dispersion branches shrink towards lower frequencies and smaller wavenumbers, and the activated frequency bands are increasingly narrowed, as shown in \figsref{fig:Mcos40d}, \ref{fig:Mcos160d} and \figsref{fig:Gcos40d}, \ref{fig:Gcos160d}.
Interestingly, for the sine input signals, with decreasing input frequency, the activated horizontal bands become denser and the band widths increasingly narrowed, as shown in \figsref{fig:Msin40d}, \ref{fig:Msin160d} and \figsref{fig:Gsin40d}, \ref{fig:Gsin160d}.

\begin{figure} [t!]
	\begin{subfigure}{0.5\textwidth}
		\centering
		\includegraphics[width=9cm]{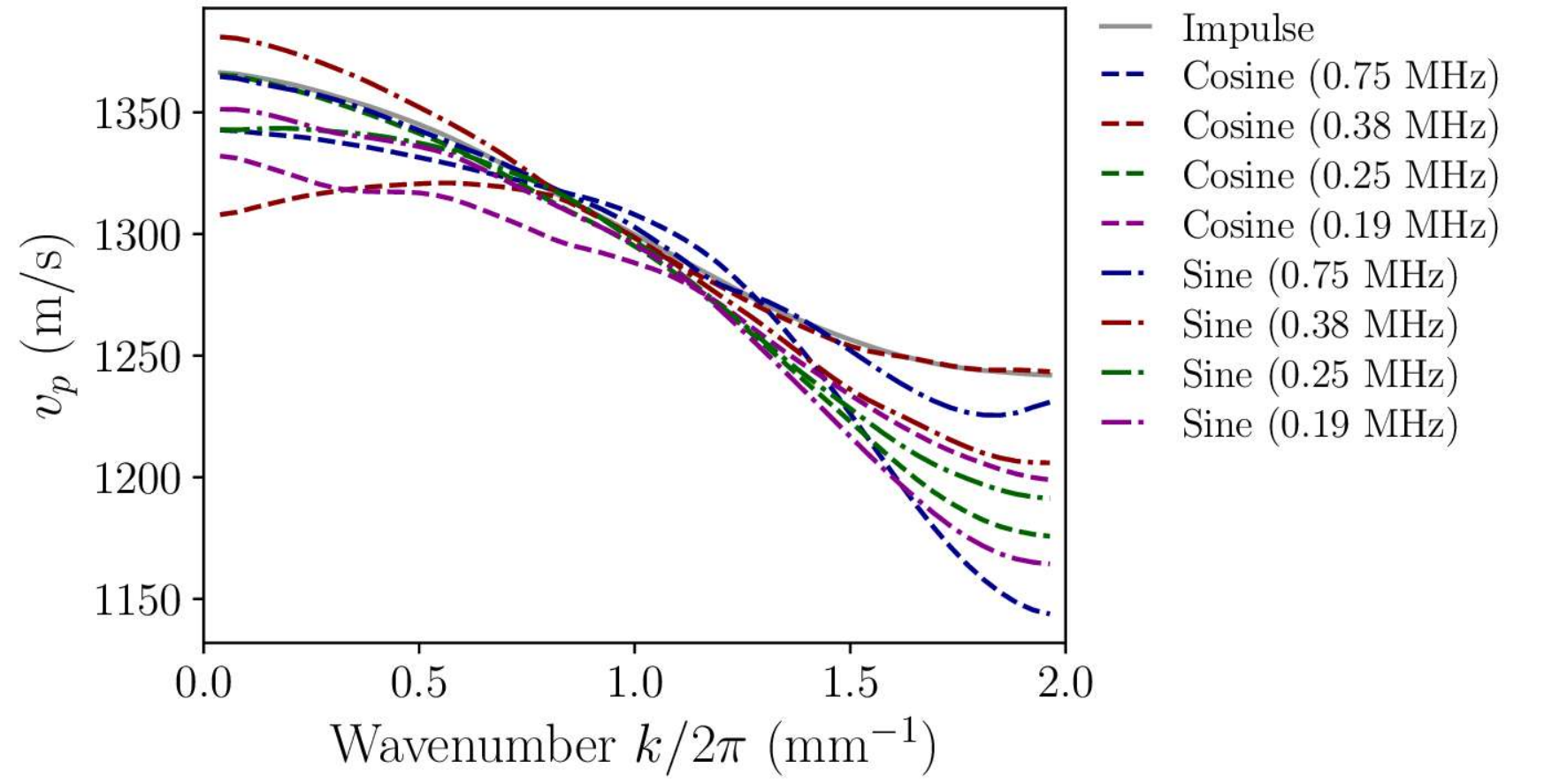}
		\caption{P-wave velocity versus wavenumber}
		\label{fig:diffSigVpKFreq}
	\end{subfigure}
	\begin{subfigure}{0.5\textwidth}
		\centering
		\includegraphics[width=9cm]{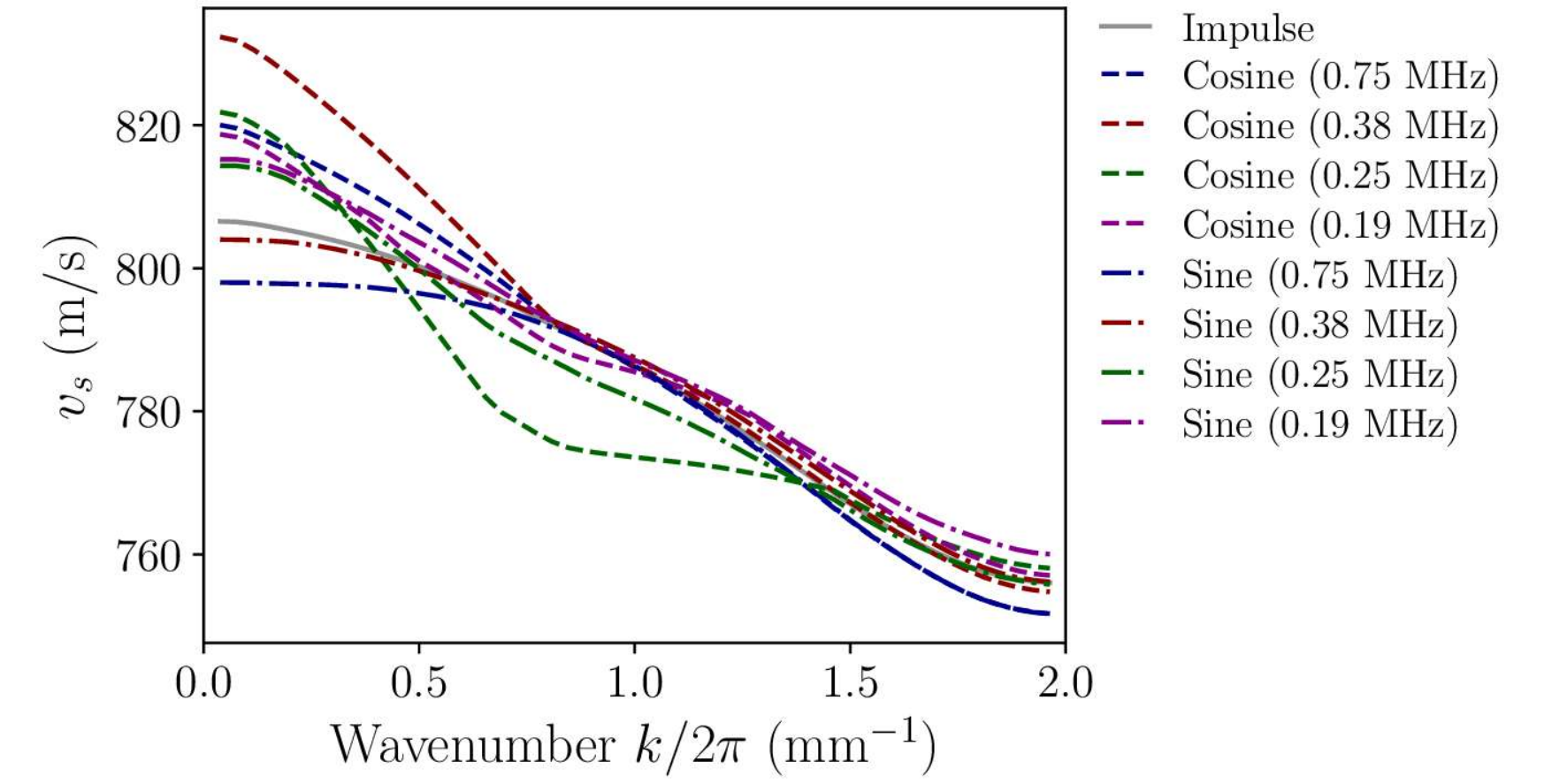}
		\caption{S-wave velocity versus wavenumber}
		\label{fig:diffSigVsKFreq}
	\end{subfigure}
	\caption{P-wave velocities (a) and S-wave velocities (b) as functions of wavenumber.}
	\label{fig:diffSigWaveKFreq}
\end{figure}

Irrespective of the different input waveforms and frequencies, the power spectra of both longitudinal and transverse particle velocities can be fitted with the Lorentzian function in \secref{sec:lenEffectFreq}, as shown in \figsref{fig:diffSigVpFreq} and \ref{fig:diffSigVsFreq}.
The dispersion relations obtained with various input waveforms and frequencies are plotted together against wavenumber in \figref{fig:diffSigWaveKFreq}.
It appears that the curves collapse at the intermediate wavenumbers, namely, $ k/2\pi \in (0.7, 1.3)$ mm$^{-1}$ for the P-wave and $ k/2\pi \in (0.7, 2.0)$ mm$^{-1}$ for the S-wave.
The worse agreement outside the intermediate wavenumber ranges can be attributed to i) the high-intensity horizontal bands in \figsref{fig:diffSigVpFreq} and \ref{fig:diffSigVsFreq} which correspond to the input frequencies and ii) the weak intensity of the power spectra in the high frequency regime.
In both cases, the quality of the Lorentzian fitting becomes worse, meaning that the dominant frequencies therein are very difficult to identify, which accounts for the underestimation of P-wave velocities at $ k/2\pi>1.3$ mm$^{-1}$ compared with the ones obtained with the impulse.
Therefore, to accurately characterize the dispersion relations from small to large wavenumbers, impulse signals will be used for the dynamic probing along the cyclic oedometric stress path.
Compared with the time domain analysis which shows a strong dependence of the wave velocities on input waveform and frequency, the frequency domain analysis estimates the wave velocities with much smaller deviation, and the dispersion curves fitted with the Lorentzian are more consistent for different input signals.

\subsection{Evolution of elastic wave velocities during cyclic oedometric compression}

Based on the discussions above, the impulse signals in \tabref{tab:source} are applied to the particles in the source layer, in order to agitate elastic waves in the granular column, at various stress states on the oedometric stress path.
The frequency domain technique is employed to determine the wave velocities from the space-time evolution of particle motion, as the influence of packing length, input waveform and frequency on the dispersion relations is negligible.

\subsubsection{Evolution of P-wave and S-wave dispersion relations}
\label{sec:evolveDispersion}
\begin{figure} [t!]
	\begin{subfigure}{0.246\textwidth}
		\centering
		\includegraphics[width=\textwidth]{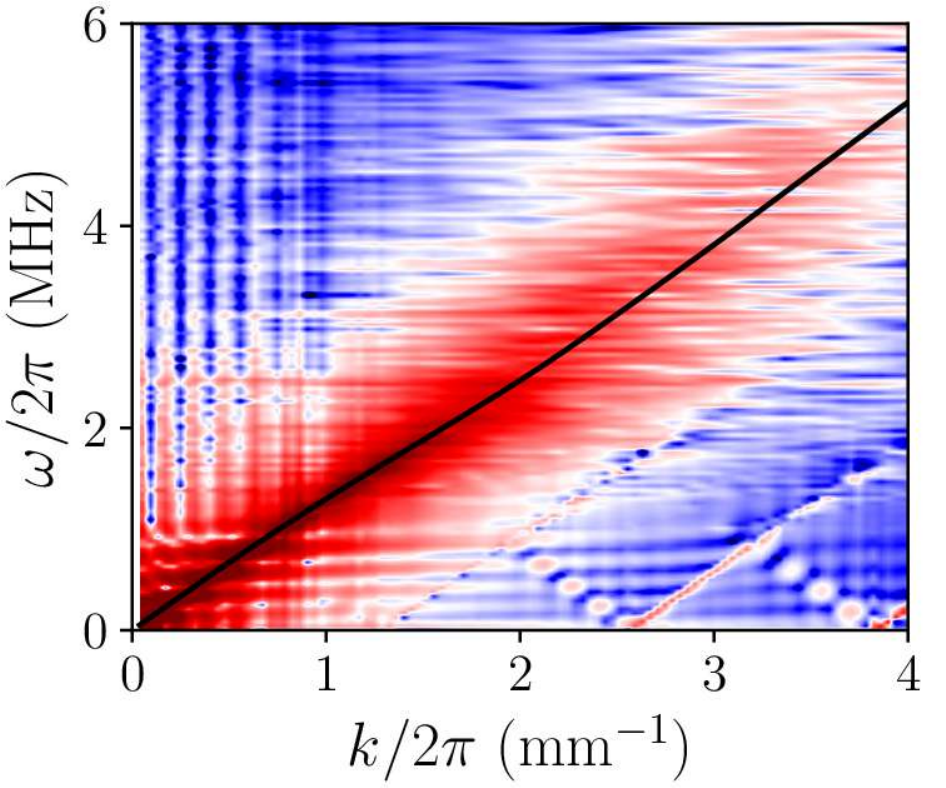}
		\caption{P-wave: $ \varepsilon_a^{\text{\tiny $\bigcirc$}} = 0.11\% $}
		\label{fig:M0.11}
	\end{subfigure}
	\begin{subfigure}{0.246\textwidth}
		\centering
		\includegraphics[width=\textwidth]{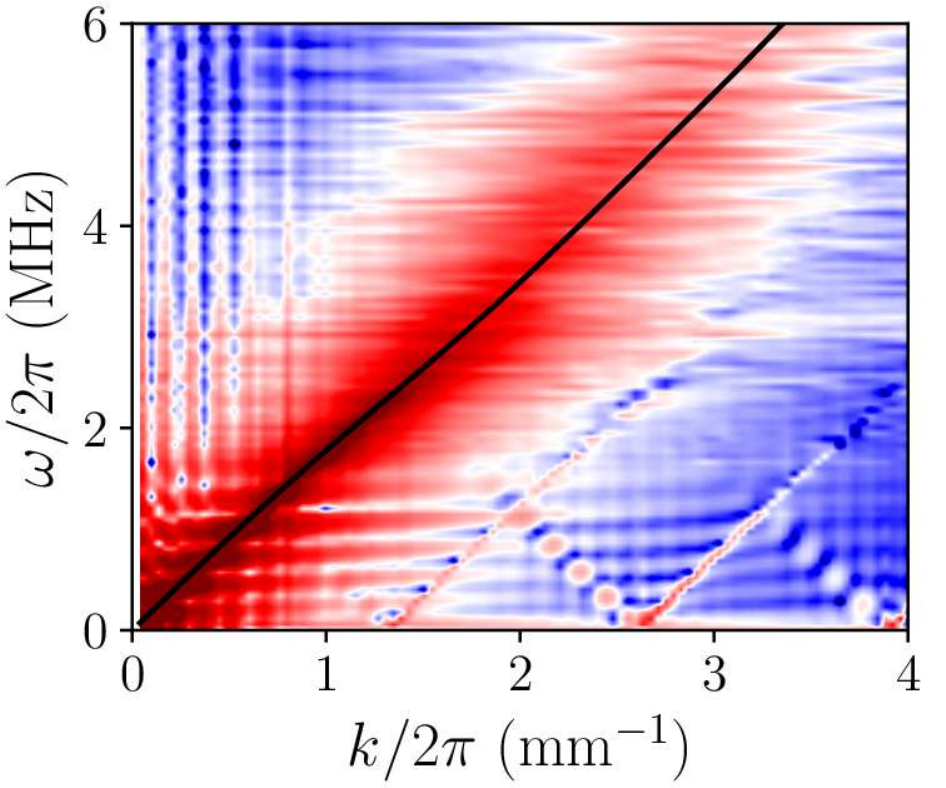}
		\caption{P-wave: $ \varepsilon_a^{\bigtriangleup} = 1.68\% $}
		\label{fig:M1.68}
	\end{subfigure}
	\begin{subfigure}{0.246\textwidth}
		\centering
		\includegraphics[width=\textwidth]{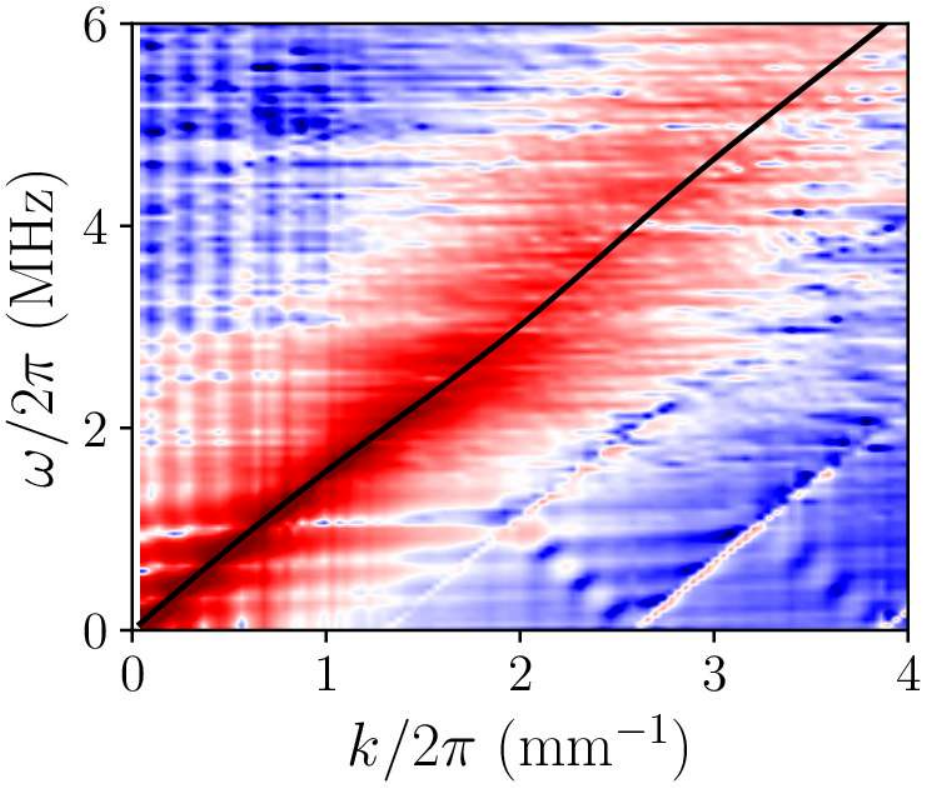}
		\caption{P-wave: $ \varepsilon_a^{\bigtriangledown} = 1.09\% $}
		\label{fig:M1.09}
	\end{subfigure}
	\begin{subfigure}{0.246\textwidth}
		\centering
		\includegraphics[width=\textwidth]{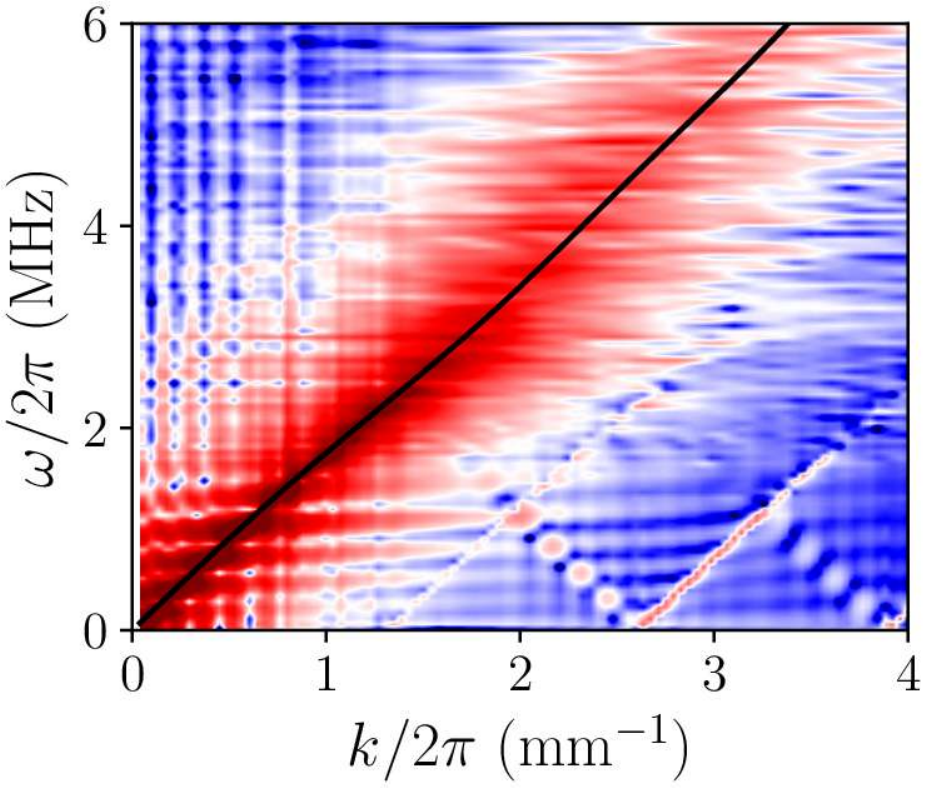}
		\caption{P-wave: $ \varepsilon_a^{\square} = 1.62\% $}
		\label{fig:M1.62}
	\end{subfigure} \\
	\begin{subfigure}{0.246\textwidth}
		\centering
		\includegraphics[width=\textwidth]{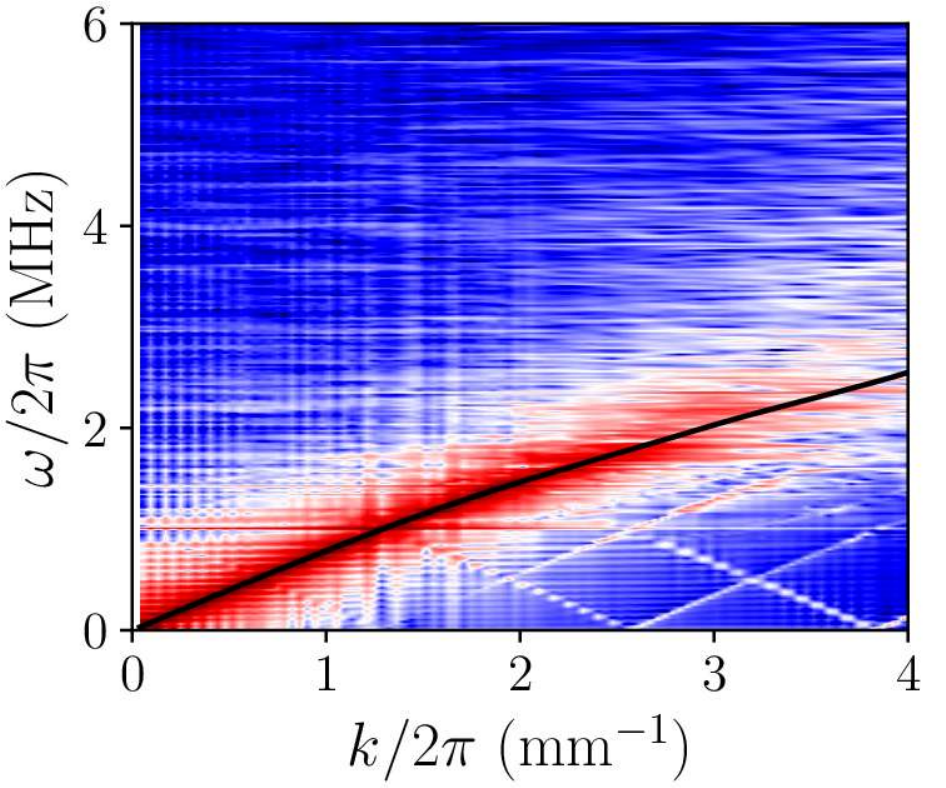}
		\caption{S-wave: $ \varepsilon_a^{\text{\tiny $\bigcirc$}} = 0.11\% $}
		\label{fig:G0.11}
	\end{subfigure}
	\begin{subfigure}{0.246\textwidth}
		\centering
		\includegraphics[width=\textwidth]{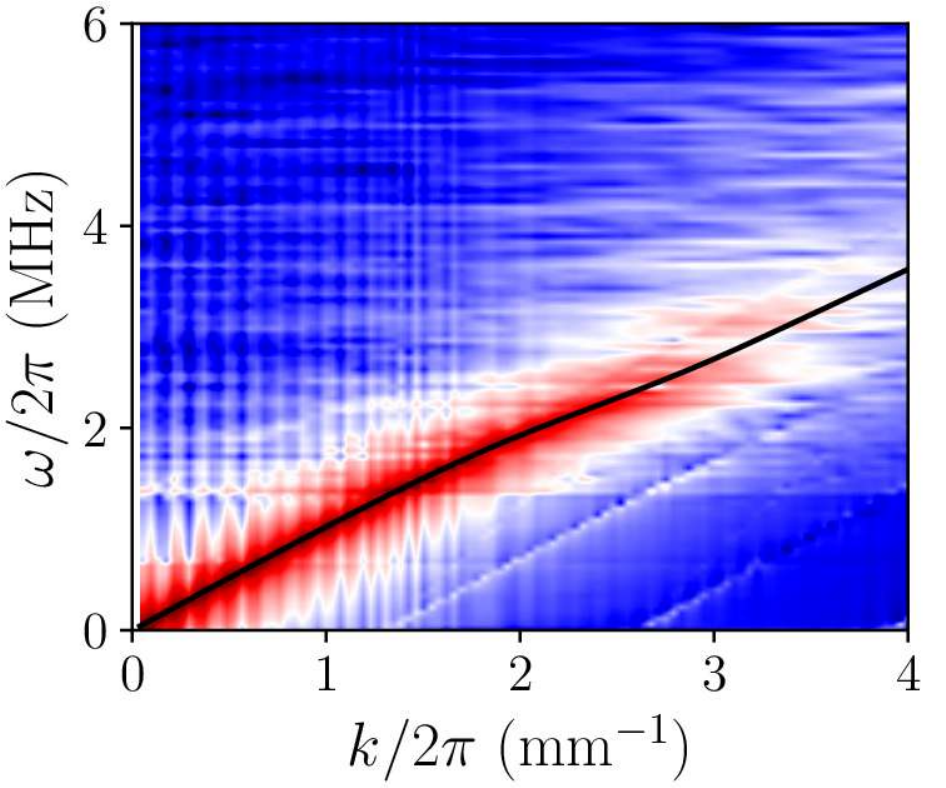}
		\caption{S-wave: $ \varepsilon_a^{\bigtriangleup} = 1.68\% $}
		\label{fig:G1.68}
	\end{subfigure}
	\begin{subfigure}{0.246\textwidth}
		\centering
		\includegraphics[width=\textwidth]{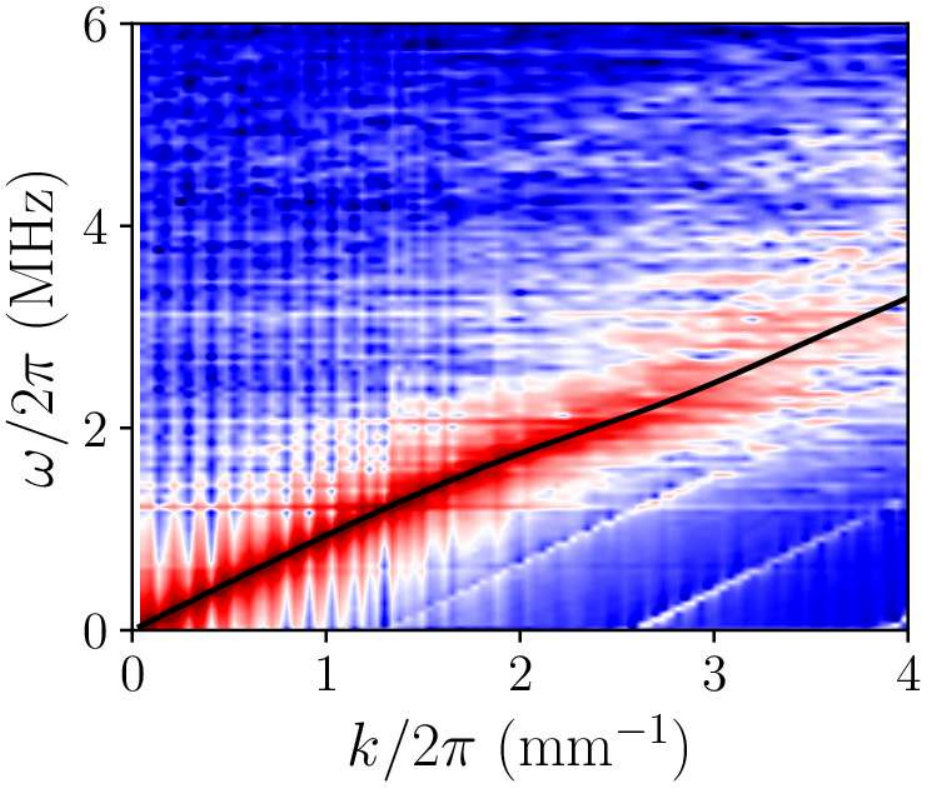}
		\caption{S-wave: $ \varepsilon_a^{\bigtriangledown} = 1.09\% $}
		\label{fig:G1.09}
	\end{subfigure}
	\begin{subfigure}{0.246\textwidth}
		\centering
		\includegraphics[width=\textwidth]{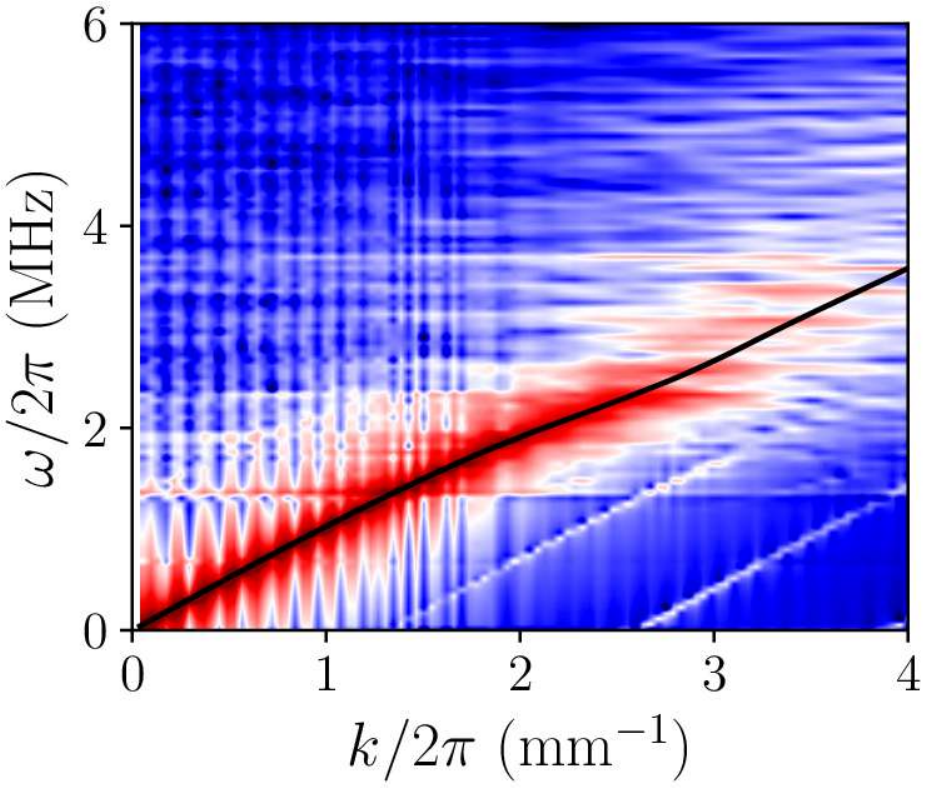}
		\caption{S-wave: $ \varepsilon_a^{\square} = 1.62\% $}
		\label{fig:G1.62}
	\end{subfigure}
	\caption{Dependence of the P-wave dispersion relation (a--d) and the S-wave dispersion relation (e--h) on the stress states during cyclic oedometric compression. Colors in each subplot show the power spectrum given by the discrete Fourier transformation.}
	\label{fig:AnisoWaveFreq}
\end{figure}

For the sake of compactness, we only show in \figsref{fig:M0.11}, \ref{fig:G0.11}, \figsref{fig:M1.68}, \ref{fig:G1.68}, \figsref{fig:M1.09}, \ref{fig:G1.09} and \figsref{fig:M1.62}, \ref{fig:G1.62} the power spectra at the \emph{beginning} of: the first load ($ \varepsilon_a^{\text{\tiny $\bigcirc$}} = 0.11\% $), 
the first unload ($ \varepsilon_a^{\bigtriangleup} = 1.68\% $), 
the first reload ($ \varepsilon_a^{\bigtriangledown} = 1.09\% $) 
and the second unload ($ \varepsilon_a^{\square} = 1.62\% $), respectively.
Although not included in \figref{fig:AnisoWaveFreq} (see supplementary material C), the power spectra at a total number of 30 stress states on the cyclic oedometric path are well fitted with the Lorentzian function.
The increase and decrease of the slopes of the dispersion curves during unload and reload can be clearly observed from the subplots in \figref{fig:AnisoWaveFreq}.
Note that the secondary increase of the phase velocities, which is a distinguished feature of granular systems \cite{Saitoh2018}, are captured as well, for both P- and S-waves.

\begin{figure} [t!]
	\begin{subfigure}{0.33\textwidth}
		\centering
		\includegraphics[width=6cm]{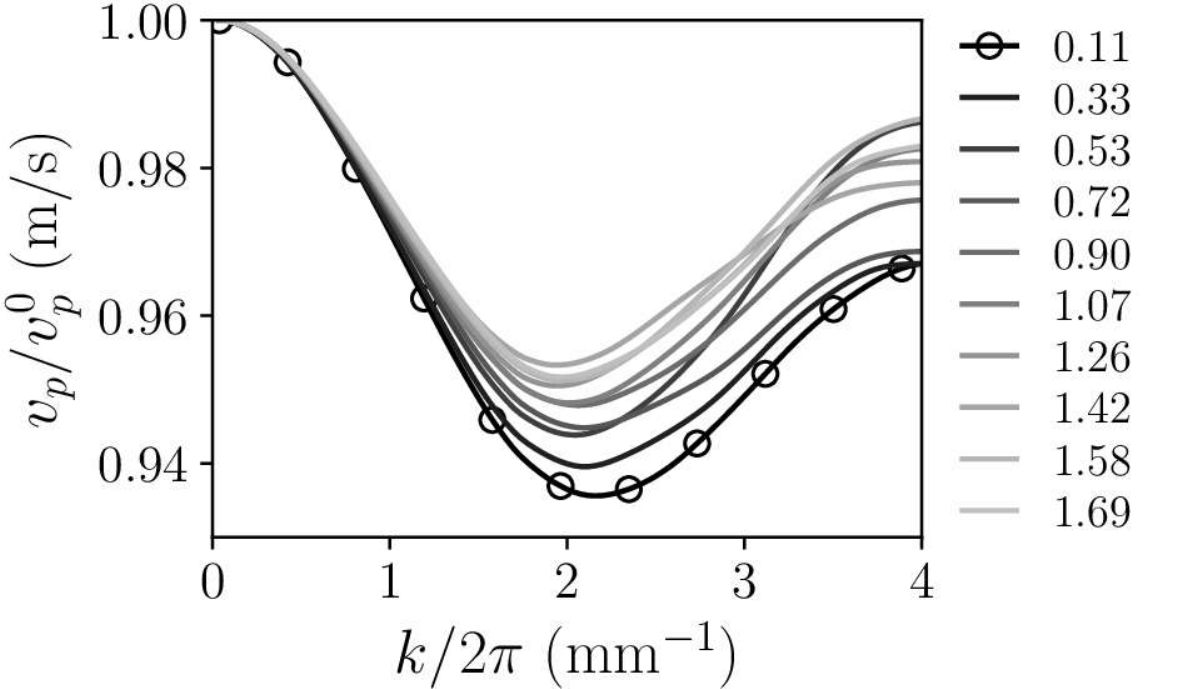}
		\caption{$ v_p $: first load}
		\label{fig:AnisoVp1stLoad}
	\end{subfigure}
	\begin{subfigure}{0.33\textwidth}
		\centering
		\includegraphics[width=6cm]{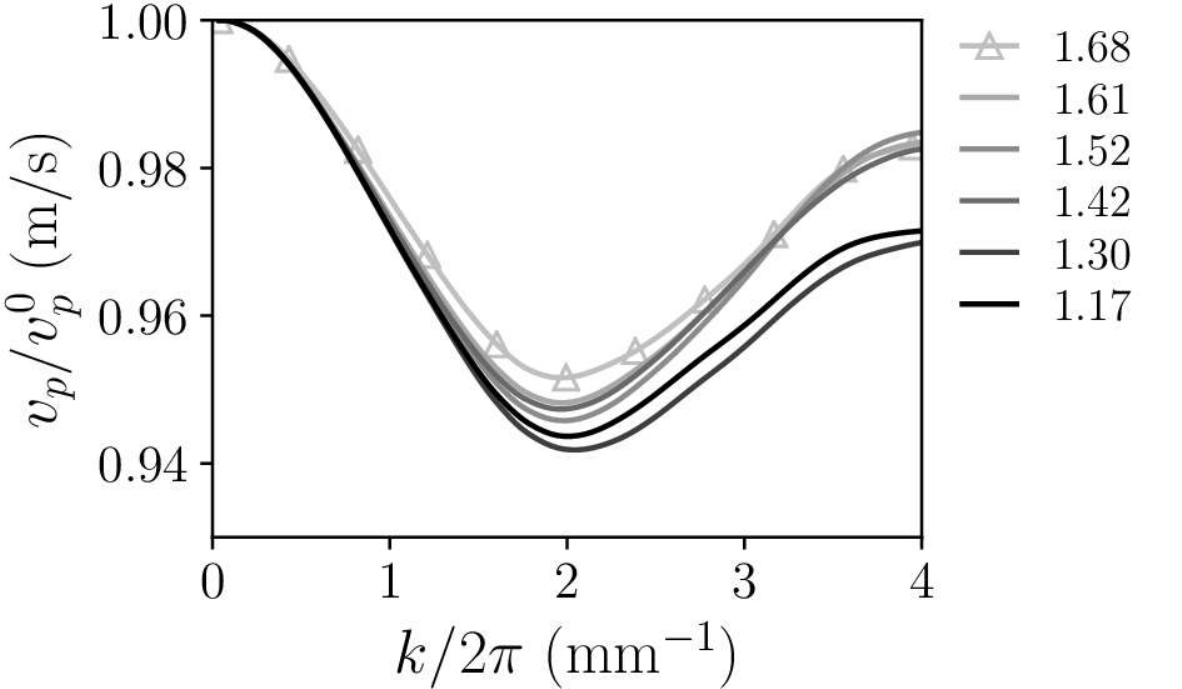}
		\caption{$ v_p $: first unload}
		\label{fig:AnisoVp1stUnload}
	\end{subfigure}
	\begin{subfigure}{0.33\textwidth}
		\centering
		\includegraphics[width=6cm]{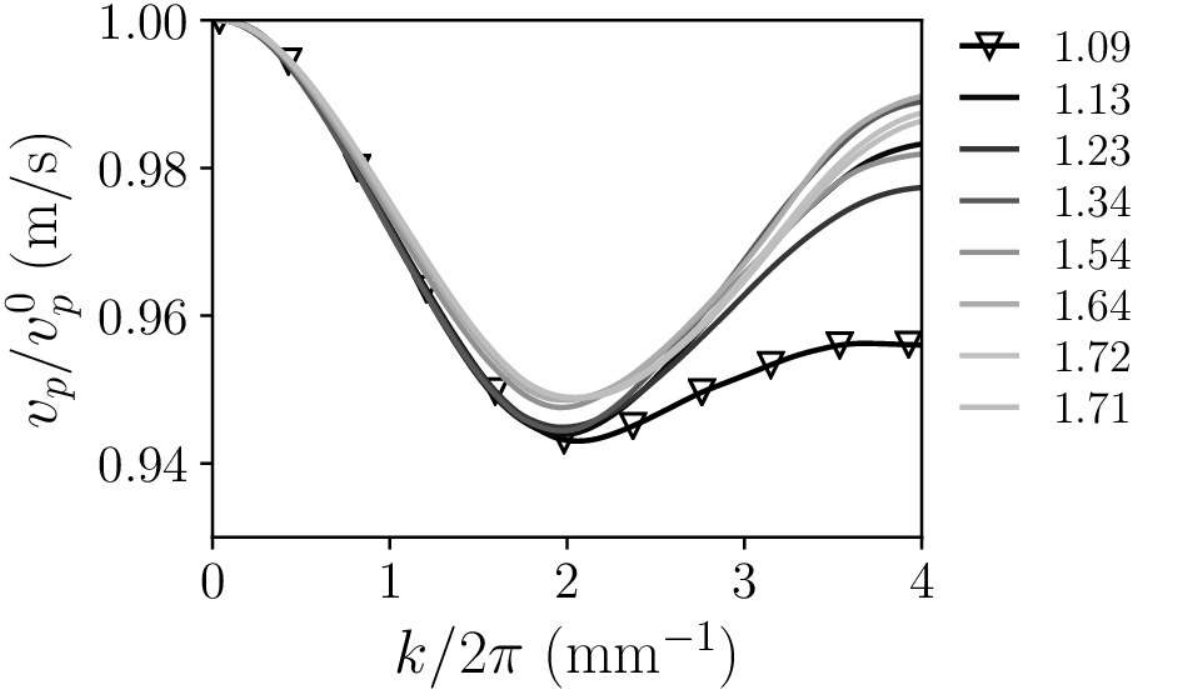}
		\caption{$ v_p $: first reload}
		\label{fig:AnisoVpsecondLoad}
	\end{subfigure}	\\
	\begin{subfigure}{0.33\textwidth}
		\centering
		\includegraphics[width=6cm]{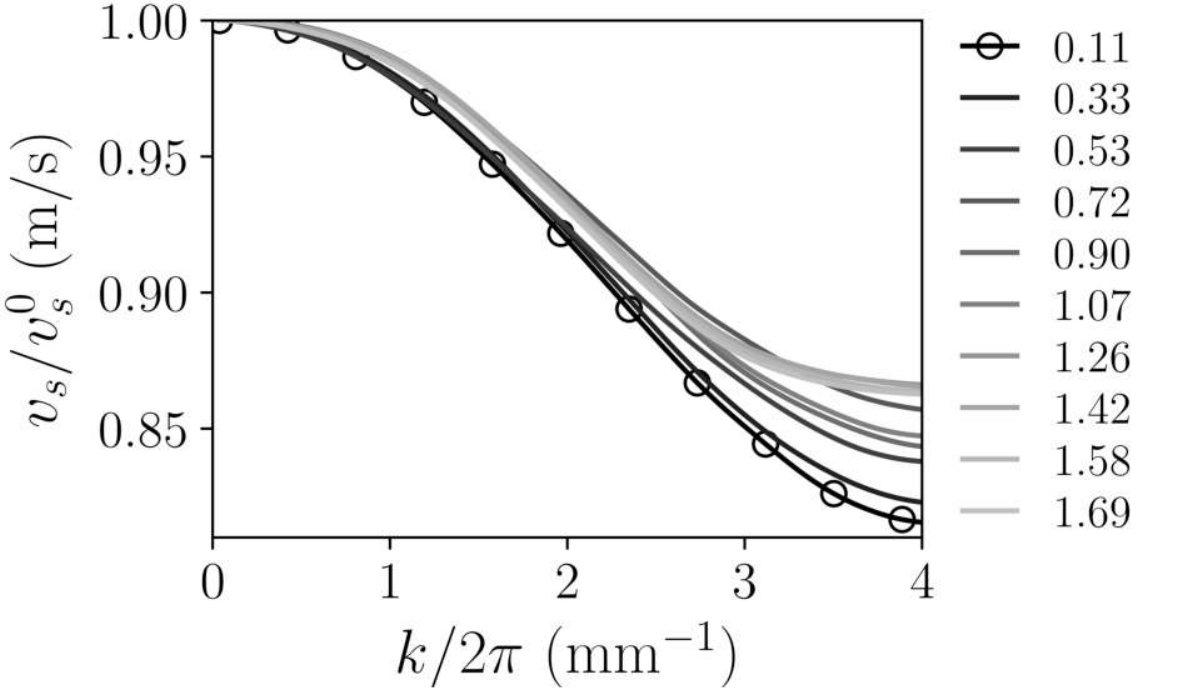}
		\caption{$ v_s $: first load}
		\label{fig:AnisoVs1stLoad}
	\end{subfigure}
	\begin{subfigure}{0.33\textwidth}
		\centering
		\includegraphics[width=6cm]{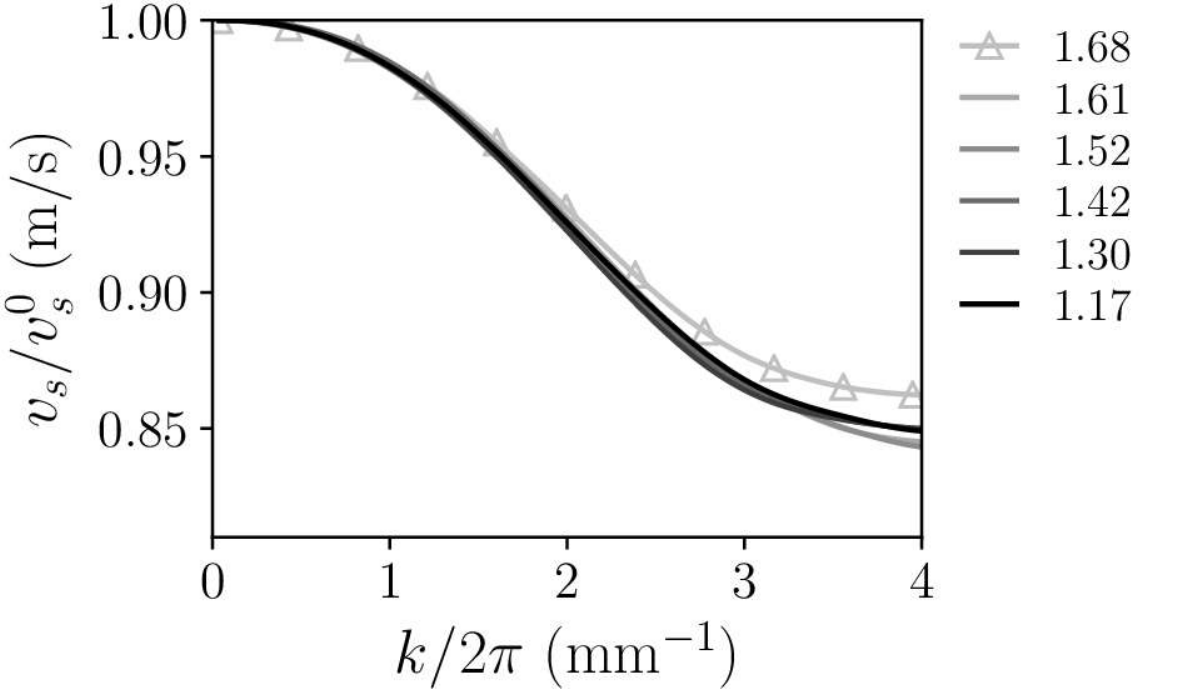}
		\caption{$ v_s $: first unload}
		\label{fig:AnisoVs1stUnload}
	\end{subfigure}
	\begin{subfigure}{0.33\textwidth}
		\centering
		\includegraphics[width=6cm]{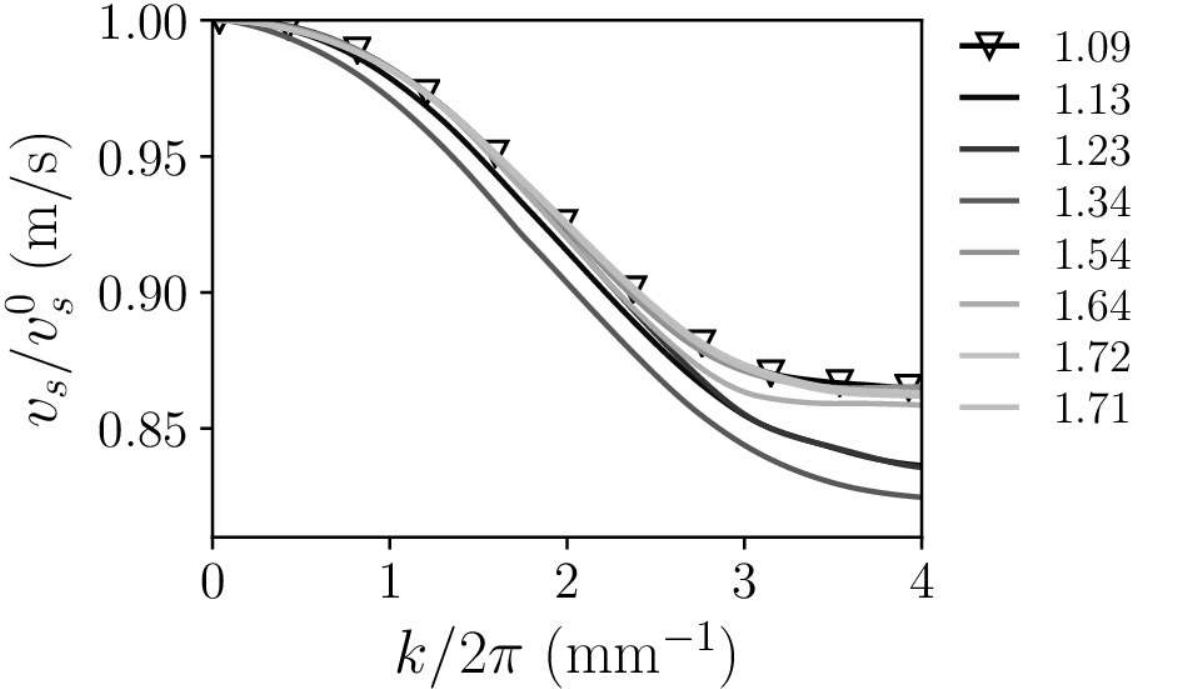}
		\caption{$ v_s $: first reload}
		\label{fig:AnisoVssecondLoad}
	\end{subfigure}
	\caption{Evolution of P-wave velocities (a--d) and S-wave velocities (e-h) as functions of wavenumber during cyclic oedometric compression. Different levels of axial strain $\varepsilon_a$ in percentage are indicated by the different gray levels.}
	\label{fig:diffAnisoWave}
\end{figure}

In order to understand how stress history and anisotropy affect the P- and S-wave dispersion relations, each dispersion curve is scaled by the long-wavelength wave velocity $ v_p^0 $ or $ v_s^0 $ for $ k\to0 $.
\figref{fig:AnisoVp1stLoad}--\ref{fig:AnisoVpsecondLoad} and \figref{fig:AnisoVs1stLoad}--\ref{fig:AnisoVssecondLoad} respectively show the normalized P- and S-wave dispersion relations at the stress states on the oedometric loading/unloading branches.
During the first load, the nonlinearity of all dispersion curves reduces, as shown in \figsref{fig:AnisoVp1stLoad} and \ref{fig:AnisoVs1stLoad}.
The largest decrease of wave velocities with respect to wavenumber , changes from $ \max(1-v_p(k)/v_p^0)=6.5\%$ to 4.5\% for the P-waves and from $ \max(1-v_p(k)/v_p^0)=20.0\%$ to 13.0\% for the S-waves, with increasing axial strain.
It appears that the variation of the scaled dispersion curves become almost negligible, after the axial strain exceeds 1.26 \% during the first load.
As expected, the dispersion curves turns to be increasingly nonlinear, with decreasing axial strain during the first unload (see \figsref{fig:AnisoVp1stUnload} and \ref{fig:AnisoVs1stUnload}).
Although the trend is reversed again during the first reload, the difference between the dispersion curves becomes less pronounced, which is similar to the first unload.
In particular, the P- and S-wave dispersion curves in \figsref{fig:AnisoVpsecondLoad} and \ref{fig:AnisoVssecondLoad} collapse excellently, for the wavenumber $ k/2\pi<2.0$ mm$^{-1}$ and $ k/2\pi<3.0$ mm$^{-1}$, respectively.
From the collapsed dispersion curves, it can be deduced that cyclic loading makes the shape of dispersion relations independent of stress history and anisotropy, which means the dispersive relations are scalable with the long-wavelength wave velocities.
Note that the increase of P-wave velocities, after $ k/2\pi$ passes approximately 2.0 mm$^{-1}$, is also well captured at each stress state on the oedometric path.
From \figref{fig:AnisoVs1stLoad}--\ref{fig:AnisoVssecondLoad}, it is also interesting to note that the S-wave dispersion curves become less nonlinear for $ k/2\pi>3.0$ mm$^{-1}$, as the granular column undergoes more compression/decompression on each loading/unloading branch.

\subsection{Comparisons of wave velocities obtained from static and dynamic probing}
\label{sec:compareProbs}

\begin{figure} [t!]
	\begin{subfigure}{0.5\textwidth}
		\centering
		\includegraphics[width=8cm]{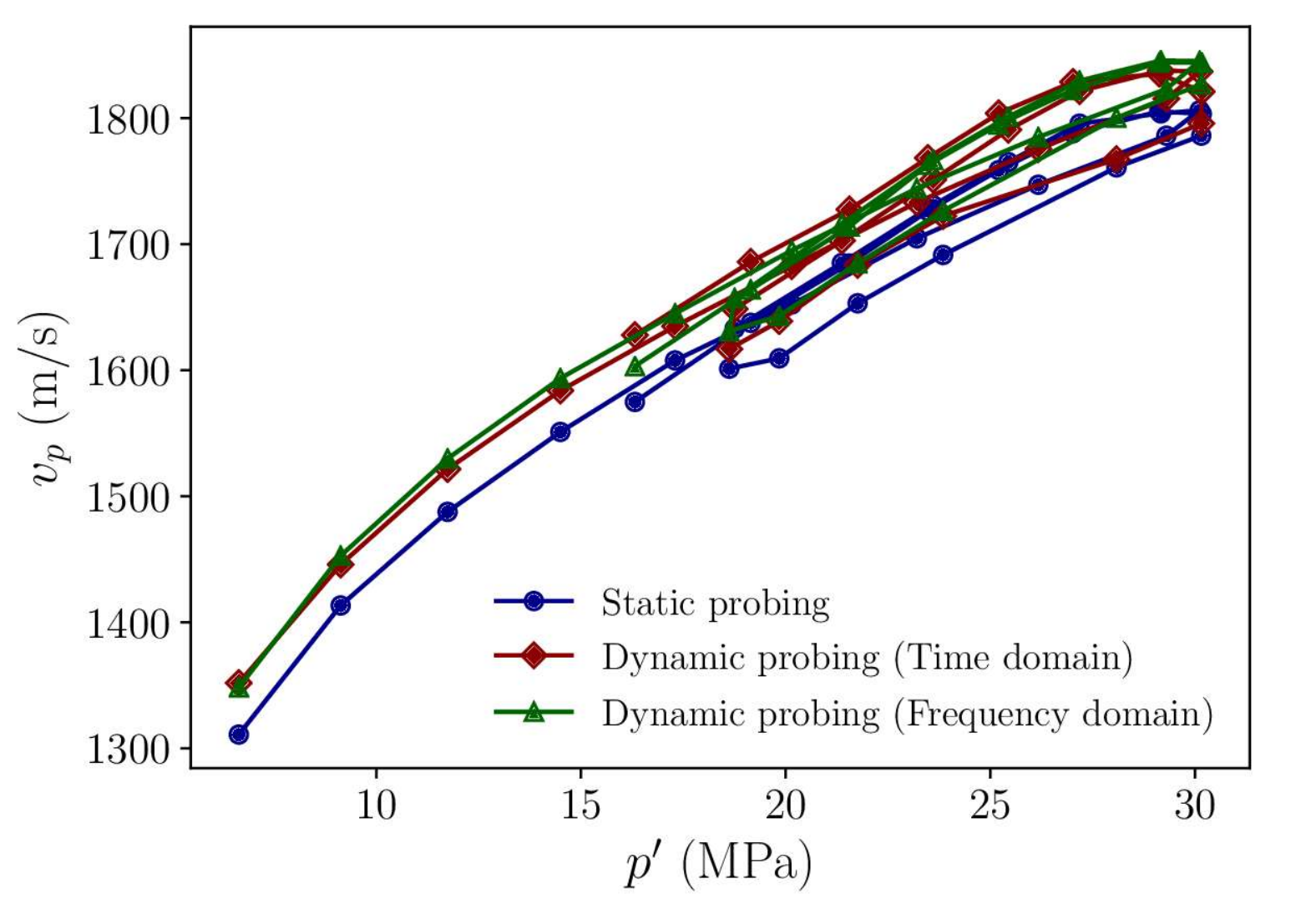}
		\caption{P-wave velocity versus mean effective stress}
		\label{fig:compareVp}
	\end{subfigure}
	\begin{subfigure}{0.5\textwidth}
		\centering
		\includegraphics[width=8cm]{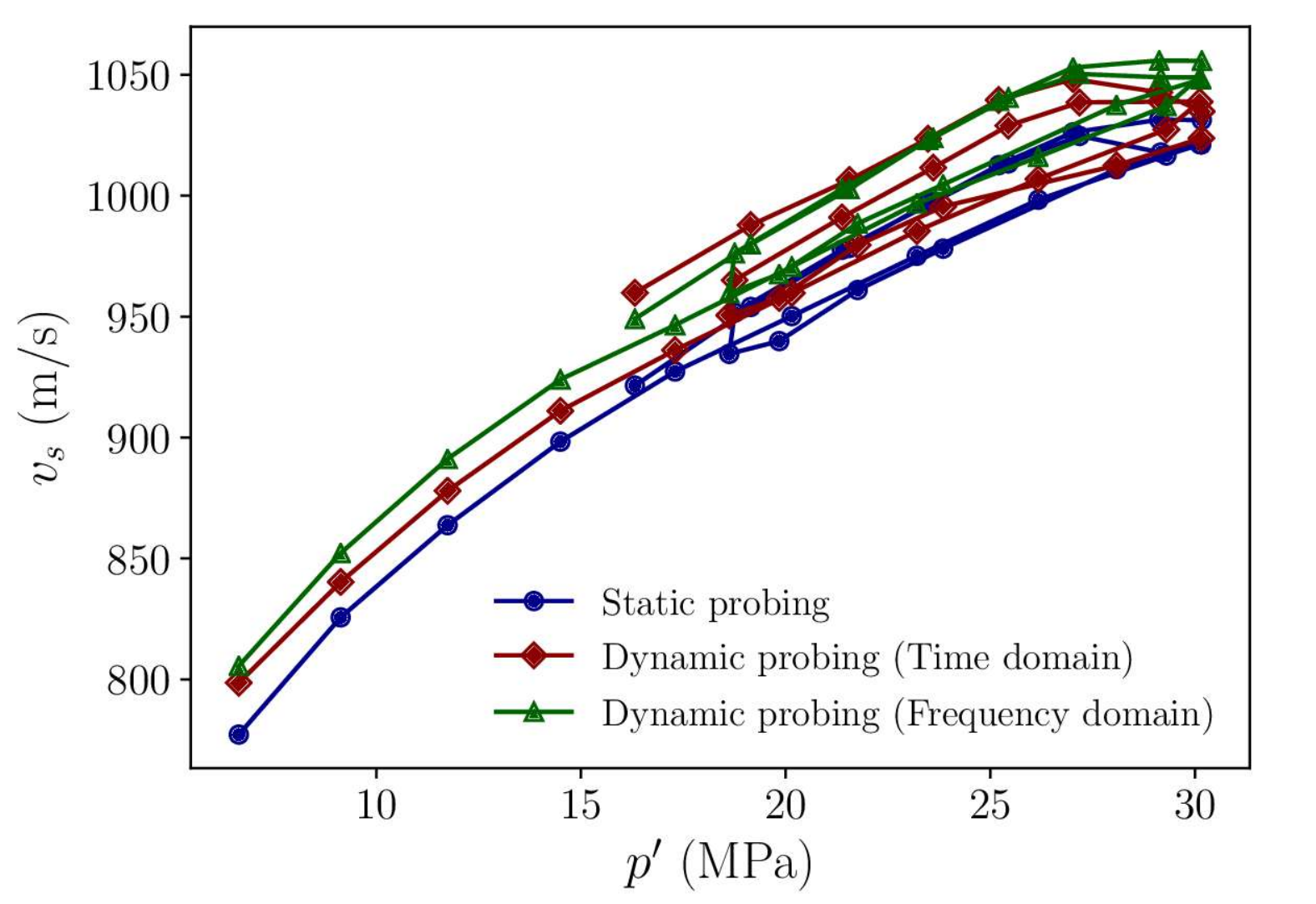}
		\caption{S-wave velocity versus mean effective stress}
		\label{fig:compareVs}
	\end{subfigure}
	\caption{Comparison of elastic wave velocities estimated by static and dynamic probing versus mean effective stress over the cyclic oedometric stress path.}
	\label{fig:compare}
\end{figure}

Finally, the wave velocities estimated by static and dynamic probing (from both the time domain and the frequency domain) are compared and plotted against the mean effective stress $p'$ in \figref{fig:compare}.
While the frequency-domain results are extracted from the dispersion curves at $k\to0$ in \figref{fig:diffAnisoWave} using the impulse, the cosine signal with $\omega/2\pi=0.38$ MHz is adopted for the time-domain analysis (see the stagnant wave velocity profile in \figsref{fig:diffSigVpTime} and \ref{fig:diffSigVsTime}).
The wave velocities obtained from static probing in \secref{sec:staModuli} are deduced from the elastic moduli, and compared with those from dynamic probing, as displayed also in \figref{fig:compare}.

The elastic wave velocities obtained from the time domain and frequency domain with the dynamic probes excellently overlap.
The well-matched pressure--wave velocity relationships confirms that the long-wavelength limits are indeed attained with the right choice of packing length, input waveform and frequency.
The accuracy between the frequency and time domain analyses is comparable.
However, the same level of accuracy, the granular packing (33$\times$RV) for the time domain analysis contains three times more particles than the packing (11$\times$RV) for the frequency domain analysis, meaning that time domain analysis is less computationally efficient.
Moreover, in determining the appropriate input signals, a variety of input waveforms and frequencies need to be investigated, which brings more computational cost.
Nevertheless, determining appropriate input signals by means of DEM simulations can facilitate the bender element testing of granular soils.
The dispersion curves obtained from the frequency domain, on the other hand, show negligible dependence on the input signals.

Both \figsref{fig:compareVp} and \ref{fig:compareVs} show that the wave velocities deduced from the elastic moduli given by \emph{static} probing are systematically lower than those inferred by dynamic probing.
It seems that the representative volumes (RVs) can only reflect the wave velocities at certain wavelengths constrained to the size of the representative volume.
Although the dynamic probing can accurately estimate the long-wavelength wave velocities and reproduce the dispersion phenomenon in granular media, the degradation of small-strain moduli with increasing perturbation magnitude can only be reproduced by means of static probing.

\subsection{Evolution and density of states}
\label{sec:rhoStates}

\begin{figure} [t!]
	\begin{subfigure}{0.33\textwidth}
		\centering
		\includegraphics[width=6cm]{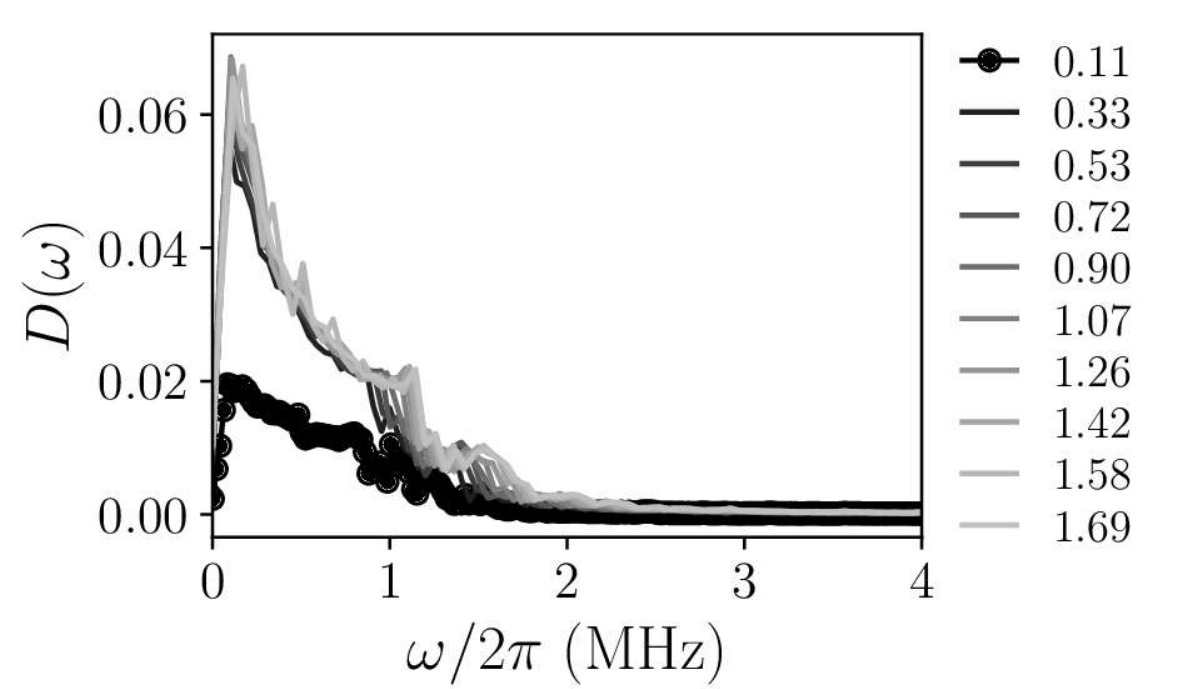}
		\caption{First load}
		\label{fig:rhoState1stLoad}
	\end{subfigure}
	\begin{subfigure}{0.33\textwidth}
		\centering
		\includegraphics[width=6cm]{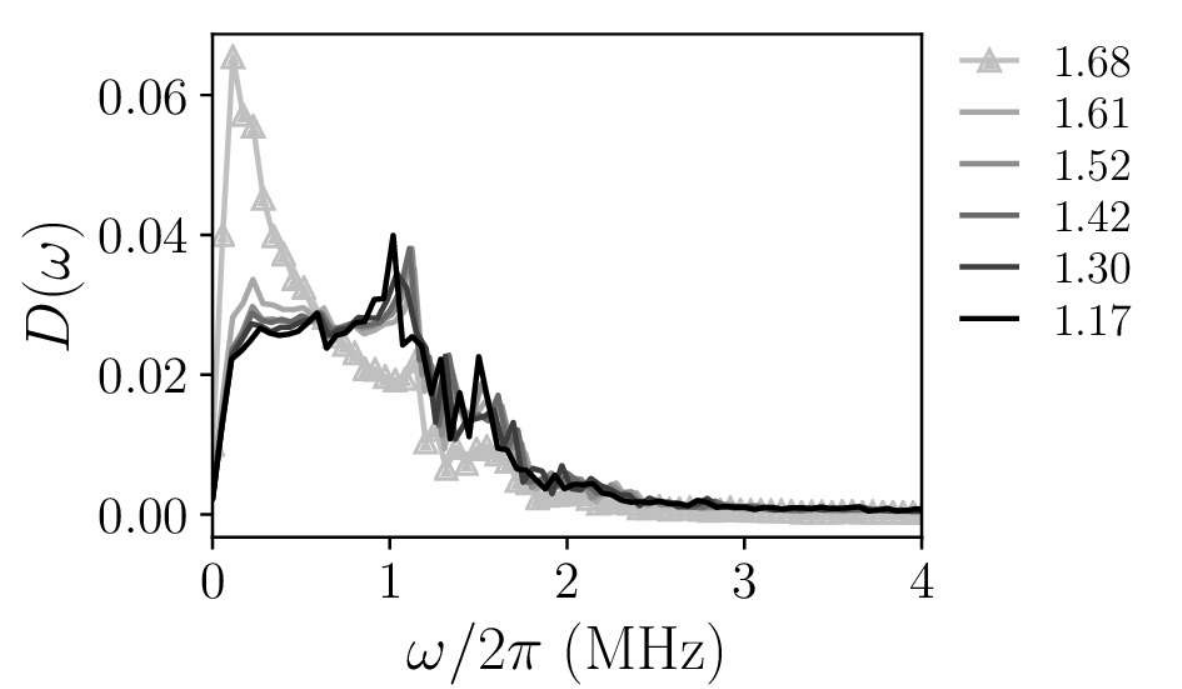}
		\caption{First unload}
		\label{fig:rhoState1stUnload}
	\end{subfigure}
	\begin{subfigure}{0.33\textwidth}
		\centering
		\includegraphics[width=6cm]{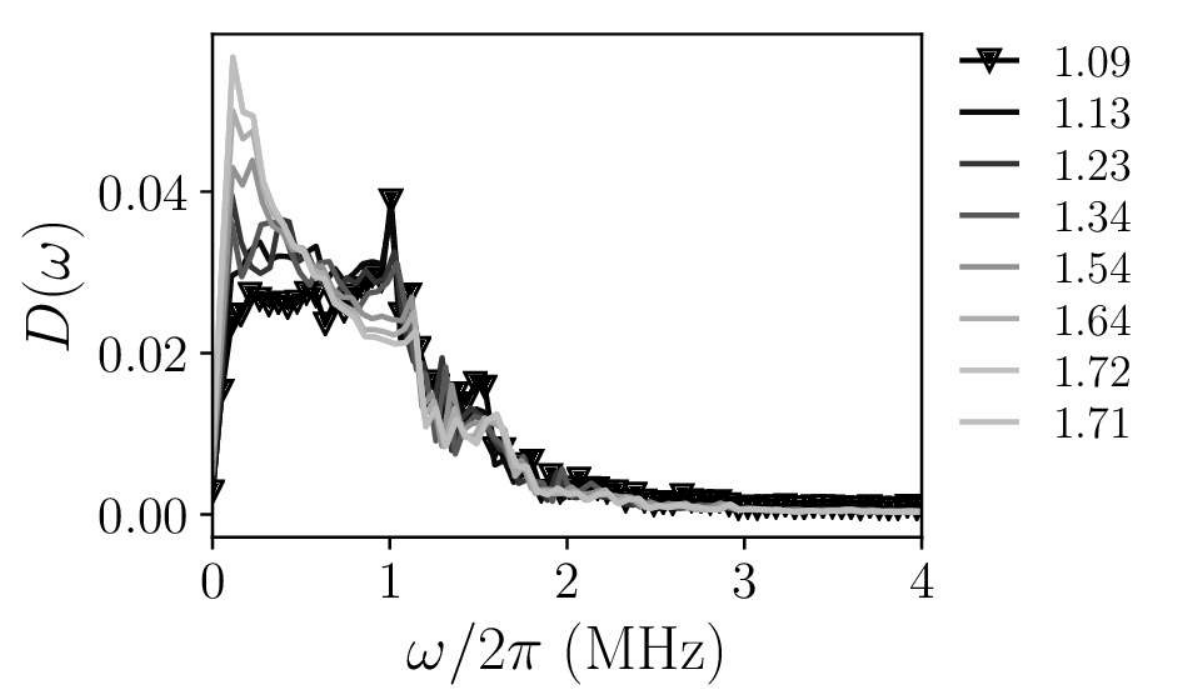}
		\caption{First reload}
		\label{fig:rhoState2ndLoad}
	\end{subfigure}
	\caption{Evolution of density of states of the 33$\times$RV granular column during cyclic oedometric compression. Different levels of axial strain $\varepsilon_a$ in percentage are indicated by the different gray levels.}
	\label{fig:diffAnisoRhoState}
\end{figure}

The vibrational density of states (vDOS) provides the distribution (weight) of energy at certain frequencies and vibrational modes.
From the power spectra of longitudinal and transverse particle motion, respectively agitated by the impulse signals in longitudinal and transverse directions (see \secref{sec:evolveDispersion}), the vDOS of the granular column is calculated as $ D(\omega) = \int_{0}^{\infty}(S_l(k,\omega)+S_t(k,\omega))\textnormal{d}k \propto \sum_{\textnormal{d}k}^{N_k\textnormal{d}k}(S_l(k,\omega)+S_t(k,\omega))\d k$, where $ \textnormal{d}k $ and $N_k\textnormal{d}k$ are the minimum and maximum wavenumbers defined by the packing length and particle size.
\figref{fig:diffAnisoRhoState} show the evolution of the vDOS of the granular column during cyclic oedometric compression.
It appears that during the initial load, the low-frequency vibrations are much stronger than the others and the distribution becomes increasingly overpopulated until a constant stress anisotropy is reached.
This is in line with the fact that the scaled dispersion curves do not vary for $ \varepsilon>1.26 \% $, as shown in \figsref{fig:AnisoVp1stLoad} and \ref{fig:AnisoVs1stLoad}.
\figref{fig:rhoState1stUnload} shows that immediately after the load reversal, the vDOS of the granular system remains almost the same as the one preceding the reversal.
As the unload proceeds, the vDOS qualitatively changes with the low-frequency peak diminishing, and the secondary peaks growing at $ k/2\pi \in (1.0, 2.0)$ mm$^{-1}$.
Similar to the collapsing dispersion curves in \figsref{fig:AnisoVp1stUnload} and \ref{fig:AnisoVs1stUnload}, the variation of the vDOS quickly becomes negligible, for $ \varepsilon_a<1.61\%$.
The trend as in \figref{fig:rhoState1stUnload} is reversed in \figsref{fig:rhoState2ndLoad}, with the vDOS at the end of the reload returning to the one at the end of the initial load, both of which have a similar level of stress and anisotropy.
From the evolution of the vDOS during cyclic compression, it can be deduced that for newly deposited granular media that do not have a history of precompression, low-period seismic waves are easer to propagate than fast oscillations, whereas precompressed granular media are more prone to high-frequency waves.

\section{Conclusions}
\label{sec:conclude}

In this paper, a DEM representative volume previous calibrated against oedometric experiments as obtained from 3DXRCT images \cite{Cheng2018b}is probed with small perturbations.
The numerical probes are performed with a wide range of magnitudes, along both forward and backward loading directions.
It is found that the backward perturbation retains bigger elastic-regimes along the forward perturbation, particularly for the oedometric moduli.
With increasing mean and deviatoric stresses during the initial oedometric loading, the elastic regimes of all small-strain moduli clearly shrink towards the minimum perturbation magnitude, whereas the sizes of the elastic regimes remain unchanged during the unloading-reloading cycles.
It is interesting to note that after each load reversal, the small-strain moduli drop dramatically with a slight increase of perturbation magnitude, which is very similar to the modulus degradation right before the reversal.
This could be macroscopically related to cyclic mobility and instability at yield surface.

For the dynamic probes, four different granular columns that have different lengths in the propagation direction are compared.
The P- and S-wave velocities obtained from the frequency domain analyses show a clear dependence on the source-receiver distance and seem to converge towards the long-wavelength limits.
With a variety of input signals, the wave velocities show dependency not only on source-receiver distance, but also on the input waveform and frequency.
The dependence of wave velocities on the travel distance are seemingly attributed to the interplay between nonlinear elasticity at contacts and dispersion effects of an initial sharp pulse.

Interestingly, the dispersion branches obtained from the frequency domain analyses exhibit almost no dependence on the packing length and input signals considered.
The dispersion relations of the granular column at stress states on the cyclic oedometric path are identified by applying the Lorentzian fitting to the power spectra.
The wave velocities at the smallest wavenumber are extracted from the dispersion relations and compared to those obtained from the time domain and the static probes.
The appropriate input signal, which gives stagnant wave velocity profiles along the granular column, is selected for the time domain analysis, so that the time-domain and frequency-domain estimations match well.
Several remarks relevant to the accuracy and efficiency of dynamic probing are:

\begin{itemize}
	\item High-frequency cosine (smooth) input signals lead to more accurate estimations of wave velocities in time domain analyses than sine signals (non-smooth).
	\item Stress history can qualitatively affect the dispersion properties of the granular material, particularly in the high wavenumber regime.
	\item As the granular column undergoes oedometric unloading-reloading cycles, dispersion relations normalized with the long-wavelength limits become increasingly identical, which means the dispersive relations are scalable, independent of stress history and anisotropy.
	\item Processing the DEM simulation results of dynamic probing in the frequency domain can give frequency-dependent wave velocities at much lower computational cost than in the time domain.
	\item Choosing input signals such that the wave velocities received at the end of a granular column approach the long-wavelength limits identified from the dispersion relations, can facilitate the determination of inserted signals for bender element experiments.
\end{itemize}

Because the calibration procedure and a posterior probability distribution of micromechanical parameters is already available from the Bayesian calibration \cite{Cheng2018b}, it would be interesting to see how the uncertainties at the microscale affect the dispersion relations at various wavelengths.
Future work also involves using existing continuum models to better approximate the dispersion branches obtained from the DEM simulations, so as to accurately capture the change of dispersion relations with respect to stress history and anisotropy in the materials.

\section*{References}

\bibliography{mybibfile}

\end{document}